\documentclass[iop]{emulateapj}
\shorttitle{Outburst of V899 Mon}
\shortauthors{Ninan et al.}

\begin{document}

\title{V899 Mon: An Outbursting Protostar With Peculiar Light Curve And Its Transition Phases}

\author{J. P. Ninan, D. K. Ojha, T. Baug}
\affil{Department of Astronomy and Astrophysics, Tata Institute of Fundamental Research, 
Homi Bhabha Road, Colaba, Mumbai 400 005, India}
\email{ninan@tifr.res.in}
\author{B. C. Bhatt}
\affil{Indian Institute of Astrophysics, Koramangala, Bangalore 560 034, India}
\author{V. Mohan}
\affil{Inter-University Centre for Astronomy and Astrophysics, Pune 411 007, India}
\author{S. K. Ghosh}
\affil{National Centre for Radio Astrophysics, Tata Institute of Fundamental Research, 
Pune 411 007, India}
\author{A. Men'shchikov}
\affil{Laboratoire AIM, CEA/DSM-CNRS-Universit\'{e} Paris Diderot, IRFU/SAp, CEA Saclay, Orme des Merisiers, 91191 Gif-sur-Yvette, France}
\author{G. C. Anupama}
\affil{Indian Institute of Astrophysics, Koramangala, Bangalore 560 034, India}
\author{M. Tamura}
\affil{National Astronomical Observatory of Japan, Mitaka, Tokyo 181-8588, Japan}
\and
\author{Th. Henning}
\affil{Max-Planck-Institute for Astronomy, K\"onigstuhl 17, 69117 Heidelberg, Germany}

\begin{abstract}

We present a detailed study of V899 Mon (a new member in the FUors/EXors family of young low-mass stars undergoing outburst), based on our long-term monitoring of the source starting from November 2009 to April 2015. Our optical and near-infrared photometric and spectroscopic monitoring recorded the source transitioning from its first outburst to a short duration quiescence phase ($<$ 1 year), and then returning to a second outburst. We report here the evolution of the outflows from inner region of the disk as the accretion rate evolved in various epochs. Our high resolution (R$\sim$37000) optical spectrum could resolve interesting clumpy structures in the outflow traced by various lines. Change in far-infrared flux was also detected between two outburst epochs. Based on our observations we constrained various stellar and envelope parameters of V899 Mon, as well as the kinematics of its accretion and outflow. The photometric and spectroscopic properties of this source fall between classical FUors and EXors. Our investigation of V899 Mon hints instability associated with magnetospheric accretion to be the physical cause of sudden short duration pause of outburst in 2011. It is also a good candidate to explain similar short duration pauses in outburst of some other FUors/EXors sources.

\end{abstract}



\section{Introduction}

Over the past few decades it is becoming increasingly evident that the process by which low-mass stars accrete gas from disk is an episodic process. Short duration increase in accretion rate capable enough to deliver a substantial fraction of the final stellar mass, has been observed in young sources over a wide spectrum of age from Class 0 to Class III \citep{safron15}. Traditionally, on the basis of light curve and spectrum, these accretion outbursts are classified as FUors (showing decades long outbursts with 4-5 mag change in optical and an absorption line spectrum) and EXors (showing few months-years long outbursts with 2-3 mag change in optical and an emission line spectrum) \citep{herbig77, hartmann96, hartmann98}. These episodic outbursts can possibly solve some of the open issues in star formation like \textquotedblleft Luminosity Problem\textquotedblright \citep{kenyon90,evans09} 
and the origin of knots in the outflows/jets from young stellar objects (YSOs) \citep{ioannidis12}. Recent discovery of these outbursts resulting in silicate crystallization in EX Lup shows the importance of this phenomenon in planet and comet formation \citep{abraham09}. Episodic accretion can also significantly change the pre-main sequence isochrones used extensively for initial mass function studies and age/mass estimation of YSOs \citep{baraffe12}.  From the number statistics it is estimated that every low-mass star  undergo $\sim$ 50 such short duration outbursts during its formation stage \citep{scholz13}. But the short duration of outbursts (months/years) with respect to million year timescale of formation, makes these events extremely rare to detect in star-forming regions. Various models based on instabilities have been proposed to explain this episodic nature of accretion, but so far every newly discovered FUors/EXors pose a new challenge for the models to explain the observed light curves.

We report here our long-term monitoring of a very peculiar outburst source \object[V899 Mon]{\textbf{V899 Monoceros}}, which gives significant insight into the mechanism of triggering and quenching of this class of outbursts. A possible FUor type eruption of \object[V899 Mon]{V899 Mon} (a.k.a. \object[IRAS 06068-0641]{IRAS 06068-0641}) located near the Monoceros R2 region (d$\sim$905 pc; \citet{lombardi11}) was first discovered by Catalina Real-Time Transient Survey (CRTS) and reported by \citet{wils09}. They announced the source as a FUor candidate based on the constant brightening it has been undergoing since 2005. The spectrum published by \citet{wils09} showed strong H$\alpha$ and Ca II IR triplet lines which identify the outbursting source as a YSO.

We started our long-term multi-wavelength observations of this source in 2009 and could track the source undergoing transition from the first outburst phase to quiescent phase, and finally returning to a second outburst within a year of quiescence. Our multi-wavelength data provides a consistent picture on the physical processes which might have led to these sudden transitions in the accretion rate of V899 Mon.

In this paper, we present the results of our long-term observations of V899 Mon. In Section \ref{Observations} we describe observational details and data reduction procedures of photometric and spectroscopic data. In Section \ref{results} we present our results and deduction procedures. In Section \ref{discuss} we discuss possible physical reasons for the short duration of quiescence phase between the first and second outbursts. We conclude in Section \ref{conclusion} with major results we have obtained on V899 Mon, as well as its implications in understanding the general FUors/EXors outburst phenomenon.

\section{Observations} \label{Observations}

\subsection{Optical Photometry} \label{OpticPhoto}
Our long-term optical monitoring of V899 Mon started on 2009 November 30. The observations were carried out using the  2-m Himalayan \textit{Chandra} Telescope (HCT) at Indian Astronomical Observatory, Hanle (Ladakh), belonging to Indian Institute of Astrophysics (IIA), India and the 2-m telescope at IUCAA (Inter-University Centre for Astronomy and Astrophysics) Girawali Observatory (IGO), Girawali (Pune), India. For optical photometry, the  central 2K $\times$ 2K CCD section of Himalaya Faint Object Spectrograph \& Camera (HFOSC) with a pixel scale of 0.296\arcsec\,  was used on the HCT and the 2K $\times$ 2K CCD of IUCAA Faint Object Spectrograph \& Camera (IFOSC) with a similar pixel scale of 0.3\arcsec\, was used on the IGO. This gave us a field of view (FoV) of $\sim 10 \times 10$ arcmin$^{2}$ on both HCT and IGO. 
 A detailed description of the instruments and telescopes are available at IAO\footnote{\url{http://www.iiap.res.in/iao/hfosc.html}} and IGO\footnote{\url{http://www.iucaa.ernet.in/\~itp/igoweb/igo\_tele\_and\_inst.htm}} websites. 
The \textbf{V899 Mon's} field $((\alpha,\delta)_{2000} = 06^h09^m19^s.28, -06^{\circ}41\arcmin\,55\arcsec\,.4)$  was observed in standard $UBVRI$ Bessel filters.
We carried out observations for 69 nights, out of which 41 nights were observed through HCT and the remaining ones from IGO.
The photometric observation log is given in Table \ref{table:Obs_Log}. Only a portion of the table is provided here. The complete table is available in machine-readable form in the online journal.

\begin{figure*}
 \includegraphics[width=0.9\textwidth]{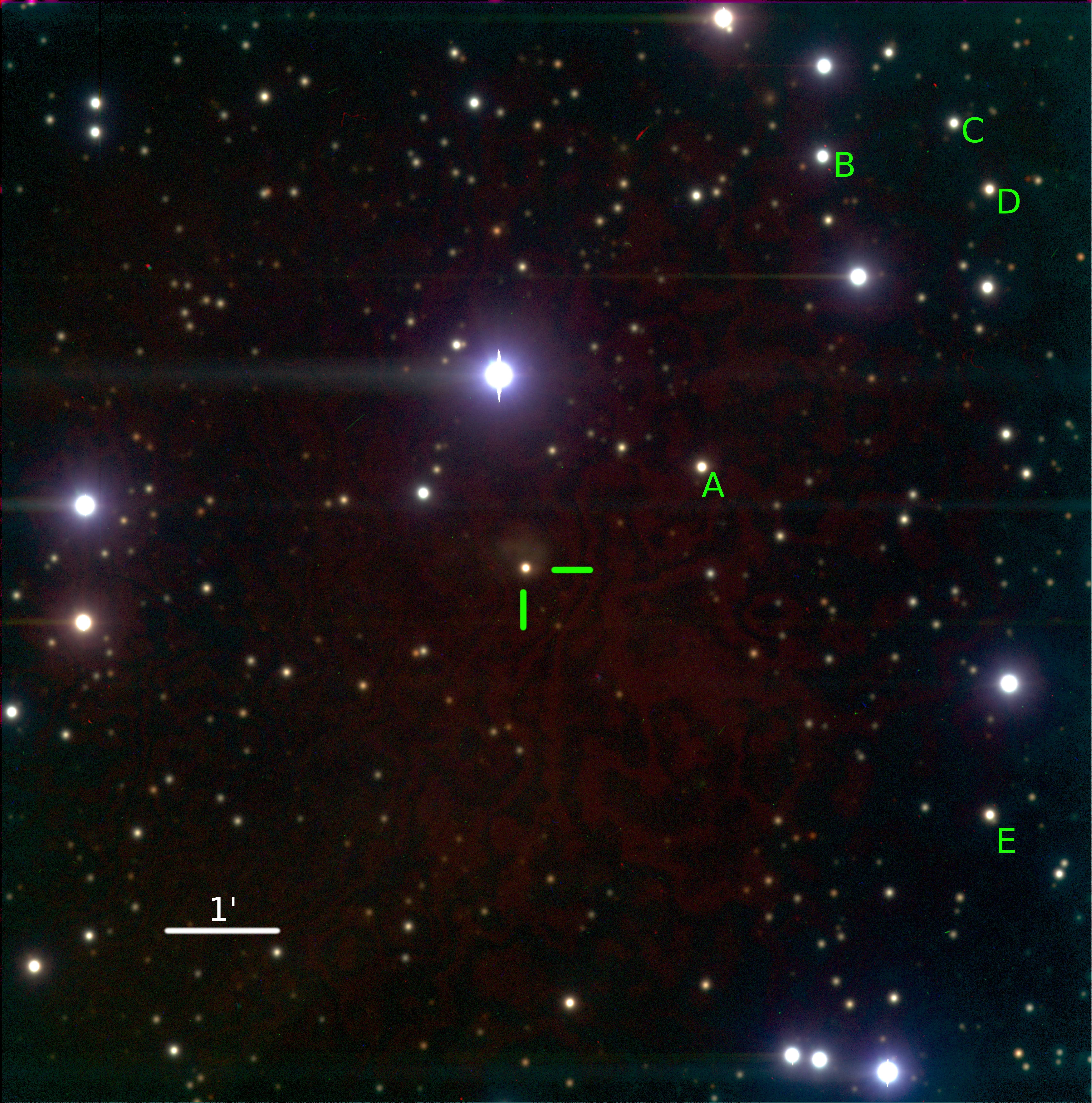}
 \caption{RGB color composite optical image of V899 Mon taken with HFOSC  from HCT ($V$: blue; $R$: green; $I$: red) on 2011 January 2. V899 Mon was in its short duration quiescent phase at this epoch. The FoV is $\sim$ 10 $\times$ 10 arcmin$^2$. North is up and east is to the left-hand side. Stars marked as A, B, C, D and E are the secondary standard stars used for magnitude calibration. Location of V899 Mon is marked with two perpendicular lines at the middle.}
 \label{img:opticalfield}
 \end{figure*}

The  standard photometric data reduction steps like bias-subtraction, median flat-fielding and finally aperture photometry of V899 Mon were done using our publicly released\footnote{\url{https://indiajoe.github.io/OpticalPhotoSpecPipeline/}} pipeline for these instruments. Since V899 Mon source is quite bright and is not affected by any bright nebulosity, we used 4 $\times$ full width half maximum (FWHM) aperture for optical photometry. The background sky was estimated from a ring outside the aperture radius with a width of 4.5\arcsec . Magnitude calibration was done by solving color transformation equations for each night using Landolt's standard star field SA 98 \citep{landolt92} and the five field stars (see Figure \ref{img:opticalfield}) that we identified to have a stable magnitude over the period of our observations.

\subsection{NIR Photometry} \label{NirPhoto}
Near-infrared (NIR) photometric monitoring of the source in \textit{J, H} and \textit{K / K$_S$} bands were carried out using HCT NIR camera (NIRCAM), TIFR Near Infrared Spectrometer and Imager (TIRSPEC) mounted on HCT, and TIFR Near Infrared Imaging Camera-II (TIRCAM2) mounted on IGO telescope. NIRCAM has a $512 \times 512$ Mercury Cadmium Telluride (HgCdTe) array, with a pixel size of 18 $\mu m$, which gives a FoV of $\sim$ $3.6 \times 3.6$ arcmin$^{2}$ on HCT. Filters used for observations were $J$ ($\lambda_{center}$= 1.28 $\mu m$, $\Delta\lambda$= 0.28 $\mu m$), $H$ ($\lambda_{center}$= 1.66 $\mu m$, $\Delta\lambda$= 0.33 $\mu m$) and $K$ ($\lambda_{center}$= 2.22 $\mu m$, $\Delta\lambda$= 0.38 $\mu m$). Further details of the instrument are available at \url{http://www.iiap.res.in/iao/nir.html}. TIRSPEC has a $1024 \times 1024$ Hawaii-1 PACE\footnote{HgCdTe Astronomy Wide Area Infrared Imager -1, Pro
ducible Alternative to CdTe for Epitaxy} (HgCdTe) array, with a pixel size of 18 $\mu m$, which gives a FoV of $\sim$ $5 \times 5$ arcmin$^{2}$ on HCT. Filters used for observations were $J$ ($\lambda_{center}$= 1.25 $\mu m$, $\Delta\lambda$= 0.16 $\mu m$), $H$ ($\lambda_{center}$= 1.635 $\mu m$, $\Delta\lambda$= 0.29 $\mu m$) and $K_S$ ($\lambda_{center}$= 2.145 $\mu m$, $\Delta\lambda$= 0.31 $\mu m$) (Mauna Kea Observatories Near-Infrared filter system). Further details of TIRSPEC are available in \citet{ninan14,ojha12}. 
TIRCAM2 has a $512 \times 512$ Indium Antimonide (InSb) array with a pixel size of 27 $\mu m$. 
Filters used for observations were $J$ ($\lambda_{center}$= 1.20 $\mu m$, $\Delta\lambda$= 0.36 $\mu m$), $H$ ($\lambda_{center}$= 1.66 $\mu m$, $\Delta\lambda$= 0.30 $\mu m$) and $K$ ($\lambda_{center}$= 2.19 $\mu m$, $\Delta\lambda$= 0.40 $\mu m$). More details of TIRCAM2 are available in \citet{naik12}. The log of NIR observations is listed in Table \ref{table:Obs_Log}. We have photometric observations for a total of 25 nights in the NIR. Our first \textit{JHK} observation was carried out during the peak of the first outburst on 2010 February 20, and the remaining observations were carried out during the current ongoing second outburst.

\begin{figure*}
 \includegraphics[width=0.9\textwidth]{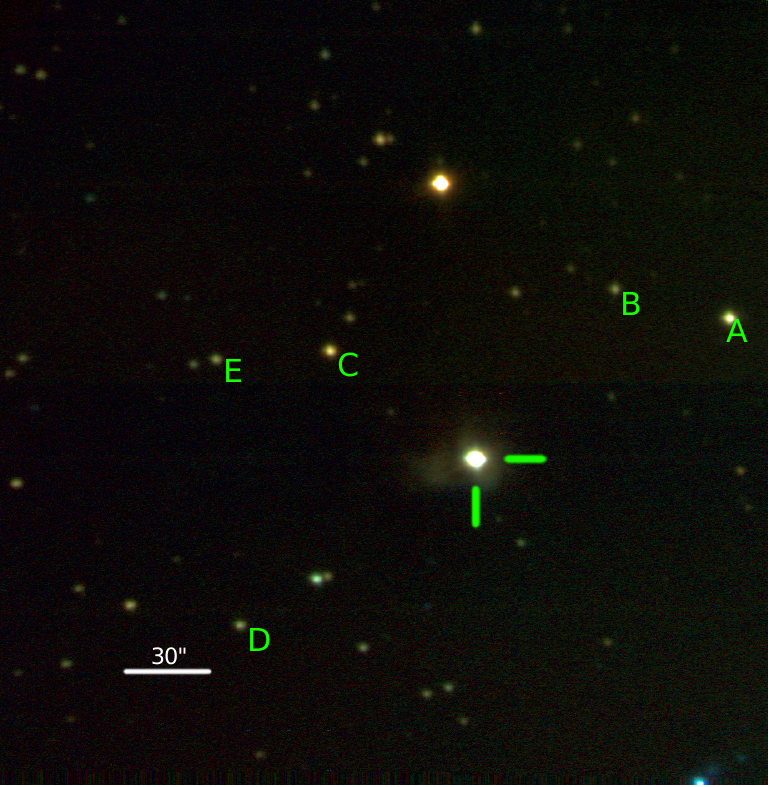}
 \caption{RGB color composite NIR image of V899 Mon taken with TIRSPEC from  HCT ($J$: blue; $H$: green; $K_S$: red) on 2013 November 14. V899 Mon was in its second outburst phase at this epoch. The FoV is $\sim$ 5 $\times$ 5 arcmin$^2$. North is up and east is to the left-hand side. Stars marked as A, B, C, D and E are the secondary standard stars used for magnitude calibration. Location of V899 Mon is marked with two perpendicular lines at the middle.}
 \label{img:nirfield}
 \end{figure*}

 NIR observations were carried out by the standard telescope dithering procedure. Five dither positions were observed. 
 All the dithered object frames were median combined to generate master sky frames for NIRCAM and TIRCAM2. We combined twilight flats and all non-extended source frames observed during the same night to create accurate flats for each night. For NIRCAM and TIRCAM2, data reduction and final photometry were performed using the standard tasks in IRAF, while the TIRSPEC data were reduced using our TIRSPEC photometry pipeline \citep{ninan14}. 
For aperture photometry we used an aperture of 3$\times$FWHM and the background sky was estimated from an annular ring outside the aperture radius with a width of 4.5\arcsec .
 Magnitude calibration was done by solving color transformation equations for each night using Hunt's standard star fields AS 13 and AS 9 \citep{hunt98}, and the five field stars (labeled in Figure \ref{img:nirfield}) that we identified to have a stable magnitude and are consistent with 2MASS\footnote{Two Micron All Sky Survey} over the period of our observations.

\subsection{Medium Resolution Optical Spectroscopy} \label{OpticSpec}
Our medium resolution ($R \sim 1000$) optical spectroscopic monitoring of V899 Mon also started on 2009 November 30. The spectroscopic observations were carried out using both HCT and IGO. The full 2K $\times$ 4K section of HFOSC CCD spectrograph  was used for HCT observations, and 2K $\times$ 2K IFOSC CCD spectrograph was used for IGO observations.
These observations were done in the effective wavelength range of $3700 - 9000$  $\mathring{A}$, using grism 7 (center wavelength 5300 $\mathring{A}$) and grism 8 (center wavelength 7200 $\mathring{A}$). The spectral resolution obtained for grism 7 and grism 8 with 150 $\mu m$ slit at IGO and 167 $\mu m$ slit at HCT was $\sim 7$ $\mathring{A}$.
The log of spectroscopic observations is listed in Table \ref{table:Obs_Log}. We have 8 spectra from the first outburst phase, 4 from the short quiescent phase and 30 from the ongoing second outburst phase.

For wavelength calibration, we have used  FeNe, FeAr and HeCu lamps. Standard IRAF tasks like \textit{APALL} and \textit{APSUM} were used for spectral reduction using our publicly released PyRAF\footnote{PyRAF is a product of the Space Telescope Science Institute, which is operated by AURA for NASA} based pipeline developed for both the HFOSC and IFOSC instruments. 

\subsection{High Resolution Optical Spectroscopy} \label{HRSSpec}
We acquired a high resolution ($R \sim 37000$) spectrum of V899 Mon during its second outburst phase on 2014 December 22 using the Southern African Large Telescope
High Resolution Spectrograph (SALT HRS) \citep{bramall10}. Both red arm ($5490 - 8810$ $\mathring{A}$) as well as blue arm ($3674 - 5490$ $\mathring{A}$) of the HRS instrument were used, to take a single exposure of 3170 seconds. 
For this medium resolution instrument mode of HRS, it uses a 2.23\arcsec\, fiber to collect light from the target source and another fiber of the same diameter is used to sample a nearby patch of the sky. Apart from target frames, ThAr arc lamp spectrum was also taken through the sky fiber for wavelength calibration.

All the SALT HRS data reduction was done by writing a reduction tool in Python, making use of \textit{scikit-image} \citep{vanderWalt14}, \textit{scipy} \citep{jones01}, \textit{numpy} \citep{vanderWalt11} and \textit{astropy} \citep{astropy13}. The additive factor to translate observed velocities in spectrum to heliocentric velocities was found to be $+$0.92 km/s using \textit{rvcorrect} task in IRAF.

\subsection{NIR Spectroscopy} \label{NirSpec}
NIR (1.02 $\mu m$ to 2.35 $\mu m$) spectroscopic monitoring of V899 Mon started on 2013 September 25 using TIRSPEC mounted on HCT. Depending on the  seeing conditions, slits with 1.48$\arcsec$ or 1.97$\arcsec$ widths were used. 
Spectra were taken in at least two dithered positions inside the slit. Spectra of Argon lamp for wavelength calibration and Tungsten lamp for continuum flat were also taken immediately after each observation without moving any of the filter wheels. For telluric line correction, bright NIR spectroscopic standard stars within a few degrees and similar airmasses were observed immediately after observing the source. Typical spectral resolution obtained in our spectra is R $\sim$ 1200.
The log of spectroscopic observations is listed in Table \ref{table:Obs_Log}.  It may be noted that all the 12 NIR spectral observations are taken during the  current ongoing second outburst phase.

NIR spectral data was reduced using TIRSPEC pipeline \citep{ninan14}. After wavelength calibration, the spectrum  of V899 Mon was divided with a standard star spectrum to remove telluric lines and detector fringes seen in $H$ and $K$ orders' spectra. 
This continuum-corrected spectrum was scaled using the flux estimated from photometry to obtain the flux-calibrated spectrum. Since we do not have $Y$-band photometry, we interpolated $I$-band and $J$-band $\lambda f_\lambda$ flux to $Y$-band and used the interpolated flux to scale $Y$-band spectrum flux.

\subsection{GMRT Radio Continuum Imaging} \label{GmrtImag}
Continuum interferometric observation of V899 Mon at 1280 MHz with 33.3 MHz bandwidth was carried out on 2014 October 17 using Giant Metrewave Radio Telescope (GMRT), Pune, India. GMRT consists of 30 dish antennas (45 m diameter each) in hybrid ``Y'' configuration \citep{swarup91}. Out of these, 25 antennas were online during our observation.  The standard flux calibrator 3C147 was observed for 15 minutes at the beginning and end of the observations. For phase calibration, nearby Very Large Array calibrator 0607-085 was observed for 10 minutes after every 20 minutes of integration on V899 Mon. The total integration on V899 Mon source was 4.5 hours.

\textit{CASAPY} software was used for the data reduction. After careful iterative bad data flagging and gain calibration, 
 we imaged large FoV of 28'$\times$28' field by dividing the field into 128 w-projection planes using the \textit{CLEAN} algorithm.

\subsection{\textit{Herschel} Far-infrared Photometry} \label{HerschelPhoto}
We obtained imaging observations of V899 Mon region in \textit{SPIRE}\footnote{Spectral and Photometric Imaging Receiver} 250, 350 and 500 $\mu m$, and \textit{PACS}\footnote{Photoconductor Array Camera and Spectrometer} 70 and 160 $\mu m$ from \textit{Herschel} data archive\footnote{\textit{Herschel} is an ESA space observatory with science instruments provided by European-led Principal Investigator consortia and with important participation from NASA.}. SPIRE data was available for two epochs, one on 2010 September 4 (Proposal ID: KPGT\_fmotte\_1) and another on 2013 March 16 (Proposal ID: OT1\_rgutermu\_1). PACS data of V899 Mon was only available in the later epoch. Data from both the epochs were reduced using the standard \textit{Herschel's HIPE} pipeline with the latest calibration file. For point source aperture photometry of V899 Mon we first tried {\it sourceExtractorSussextractor} routine in HIPE on level 2 SPIRE data, and {\it multiplePointSourceAperturePhotometry} routine on PACS data. We also tried 
 point spread function (PSF) photometry using \textit{daophot} tool in IRAF. 
 Since all these algorithms use an annular ring for background estimation and the background is highly non-uniform, the results were found to be very sensitive to aperture and background estimates. We finally 
estimated fluxes using \textit{getsources} \citep{menshchikov12} which uses a linearly interpolated background estimate. These estimates were found to be 
consistent with the differential flux between two epochs (differential flux could be estimated more accurately since the subtracted image has a uniform background). The HIPE pipeline assumes a spectral energy distribution (SED) color,  $\alpha=-1$ (where $F_{\nu} = \nu^{\alpha}$). The actual color $\alpha$ of V899 Mon was obtained from our flux estimates, and the corresponding color correction factor was multiplied to the flux values for obtaining color corrected flux estimates.

\subsection{WISE Photometry} \label{WISEPhoto}
We obtained imaging observations of V899 Mon region in W1 (3.4 $\mu m$), W2 (4.6 $\mu m$), W3 (12 $\mu m$) and W4 (22 $\mu m$) bands from \textit{Wide-field Infrared Survey Explorer (WISE)} data archive \citep{wright10}. Photometric magnitude estimates were also available in \textit{WISE All sky catalog}. W1 and W2 bands were observed at two epochs in 2010 March 17 and 2010 September 24; whereas W3 and W4 were observed only on 2010 March 17. 
 Since W1 and W2 intensity maps looked severely saturated, we have used only W3 and W4 WISE catalog magnitudes for our study.

In this paper, we have also compiled all other publicly available archival data of this region at different epochs. Data were obtained from the following surveys and instruments: CRTS DR2\footnote{Catalina Real-Time Transient Survey Data Release 2},  POSS survey\footnote{Palomar Observatory Sky Survey}, USNO-B catalog\footnote{U.S. Naval Observatory B1.0 Catalog}, IRAS survey\footnote{Infrared Astronomical Satellite, Survey}, DENIS survey\footnote{Deep Near Infrared Survey of the Southern Sky}, 2MASS survey\footnote{Two Micron All-Sky Survey}, AKARI\footnote{Akari All-Sky Survey Point Source Catalog}, IRAC\footnote{Infrared Array Camera on Spitzer} and MIPS\footnote{Multi-Band Imaging Photometer on Spitzer}.

\section{Results and Analysis} \label{results}
\subsection{Photometric Results: Light Curves \& Color Variations} \label{Photresults}
 An extended fan-shaped optical reflection nebula is typically seen around FUors. In case of V899 Mon we detect only a small faint fan-shaped reflection nebula towards north 
(see Figure \ref{img:opticalfield}). \textit{U, B, V, R, I, J, H} and \textit{K$_S$} magnitudes of V899 Mon from our long-term continuous monitoring are listed in Table \ref{table:PhotMags}. Typical magnitude errors are less than 0.03 mag. Only a portion of the table is provided here. The complete table is available in machine-readable form in the online journal.

\begin{figure*}
 \includegraphics[width=\textwidth]{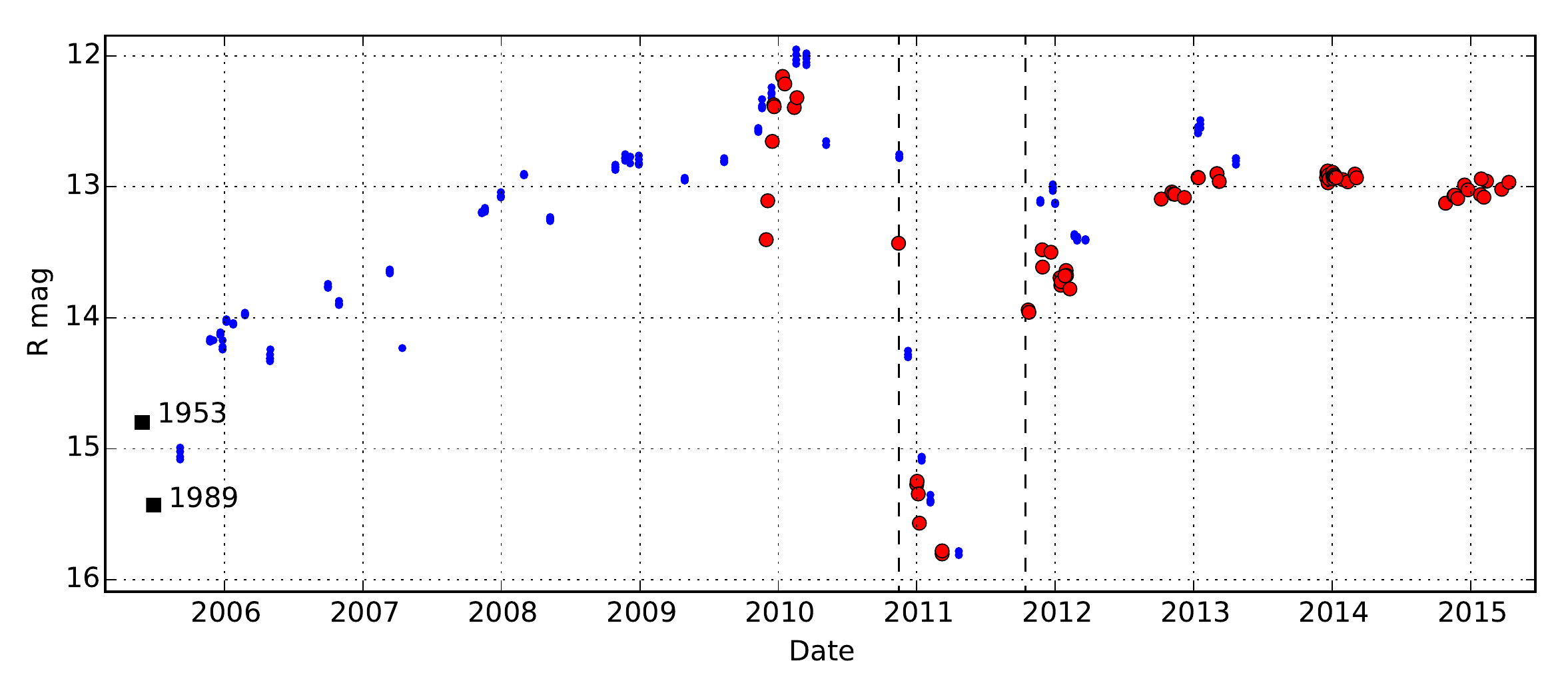}
 \caption{Light curve of V899 Mon showing its first outburst, short quiescent phase and current ongoing second outburst phase. Red circles are $R$-band magnitudes from our observations and blue circles are CRTS magnitudes converted to $R$ magnitudes. POSS-1 and POSS-2 epochs' $R$ magnitudes are also shown for comparison as black squares.}
 \label{LightCurve}
 \end{figure*}

\textit{R}-band magnitude light curve is shown in Figure \ref{LightCurve}. Effective R-band magnitudes\footnote{CRTS DR2 $V_{CSS}$ magnitudes of V899 Mon, were converted to effective $R$ magnitude using color correction equation $V = V_{CSS} + 0.91*(V-R)^2 + 0.04$ and $V-R = 1.11$. \url{http://nesssi.cacr.caltech.edu/DataRelease/FAQ2.html\#improve}} from CRTS survey are over-plotted in the same graph for displaying the complete picture of the initial rise of the first outburst. Images of V899 Mon field observed on 1953 December 29, 1983 December 29, 1989 January 9 and 2000 February 8 are available from digitized POSS-1 and POSS-2 archives. 
Magnitudes at the two epochs (1953 and 1989) which used $R$-band filter are also shown in Figure \ref{LightCurve}. It should be noted that even though these magnitudes available in USNO-B1.0 catalog are corrected for non-standard bandpass filter used in POSS-I survey, their variation seen between 1953 and 1989 might still be due to systematics in calibration.  

Few other outburst sources like V1647 Ori, V582 Aur and V2492 Cyg have also shown such a sudden brief transition to quiescent phase after reaching peak of first outburst \citep{ninan13,semkov13,kospal13}. V582 Aur and V2492 Cyg's light curves were very unstable, but V1647 Ori's light curve was very similar to V899 Mon in terms of the stability during the phase transitions.

One possible scenario which could explain the sudden dimming of the source in 2011 is occultation by a dust clump. 
By taking the dust size parameter R$_V$ and change in extinction A$_V$ as free parameters, we tried to fit the observed change in magnitudes of \textit{V, R, I, J, H} and \textit{K$_S$} bands\footnote{Even though we do not have NIR photometric measurements during quiescent phase in 2011, our quiescent phase \textit{I}-band magnitude matches with DENIS \textit{I} magnitude observed on 1999 January 25, and DENIS \textit{J} and \textit{K} magnitudes of this epoch matches with 2MASS \textit{J} and \textit{K$_S$} magnitudes observed in 1998. Hence in this study, we have taken 2MASS magnitudes as the quiescent phase \textit{J, H} and \textit{K$_S$} magnitudes of the V899 Mon source.} 
between first outburst and quiescent phases (i.e., 2.86,  2.90,  3.06,  2.96,   2.75 and  2.36 $\Delta$ mags respectively in each band). Coefficients in Table 2 of \citet{cardelli89} were used for calculation, and no positive value of R$_V$ could fit the observed changes in magnitudes. Hence we can safely conclude that the quiescent phase in 2011 was due to actual break in outburst and is not due to dust occultation. Even though we have not considered scattered excess blue light (like in UX Ori systems) in above analysis, almost similar magnitude drop ($\sim$ 2.8) in optical and NIR magnitudes to pre-outburst phase makes it unlikely to be a case of dust obscuration. A stronger argument against dust obscuration scenario comes from the observed variation in spectral line fluxes during this transition (see section \ref{specresults}). The relative flux changes observed in various spectral lines cannot be explained by a simple change in extinction.

\begin{figure}
 \includegraphics[width=0.5\textwidth]{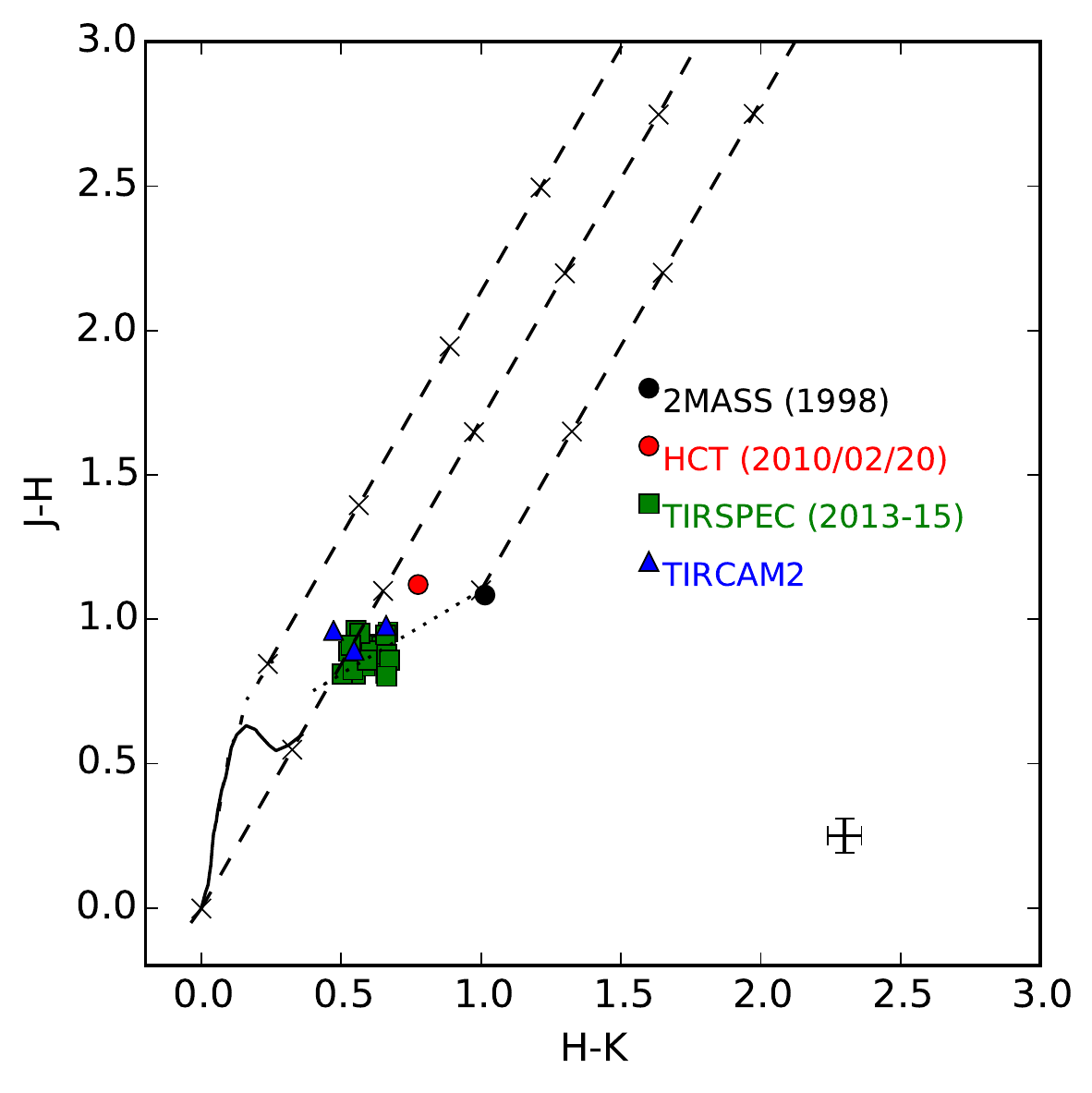}
 \caption{Positions of V899 Mon in the $J$-$H$/$H$-$K$ CC diagram during the quiescent phase (black circle), first outburst (red circle) and the second outburst phase (green squares and blue triangles). The solid curve shows the locus of field dwarfs, and the dash-dotted curve shows the locus of giants \citep{bessell88}. The dotted line represents the locus of classical T-Tauri (CTT)  stars \citep{meyer97}. The diagonal straight dashed lines show the reddening vectors \citep{rieke85}, with crosses denoting an A$_V$ difference of 5 mag. Typical error bars of each data point are shown in the bottom right corner.}
 \label{NIRCCdiag}
 \end{figure}

 Figure \ref{NIRCCdiag} shows the NIR \textit{J-H/H-K} color-color (CC) diagram. 
Position of V899 Mon in it shows no significant extinction to the source, and it falls on Classical T Tauri (CTT) locus. The positions in CC diagram during first and second outbursts are also different from each other. But, since we have only one NIR observation from the rapidly varying ending phase of first outburst, it might not be a good representation of the NIR color of first outburst. 

\begin{figure}
 \includegraphics[width=0.5\textwidth]{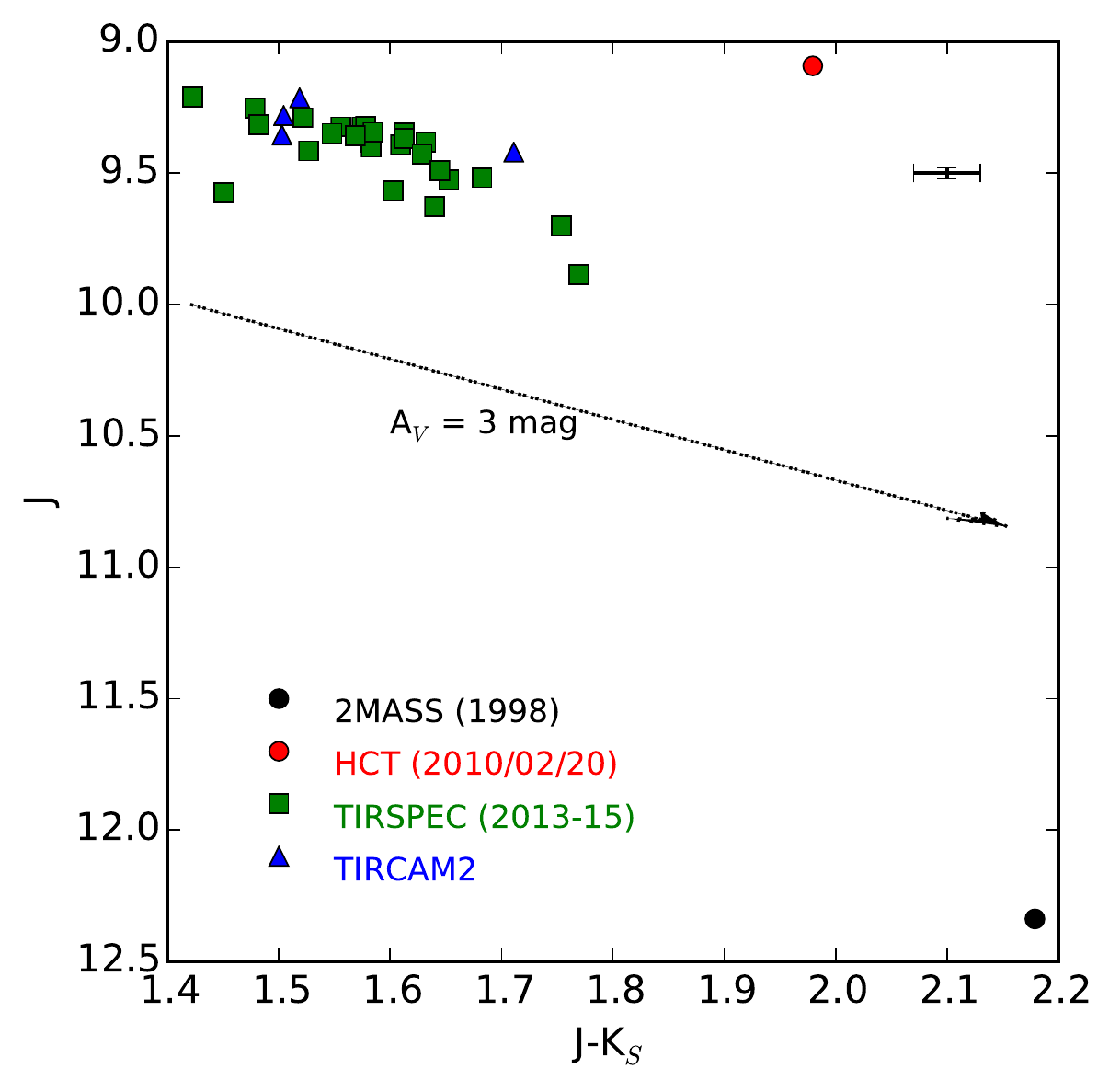}
 \caption{NIR $J$/$J$-$K$ color magnitude diagram showing the movement of V899 Mon during its quiescent (black circle), first outburst phase (red circle) and second outburst phase (green squares and blue triangles). The arrow shows the reddening vector \citep{rieke85}, which corresponds to an A$_V$ of 3 mag. Typical error bars of each data point are shown in the top right corner. }
 \label{NIRCMagdiag}
 \end{figure}

The green squares measured during the 2013 - 2015 period of second outburst show small movements along the reddening vector. Such short period, small amplitude fluctuations during the outburst phase are also seen in other FUor/EXor objects \citep{audard14,semkov13,ninan13}. 
Unlike the quiescence in 2011, these short variations in second outburst could be due to small dust clump occultations or minor fluctuations in the accretion. NIR \textit{J/J-K} color-magnitude diagram (CMD) (see Figure \ref{NIRCMagdiag}) also indicates the general trend of V899 Mon getting redder (bluer) when it dims (brightens).

\subsubsection{Outburst - quiescent transition phase}
\begin{figure*}
 \includegraphics[width=\textwidth]{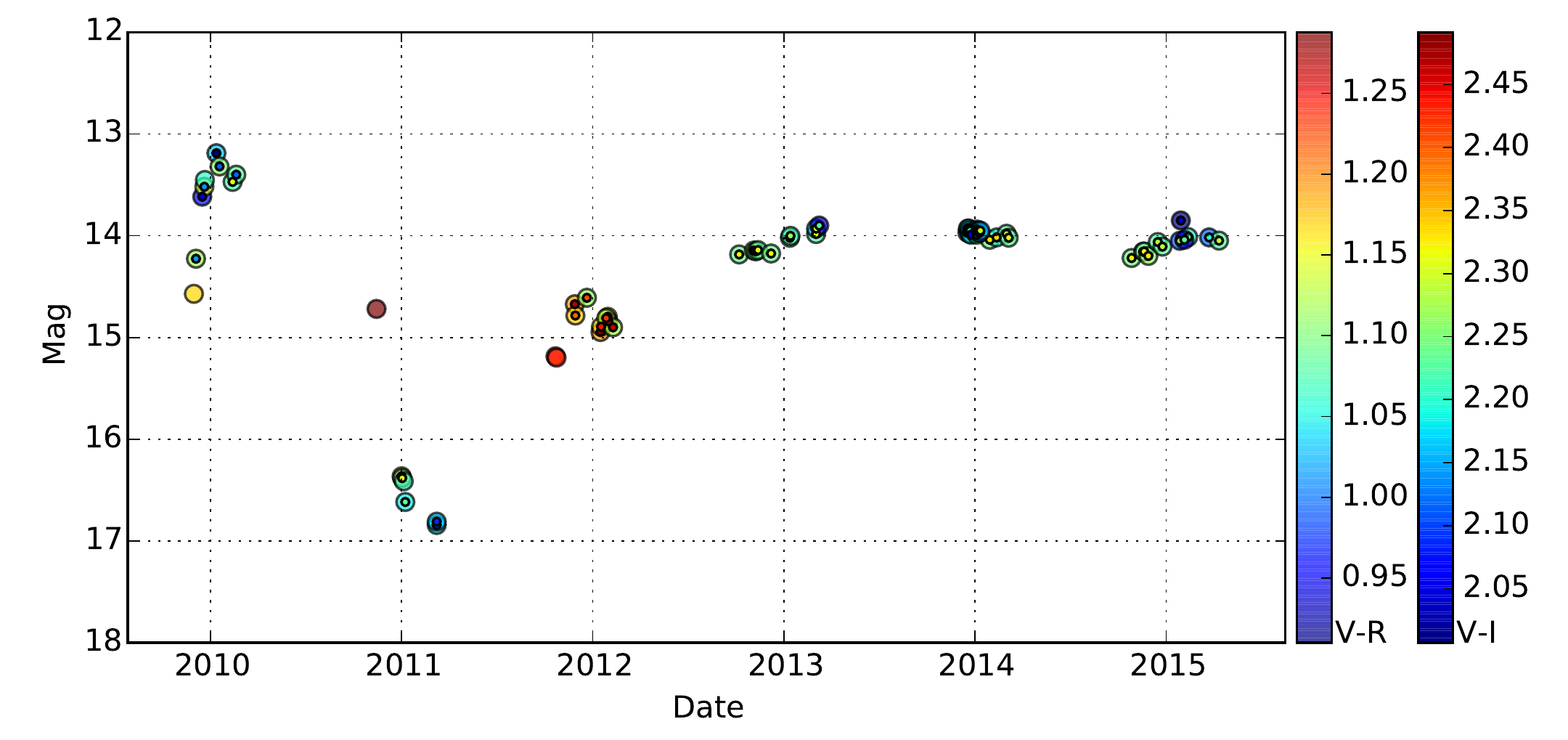}
 \caption{\textit{V}-band light curve of V899 Mon showing evolution of its \textit{V-R} and \textit{V-I} colors. Color of the outer ring shows \textit{V-R} and the color at the center shows \textit{V-I}. We see both the colors were reddest during the transition from the quiescent phase to the outburst phase. }
 \label{MagColorTransition}
 \end{figure*}

\begin{figure*}
 \includegraphics[width=\textwidth]{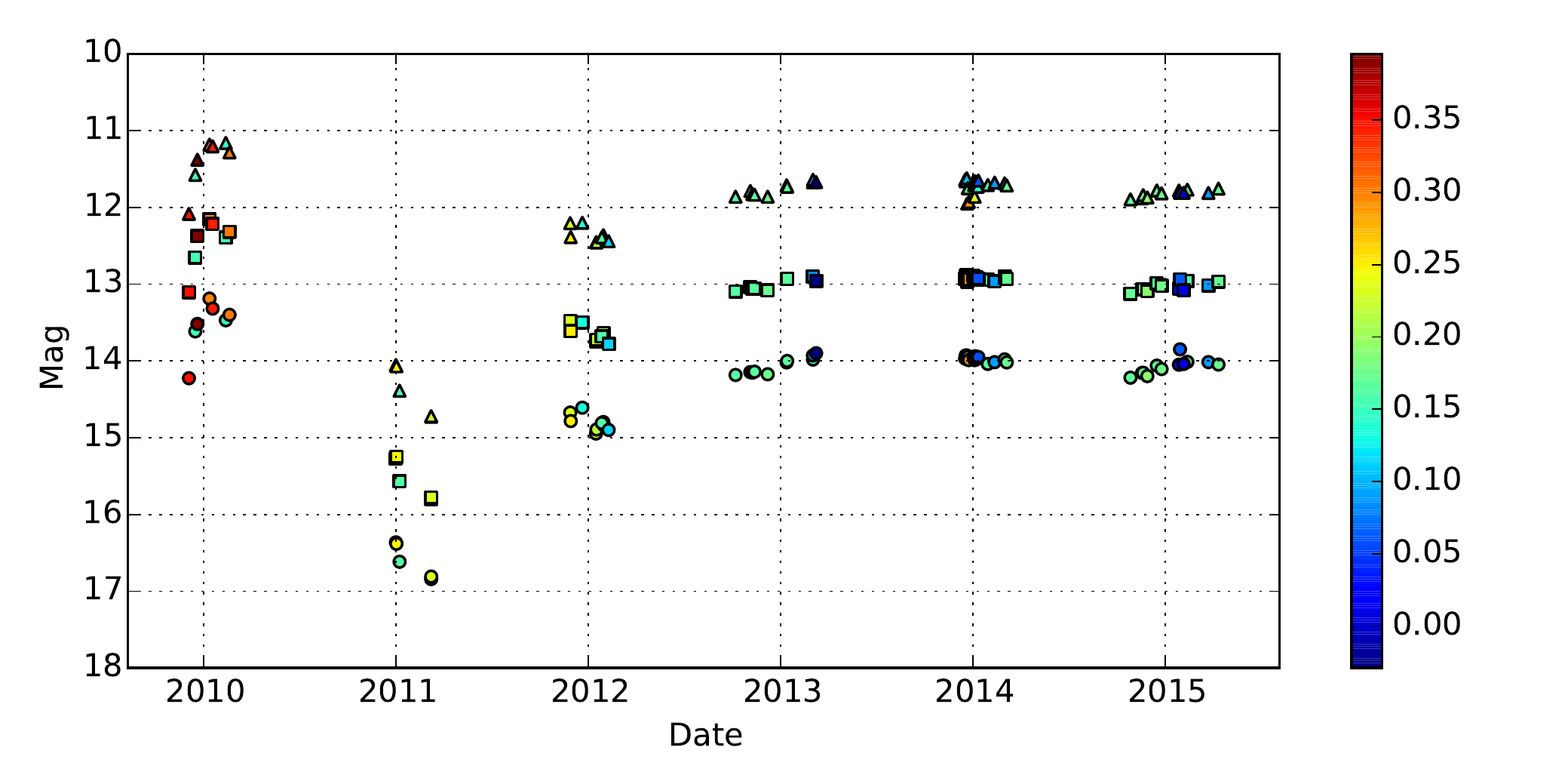}
 \caption{$V$, $R$ and $I$ light curves of V899 Mon (circles, squares and triangles respectively), color-coded with the extinction invariant color C$_{VRI}$.}
 \label{MagCxyz}
 \end{figure*}

In contrast to existing FUor/EXor observations in literature, we could carry out several observations of V899 Mon during its transition from outburst phase to quiescent phase and back.
Figure \ref{MagColorTransition} shows the $V$-band magnitude light curve of the source, color coded with \textit{V-R} color in the outer ring and \textit{V-I} color at the center. These colors indicate that the object was reddest during the transition stages rather than during the outburst or quiescent phases. It could imply the temperature of the extra flux component of the outburst was cooler at the beginning and it became hotter as the outburst reached its peak. Since the transition phase flux was cooler 
than the quiescent phase, outburst flux could not have initially originated on the surface of the hot central star, rather,  might have originated in the disk. This is the first direct observational evidence to support triggering of outburst in the disk of FUors/EXors.
The color change seen in this plot could be a combined effect of both change in extinction and temperature. Figure \ref{MagCxyz} shows the \textit{V, R} and \textit{I} band light curves color-coded  with an extinction invariant quantity. The y-intercept along the reddening vector in any CC diagrams, for example, C$_{VRI}$ = $(V-R)-(R-I)*E(V-R)/E(R-I)$ \citep{mcgehee04}, is a weighted difference of two colors, which is invariant to change in extinction A$_V$. While it is not possible to completely separate out the degenerate SED slope change due to temperature and extinction change, any change in this extinction invariant color implies real change in the intrinsic SED slope (temperature). V899 Mon shows significant change in C$_{VRI}$ between the end of first outburst phase, quiescent phase, transition stage and ongoing second outburst phase. Significant variation is also seen during the peak of first outburst, implying the temperature of the source during those epochs were different than the ongoing second outburst. However, since C$_{VRI}$ is a weighted difference between two colors and could not tell unambiguously whether the intrinsic slope of SED became red or blue, Figure \ref{MagCxyz} only indicates change in the intrinsic slope of SED (temperature). It does not clarify whether reddening of flux during transition stage was due to temperature change alone or was due to the combined effect of a short term increase in extinction and change in temperature.

\subsection{Stellar Properties}
V899 Mon was observed in WISE 12 $\mu m$ and 22 $\mu m$ bands on 2010 March 17, within one month after our NIR observations in $J, H$ and $K$ bands from HCT on 2010 February 20 (first outburst peak). We utilized them to estimate the conventional infrared logarithmic slope of the spectrum $\alpha$  (where $\lambda F_\lambda = \lambda^{\alpha}$) used for age classification of YSOs. During the peak outburst phase (2010 March) we obtained a slope $\alpha = -0.27$ between 2.16 $\mu m$ and 11.56 $\mu m$ and $\alpha = -0.51$ between 11.56 $\mu m$ and 22 $\mu m$. These values classify V899 Mon as a flat-spectrum or an early Class II YSO \citep{greene94} during its peak outburst phase.  

\begin{figure}
 \includegraphics[width=0.5\textwidth]{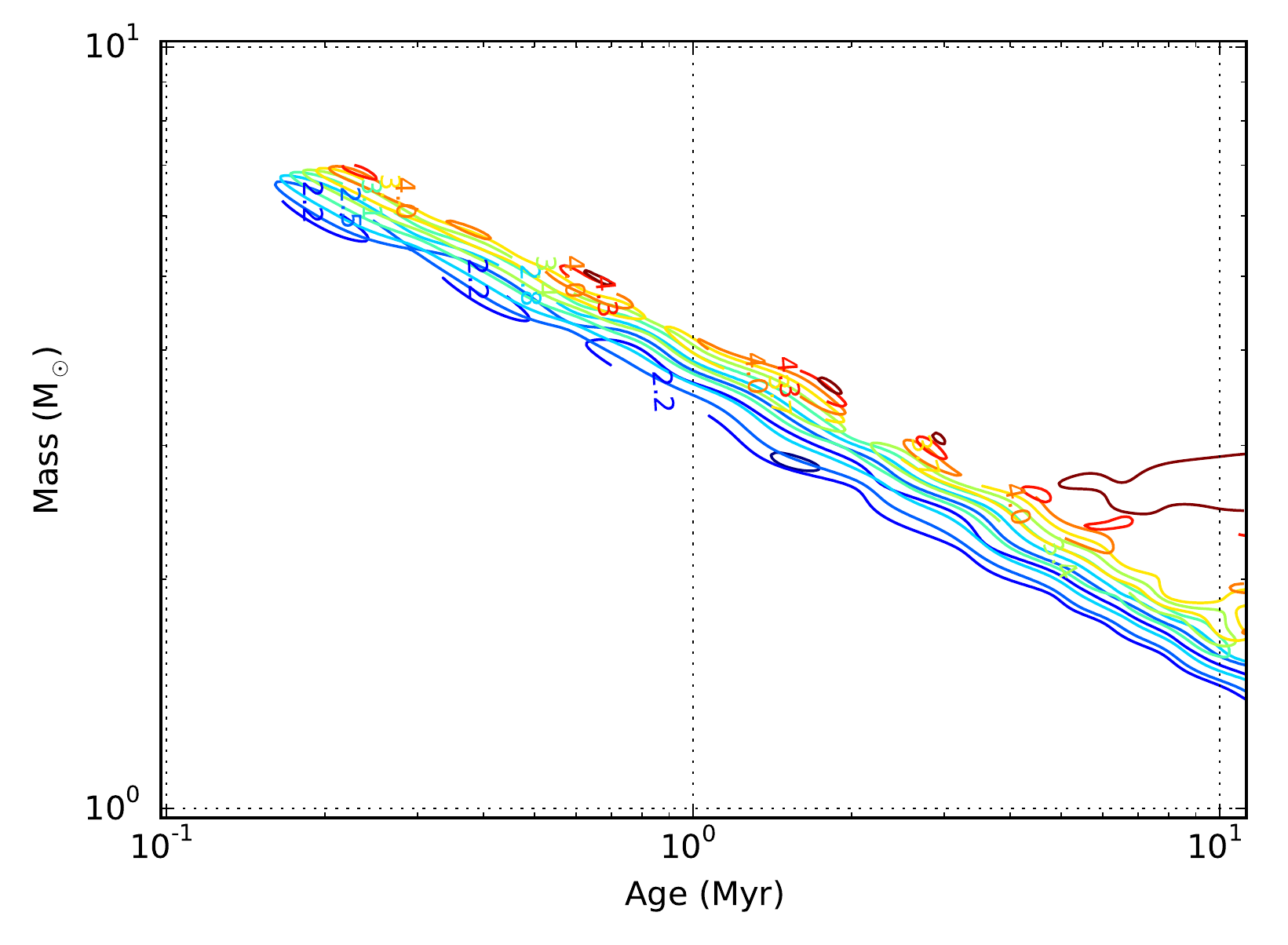}
 \caption{Contours showing 1$\sigma$ constrain on mass and age of V899 Mon assuming \citet{siess00} isochrones to be a valid model for V899 Mon in four dimensional $U$-$B$, $B$-$V$, $V$-$R$ and $R$-$I$ color-color space. Contours corresponding to a range of admissible reddening value A$_V$ are shown in different colors and corresponding A$_V$ is labeled on the contours.}
 \label{MassAgeContour}
 \end{figure}

$U$-$B$, $B$-$V$, $V$-$R$ and $R$-$I$ colors of \citet{siess00} isochrones can be used to constrain photometrically the  mass, age and extinction of the source. This is done by finding a region in mass, age and extinction parameter space in which colors predicted by \citet{siess00} model is consistent with all the observed optical colors of V899 Mon. For this analysis we sampled \citet{siess00} isochrones in its valid age range 0.01 Myr to 100 Myr, and mass range 0.1 to 7 M$_\odot$ in log space. Non-uniform rectangular mass versus age grids were generated for each of the $U$-$B$, $B$-$V$, $V$-$R$ and $R$-$I$ colors by interpolating (using 2D Bi-spline) colors predicted by \citet{siess00} isochrones. $U$-$B$ and $B$-$V$ colors of V899 Mon during second outburst, and $V$-$R$ and $R$-$I$ quiescent phase colors were used for fixing the position of V899 Mon in this four dimensional color space. The errors of the color estimates were taken to be 0.1 mag, which correspond to a symmetric 4D Gaussian in color space. Radial Basis Function (RBF) was used to calculate the distance to the source for each point in the mass-age grid. A contour was plotted for regions within 1$\sigma$ distance from V899 Mon's position. These contours were repeatedly estimated for various values of A$_V$ ranging from 0.5 to 8 mag (see Figure \ref{MassAgeContour}). As seen in the figure, A$_V < 2.2$ mag and A$_V > 4.0$ mag will make all \citet{siess00} isochrone colors incompatible with V899 Mon, and hence can be ruled out. Due to the correlation between age and mass, we obtain tight constrains only for A$_V$ from this analysis. Even though the analysis was done for all possible age ranges, since V899 Mon is a heavy disk accreting source, we expect its position inside contours to be only around the region where age is $\sim$ 1 - 5 Myr.

Another independent way to estimate interstellar extinction is from photometry of field stars. The region around Monoceros R2 was studied in detail using 2MASS data by \citet{lombardi11} (Figure 4 in their paper), and the extinction to V899 Mon location is A$_V$ = 2.6 mag. This value is consistent with our previous color space analysis. 

\begin{figure}
 \includegraphics[width=0.5\textwidth]{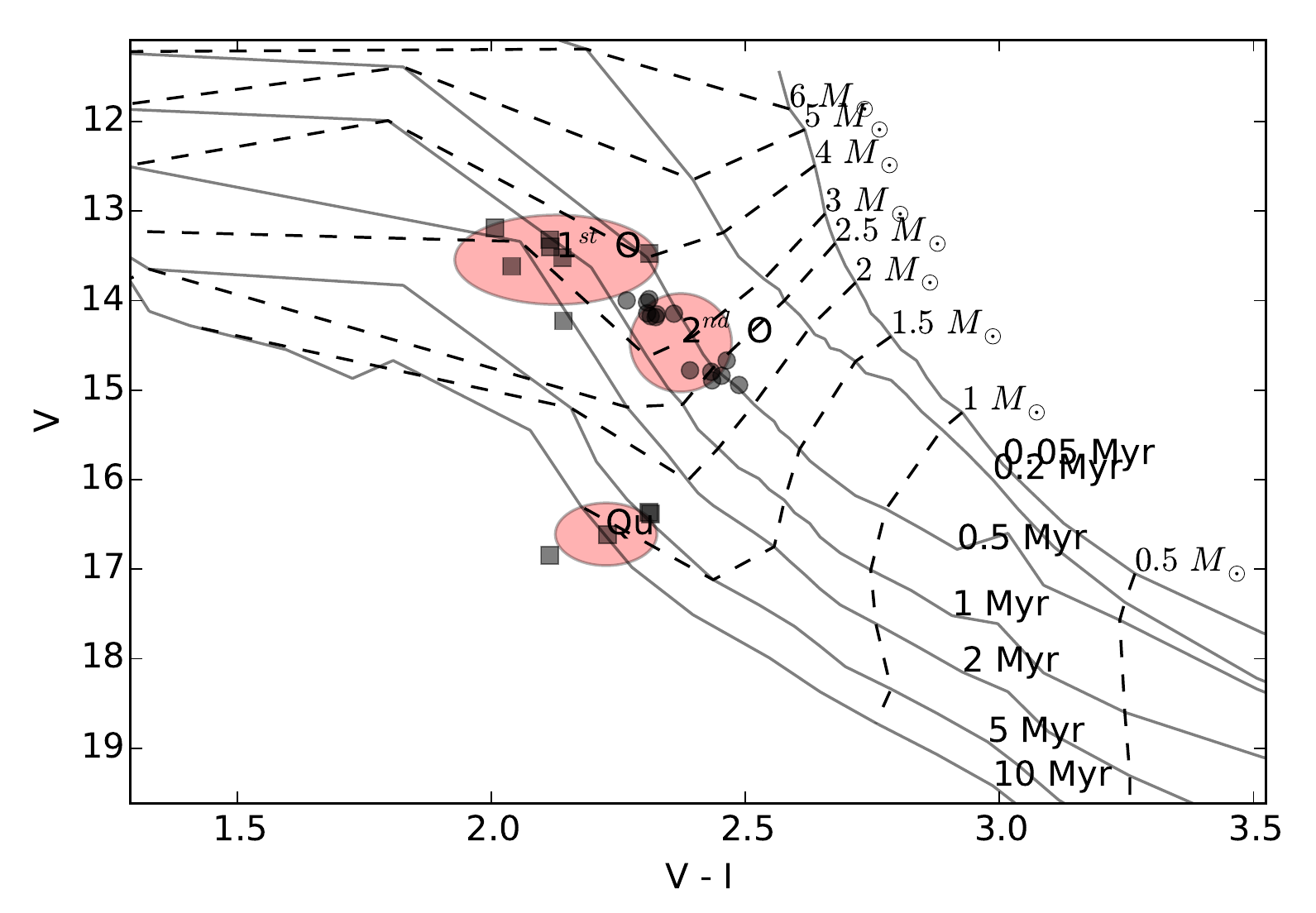}
 \caption{Optical $V$/$V$-$I$ color magnitude diagram of V899 Mon during its first outburst (1$^{st}$ O), second outburst (2$^{nd}$ O) and quiescence (Qu) phases. Solid lines are \citet{siess00} isochrones corresponding to 0.05 Myr, 0.2 Myr, 0.5 Myr, 1 Myr, 2 Myr, 5 Myr and 10 Myr. Dashed lines are evolutionary tracks of masses 0.5 M$_\odot$, 1 M$_\odot$, 1.5 M$_\odot$, 2 M$_\odot$, 2.5 M$_\odot$, 3 M$_\odot$, 4 M$_\odot$, 5 M$_\odot$ and 6 M$_\odot$.  Extinction correction for an A$_V$ = 2.6 mag, and magnitude correction for distance of 905 pc are applied to all isochrones.}
 \label{OpticalCMagdiag}
 \end{figure}

Figure \ref{OpticalCMagdiag} shows the position of V899 Mon in the  $V$/$V$-$I$ optical CMD during various epochs. \citet{siess00} isochrones after correcting for extinction of A$_V$=2.6 mag are also over-plotted. Since the major component of the outburst phase flux is non-stellar in origin, the closest representation of the true position of V899 Mon might be its position during quiescent phase. The scatter forbids the estimation of age, but we get a rough estimate of the mass of the V899 Mon as 2 M$_\odot$. However, it should be noted that this estimate is sensitive to the  estimated interstellar extinction value of A$_V$ = 2.6 mag. 

\subsection{Spectral Energy Distribution} \label{sed}
\begin{figure}
  \centering
  \begin{tabular}[b]{@{}p{0.45\textwidth}@{}}
    \centering\includegraphics[width=\linewidth]{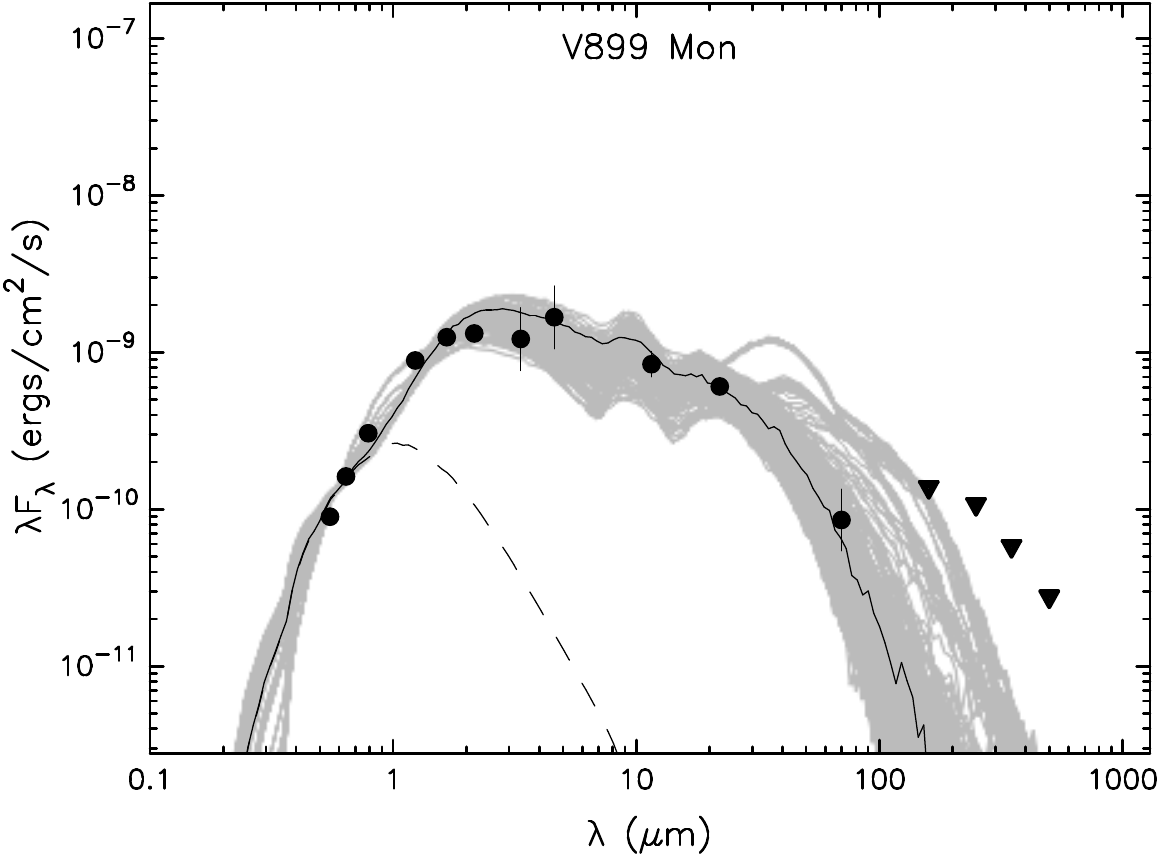} \\
    \centering\small (a) First outburst
  \end{tabular}%
  \quad
  \begin{tabular}[b]{@{}p{0.45\textwidth}@{}}
    \centering\includegraphics[width=\linewidth]{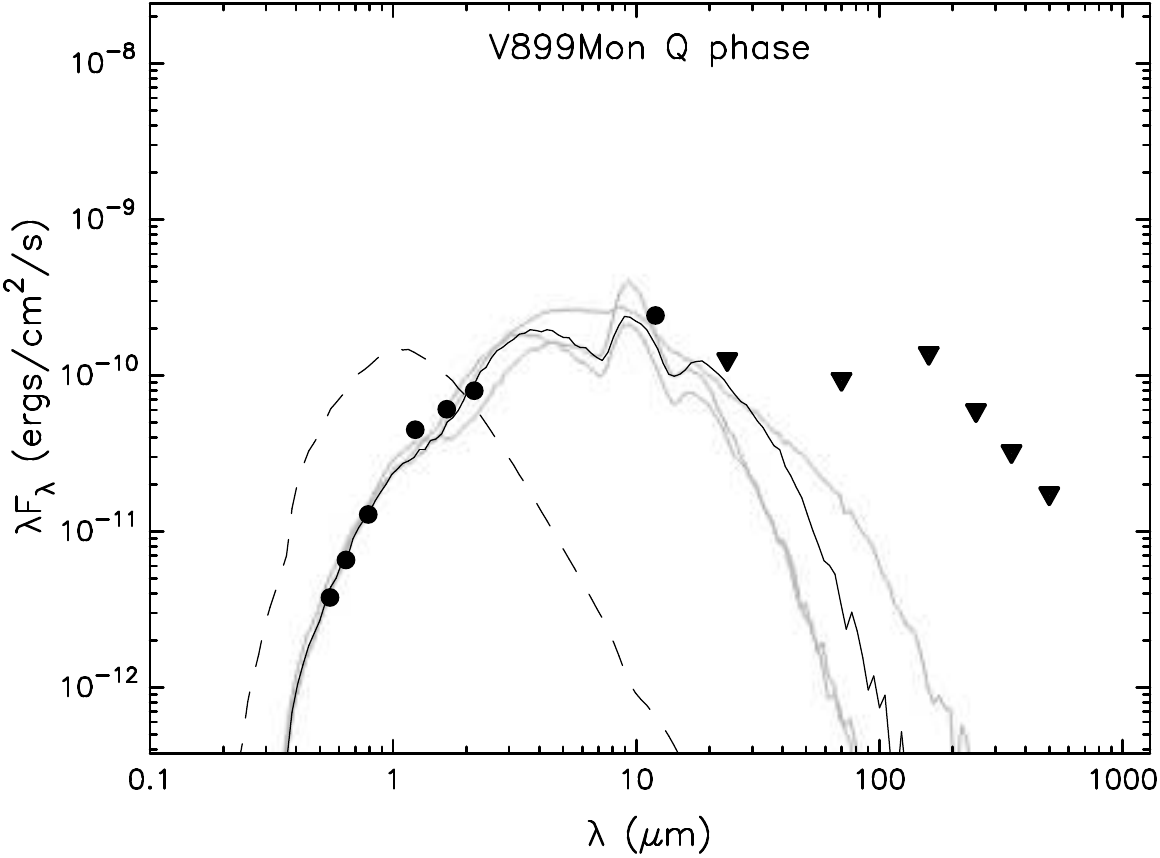} \\
    \centering\small (b) Quiescent phase
  \end{tabular}
  \caption{SED fit of V899 Mon (a) during its peak of the first outburst using \textit{V, R, I, J, H, K$_S$, W1, W2, W3, W4} fluxes from 2010 and PACS 70 $\mu m$ flux from 2013, (b) during its quiescent phase using \textit{V, R, I,} 2MASS \textit{J, H, K$_S$,} and \textit{IRAS 12} fluxes. MIPS1, PACS and SPIRE fluxes were used only as upper limits. The filled circles represent the data points used for the fit and filled triangles are upper limits. The solid black line is the best fitted model and the gray lines show subsequent good fits. The dashed gray line shows the stellar photosphere of V899 Mon in the best fitted model (in the absence of circumstellar dust, but including interstellar extinction).}
  \label{SED1stOutpeak2010Qphase}
\end{figure}

Flux estimates available of the V899 Mon over a wide wavelength range from optical to sub-mm were measured at random epochs. 
Since we had to use multi-wavelength data of nearby epochs for the construction of the SED, we chose the epoch near peak of the first outburst (February-March 2010), during which maximum observations were available.  For fitting YSO SED models, we used the online SED Fitter tool by \citet{robitaille07}.  Flux estimates from both IRAS and PACS show the flux of V899 Mon starts rising in 90 $\mu m$ to 200 $\mu m$ range after a gradual drop of SED in mid-infrared region (see Figure \ref{SED1stOutpeak2010Qphase}). No SED models in \citet{robitaille07} could fit this double peaked SED. Possible scenarios of such an SED are discussed in Section \ref{dip70um}. To avoid the second far-infrared peak while fitting SED of V899 Mon, we used all the flux estimates above 90 $\mu m$ as upper limits. For constraining the SED better we also used 70 $\mu m$ flux measured during second outburst in 2013, but to account for the change in 70 $\mu m$ flux between two outbursts we used 1 Jy error bar (which is consistent to the flux change we have seen in 250 $\mu m$ SPIRE data from two epochs). Figure \ref{SED1stOutpeak2010Qphase} (a) shows the SED fit obtained using \textit{V, R, I, J, H, K$_S$, W1, W2, W3, W4} fluxes from 2010 and PACS 70 $\mu m$ flux from 2013 observations.

For fitting the SED of V899 Mon's quiescent phase, we used flux estimates from different quiescent phase epochs. Figure \ref{SED1stOutpeak2010Qphase} (b) shows the SED fit using \textit{V, R, I, 2MASS J, H, K$_S$ and IRAS 12 $\mu m$} fluxes. MIPS 1 (24 $\mu m$), PACS (70 $\mu m$, 160 $\mu m$) and SPIRE (250 $\mu m$, 350 $\mu m$, 500 $\mu m$) fluxes were used only as upper limits.
 
\begin{figure}
 \includegraphics[width=0.5\textwidth]{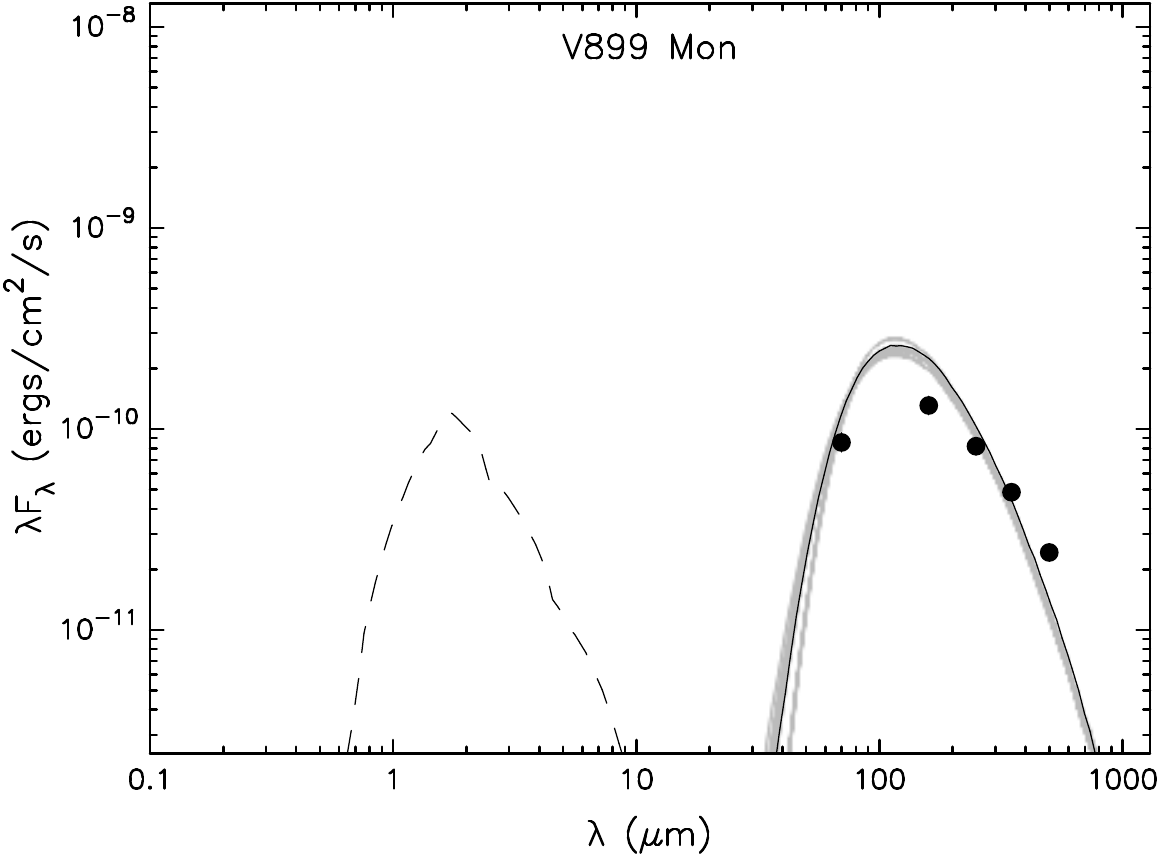}
 \caption{SED fit of V899 Mon during its second outburst phase using only \textit{Herschel} PACS 70 and 160 $\mu m$, and SPIRE 250, 350 and 500 $\mu m$ fluxes. The filled circles represent the data points used for the fit. The solid black line is the best fitted model and the gray lines show subsequent good fits. The dashed gray line shows the stellar photosphere of V899 Mon in the best fitted model (in the absence of circumstellar dust, but including interstellar extinction).}
 \label{robetFIRsedFit}
 \end{figure}

Treating far-infrared PACS and SPIRE fluxes of 2013 to be a separate independent clump, we also fitted the SED by taking these fluxes alone (see Figure  \ref{robetFIRsedFit} and section \ref{FIRclumb}).

Stellar, disk and envelope parameters obtained from all the three SED fittings are tabulated together in Table \ref{table:SEDfitResults}.

\subsubsection{70 $\mu m$ dip in SED}\label{dip70um}
SED of V899 Mon shows a dip around 70 - 90 $\mu m$ in both IRAS as well as in PACS data taken at two different epochs. This shape of SED is very unusual and not seen in any other FUors/EXors. Objects with extended circumstellar envelopes typically show a smooth flat SED. We also do not see heavy extinction in optical bands as expected from a source surrounded by a large extended envelope. One possible scenario is that this object, instead of having a steady state infall density distribution of envelope, has a huge spherical cavity around it to a certain radius in envelope. Significant optical light that we detect can be explained if our line of sight is along the opening created by the outflow. 

\begin{figure}
 \includegraphics[width=0.5\textwidth]{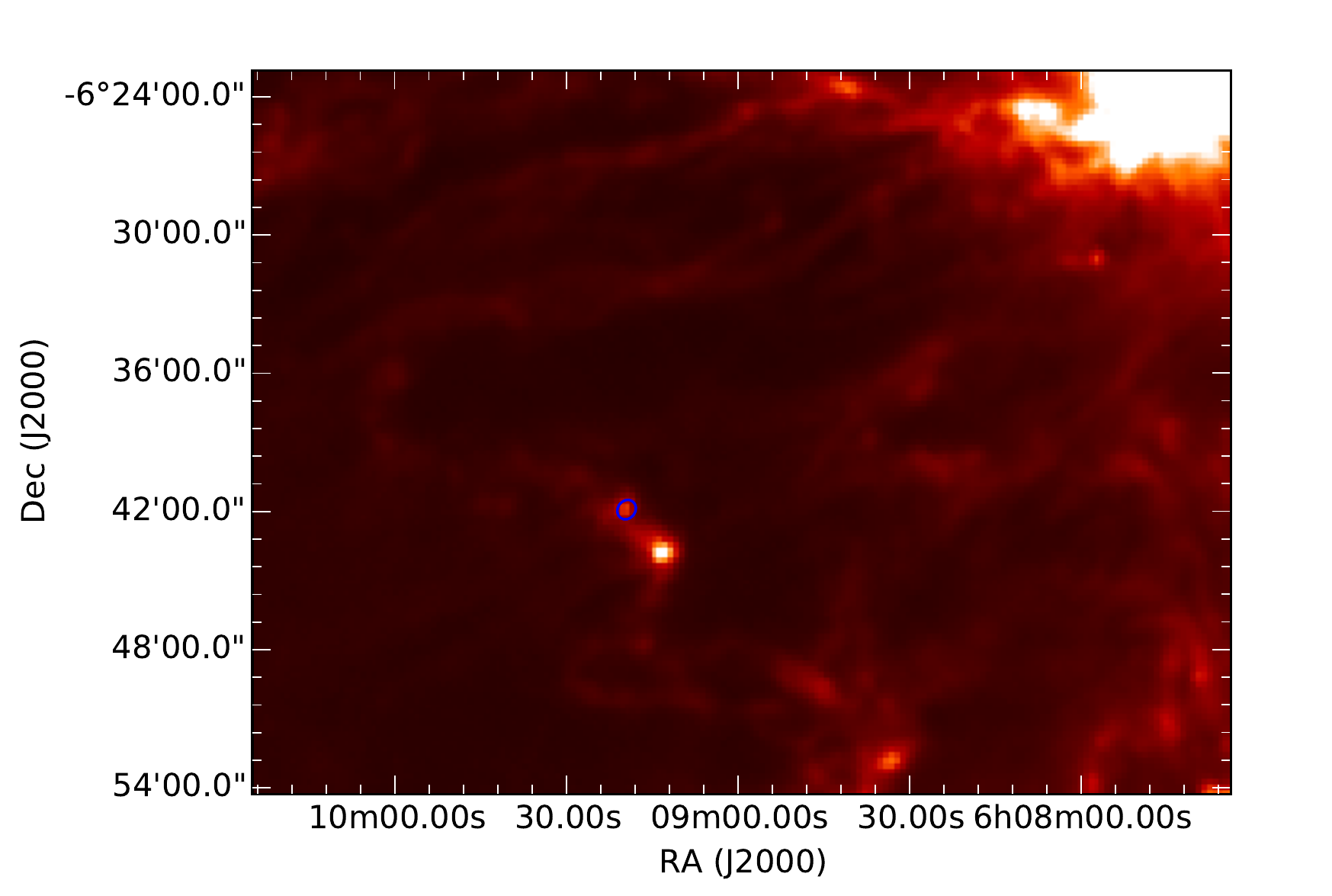}
 \caption{SPIRE 500 $\mu m$ image of V899 Mon field taken on 2013 March 16. Location of V899 Mon is marked with an ellipse to show the major and minor FWHM sizes determined for V899 Mon by the source extraction algorithm \textit{getsources} \citep{menshchikov12}. The brighter source IRAS 06068-0643, south-west of V899 Mon, is also a more embedded outburst source. The bright blob in the north-west corner of the image is the ionizing source in Monoceros R2.}
 \label{spireimage}
 \end{figure}

Another possible geometric scenario is an envelope/clump with most of its mass behind V899 Mon. In SPIRE 500 $\mu m$ image (see Figure \ref{spireimage}), we see the bright blob marked with an ellipse at the position of V899 Mon is at the edge of a cloud, possibly being pushed back by the ionizing source of Monoceros R2.

We plan to carry out a detailed 3D radiative transfer modeling of such geometric structures to see whether this seemingly two-component SED can be explained.

\subsubsection{Far-infrared component properties}\label{FIRclumb}
Considering far-infrared component part of the SED as a separate clump, we analyzed PACS-SPIRE data alone to obtain the characteristics of the clump.
L$_{bol}$ of the clump obtained by fitting \citet{robitaille07} SED model for the far-infrared flux ($>$ 70 $\mu m$) is $\sim$ 8.6 L$_\odot$ (Table \ref{table:SEDfitResults}).

We have SPIRE 250, 350 and 500 $\mu m$ observations at two epochs; during the peak of first outburst on 2010 September 4 and later during the second outburst in 2013 March 6. PACS 70 and 160 $\mu m$ observations were available only during the second epoch. Photometric flux values estimated for V899 Mon from PACS and SPIRE data are given in Table \ref{table:herschelfluxs}.

\begin{figure}
 \includegraphics[width=0.5\textwidth]{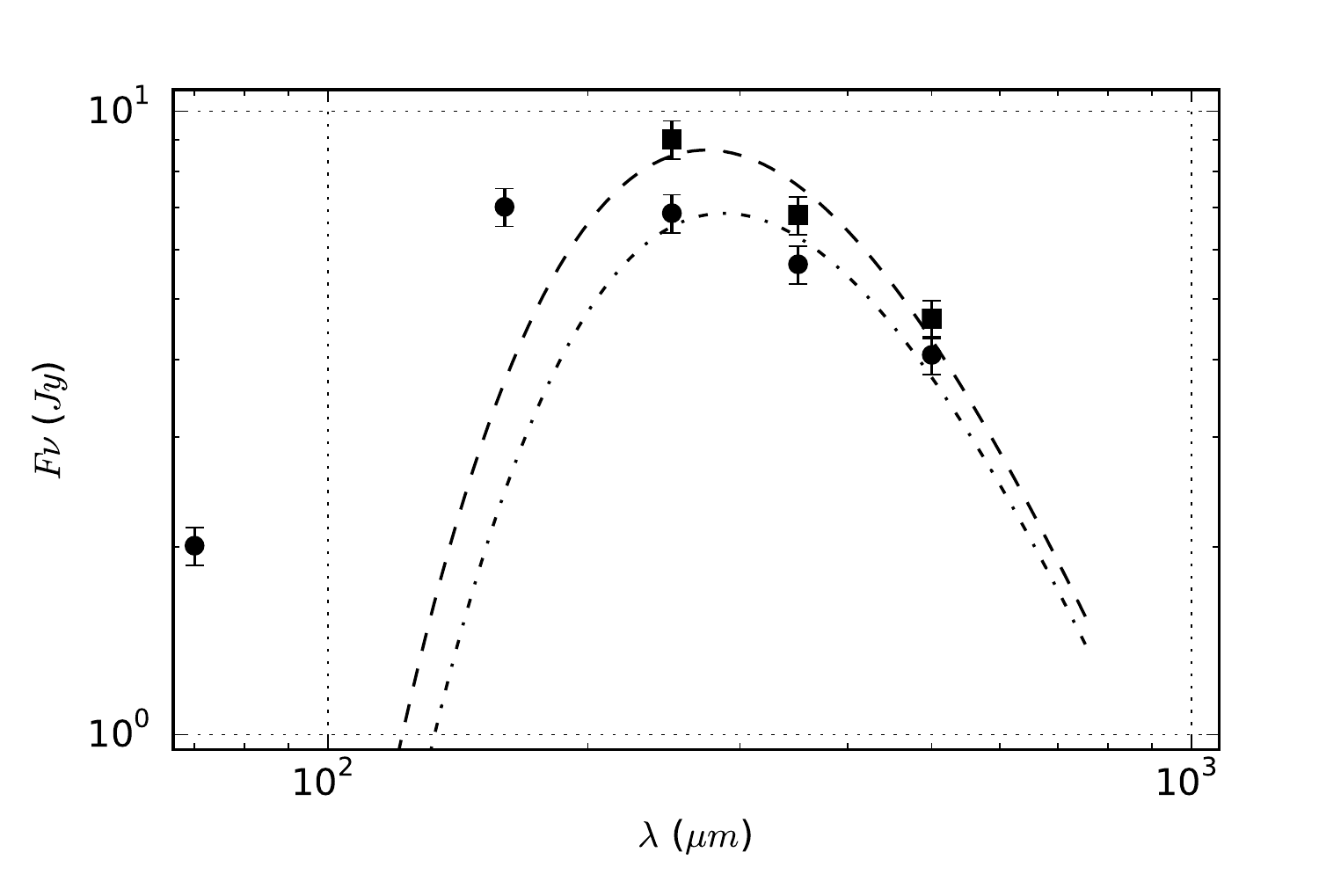}
 \caption{Gray body fit of V899 Mon's far-infrared SPIRE wavelength fluxes at two epochs. Filled squares are from 2010 epoch during the peak of the first outburst and filled circles are from 2013 during the second outburst. The dashed line is the gray body fit to the SPIRE data points of the first outburst in 2010, and dashed-dotted line is the fit to the SPIRE data points observed in 2013 during the second outburst. PACS data points of second epoch are also shown, but were not used for the gray body fit.}
 \label{GraybodyFitHer}
 \end{figure}

During the peak of first outburst, fluxes were brighter than the second outburst by 2.8 Jy, 1.5 Jy and 0.3 Jy in 250, 350 and 500 $\mu m$, respectively. V899 Mon was also optically brighter during the peak of first outburst than second outburst epoch by 0.75 magnitude (factor of 2) in $R$-band. This implies the far-infrared clump is heated by V899 Mon and is not spatially far from V899 Mon. We fitted gray body model to SPIRE data from the two epochs (Figure \ref{GraybodyFitHer}) and estimated the mass to be 20 M$_\odot$ and temperature of the far-infrared clump to be 10.6 K and 10.0 K at each epoch. This mass estimate is consistent with the envelope mass obtained from \citet{robitaille07} SED fitting (see Table \ref{table:SEDfitResults}). Gray body was fitted using the following formulation: 
\begin{eqnarray}
S_\nu(\nu) = \frac{M  (0.01 ({\frac{\nu}{1000 GHz}})^\beta) B(T,\nu)}{D^2}
\end{eqnarray}
where S$_\nu(\nu)$ is the observed flux density, $M$ is the mass of the clump, $B(T,\nu)$ is the Planck's black body function for temperature $T$ \& frequency $\nu$, $D$ is the distance (taken to be 905 pc) and the dust opacity factor ($\kappa_\nu$) was taken to be $0.01 ({\frac{\nu}{1000 GHz}})^\beta$ $m^2 kg^{-1}$, where $\beta$ was fixed to be 2 \citep{andre10}.

To obtain a lower limit of A$_V$, we assumed the entire mass $M$ is in a uniform density sphere of diameter $2R$ = 50\arcsec\, (FWHM of the source in SPIRE 500 $\mu m$ image), then the mass column density to the center of the sphere should be $3M/(4\pi R^2)$. Since most of the gas is molecular, we can equate this column density to N($H_2$)$ \mu_{H_2} m_H$, where N($H_2$) is the $H_2$ column number density, $\mu_{H_2}$ is the mean molecular weight (taken to be 2.8) and $m_H$ is the mass of Hydrogen. Our mass estimate of 20 M$_\odot$ then corresponds to N($H_2$) = $4.6 \times 10^{21}$ molecules cm$^{-2}$. Using the relation $\langle N(H_2)/A_V \rangle = 0.94 \times 10^{21}$ molecules cm$^{-2}$ mag$^{-1}$ \citep{ciardi98}, we obtain A$_V$ to the center of the clump to be 19.7 mag.

If we use PACS 160 $\mu m$ flux also in the gray body fitting then we obtain mass to be 12 M$_\odot$ and temperature of the far-infrared clumps at the two epochs to be 11.9 K and 12.2 K.
This mass corresponds to an A$_V$ = 10.7 mag to the center of the uniform spherical clump.
Both of these estimates are substantially higher than the actual A$_V$  estimated from optical and NIR data. This implies that the heating source V899 Mon is not embedded at the center of this clump; it might be partially located to the front section of it (or any other geometry as discussed in Section \ref{dip70um}).
 Even though the mass estimate has some systematic uncertainties from the distance, dust opacity factor and gray body model assumption, they are unlikely to cause a difference of 8 - 15 mag in A$_V$ value.

\subsection{Spectroscopic Results}\label{specresults}
\begin{figure*}
  \begin{tabular}[b]{@{}p{1\textwidth}@{}}
    \centering \includegraphics[width=1\textwidth]{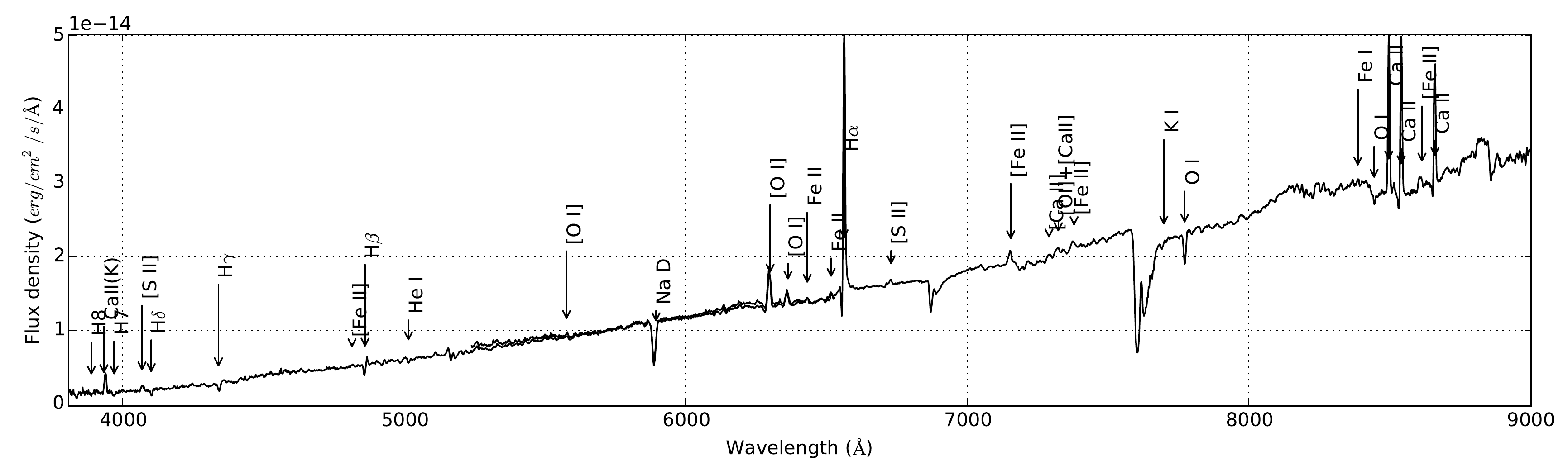}
 \\
    \centering\small (a) optical
  \end{tabular} \\

  \begin{tabular}[b]{@{}p{1\textwidth}@{}}
    \centering\includegraphics[width=1\linewidth]{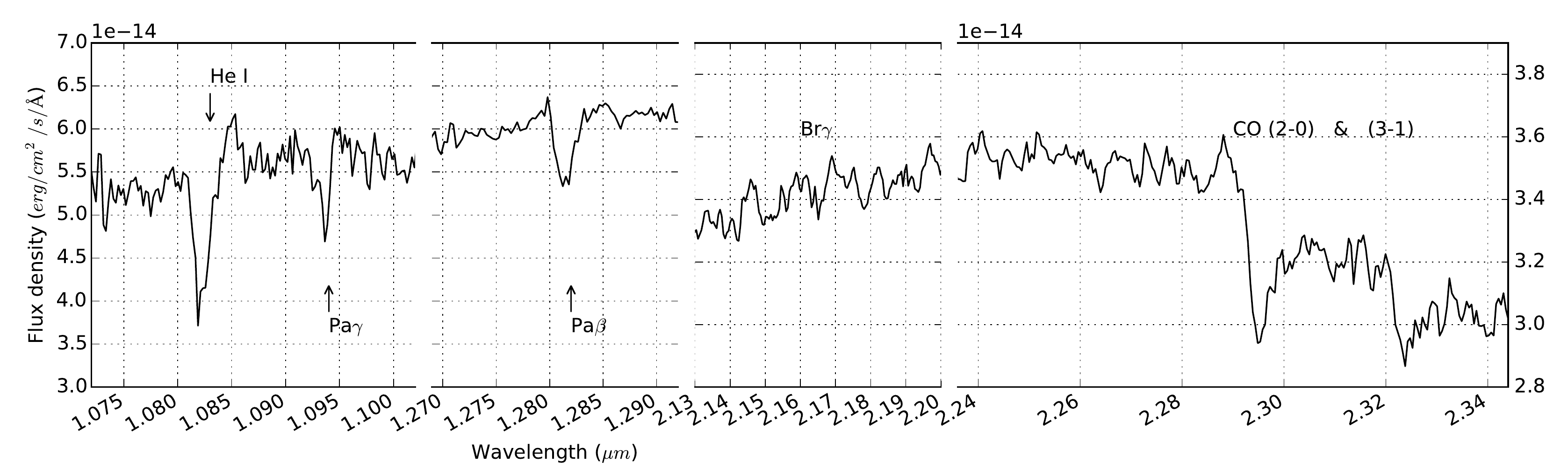} \\
    \centering\small (b) NIR
  \end{tabular}
 \caption{Flux-calibrated optical and NIR spectra of V899 Mon during its second outburst. All spectra observed during the second outburst were median combined to improve signal to noise ratio. Identified lines in the spectra are labeled. The optical spectrum (a) is not corrected for atmospheric absorption lines. The absorption lines that are not labeled in optical spectrum are atmospheric lines.}
 \label{opticalnirSpectrum}
 \end{figure*}

V899 Mon has a rich spectrum of emission as well as absorption lines in optical and NIR.
Figure \ref{opticalnirSpectrum} shows the flux-calibrated optical and NIR spectra taken during the second outburst phase of the source. Our spectroscopic observations in optical cover first outburst, transition, quiescent phases, as well as the second outburst. The fluxes and equivalent widths of the detected lines at various epochs are tabulated in Table \ref{table:linefluxeqw}. The strong Ca II IR triplet emission lines in spectrum confirm V899 Mon source to be a YSO, and the detection of CO (2-0) and (3-1) band heads absorption starting at 2.29 $\mu m$ confirm this source to be an outburst family of FUors/EXors. These overtone band head lines are seen only in giants and FUors/EXors. In FUors/EXors they are believed to be forming in the accretion heated inner regions of the disk, where temperature is in the range 2000 K $<$ T $<$ 5000 K and density n$_H$ $>$ 10$^{10}$ cm$^{-3}$ \citep{kospal11}. One notable line not detected in V899 Mon spectrum is $Br\gamma$ at 2.16 $\mu m$. While $Br\gamma$ is typically found in strong accreting T-Tauri stars, it is not detected in many FUors. 

\subsubsection{Extinction estimates}
To obtain de-reddened line fluxes, we need to know the extinction to the source. In Section \ref{Photresults} our NIR CC diagram does not show any significant extinction to the source. We also have an estimate of the interstellar extinction to the source to be A$_V$ = 2.6 mag from Figure 4 in \citet{lombardi11}. 
H$\alpha$ (6563 $\mathring{A}$) and H$\beta$ (4861 $\mathring{A}$) lines show an emission component in their P-Cygni profile. If we assume an optically thin case B emission, the Balmer decrement ratio of H$\alpha$/H$\beta$ emission component should be in the range 2.8 - 3 \citep{osterbrock06}. Comparing this ratio with the ratios measured from our spectra, we obtained various estimates of A$_V$ during outburst phase, to be ranging from 5.9 to 9.2 mag. Our H$\beta$ profiles show stronger absorption component than emission component; this might significantly cause an underestimation of the H$\beta$ flux which was estimated by fitting a two-component Gaussian model to the line profile. This implies that A$_V$ estimate from Balmer decrement is an over-estimate. On the other hand, it is quite likely that the emission is not optically thin since it is believed to be originating in magnetospheric accretion column, which will make our estimate of A$_V$ an underestimate. To be consistent with our other extinction estimates from CC diagram and SED fitting in previous sections, we take A$_V$ during the outburst phase to be 2.6 mag for the remaining spectral line calculations.

\subsubsection{Accretion rate}
\begin{figure}
 \includegraphics[width=0.5\textwidth]{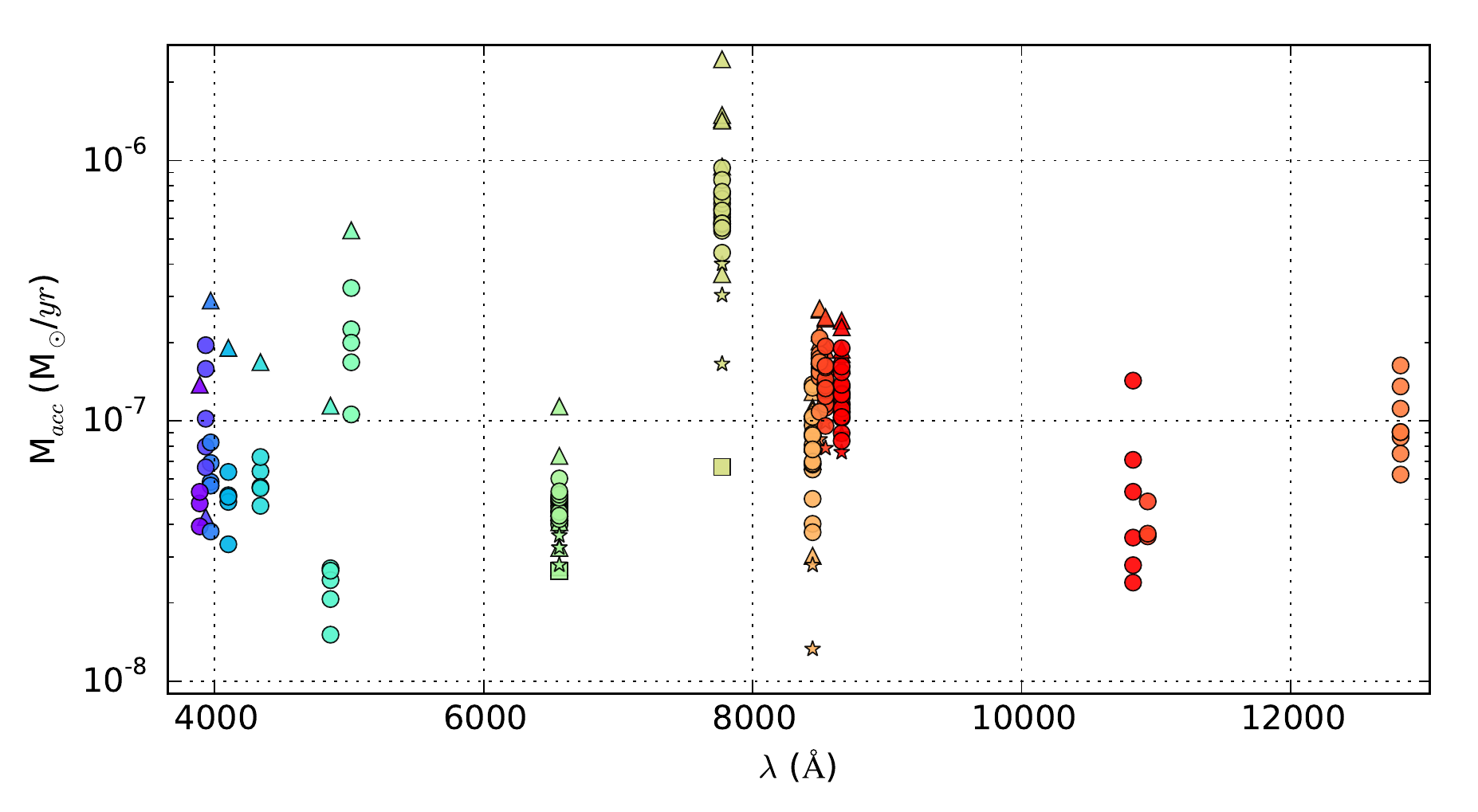}
 \caption{Accretion rates estimated from various lines from the V899 Mon's optical and NIR spectra. Triangle symbols show the first outburst phase, square symbols show the quiescent phase, star symbols show the transition phase, and circle symbols show the second outburst phase. There is a significant difference in accretion rates estimated using different lines.}
 \label{AccRatePlot}
 \end{figure}

\begin{figure}
 \includegraphics[width=0.5\textwidth]{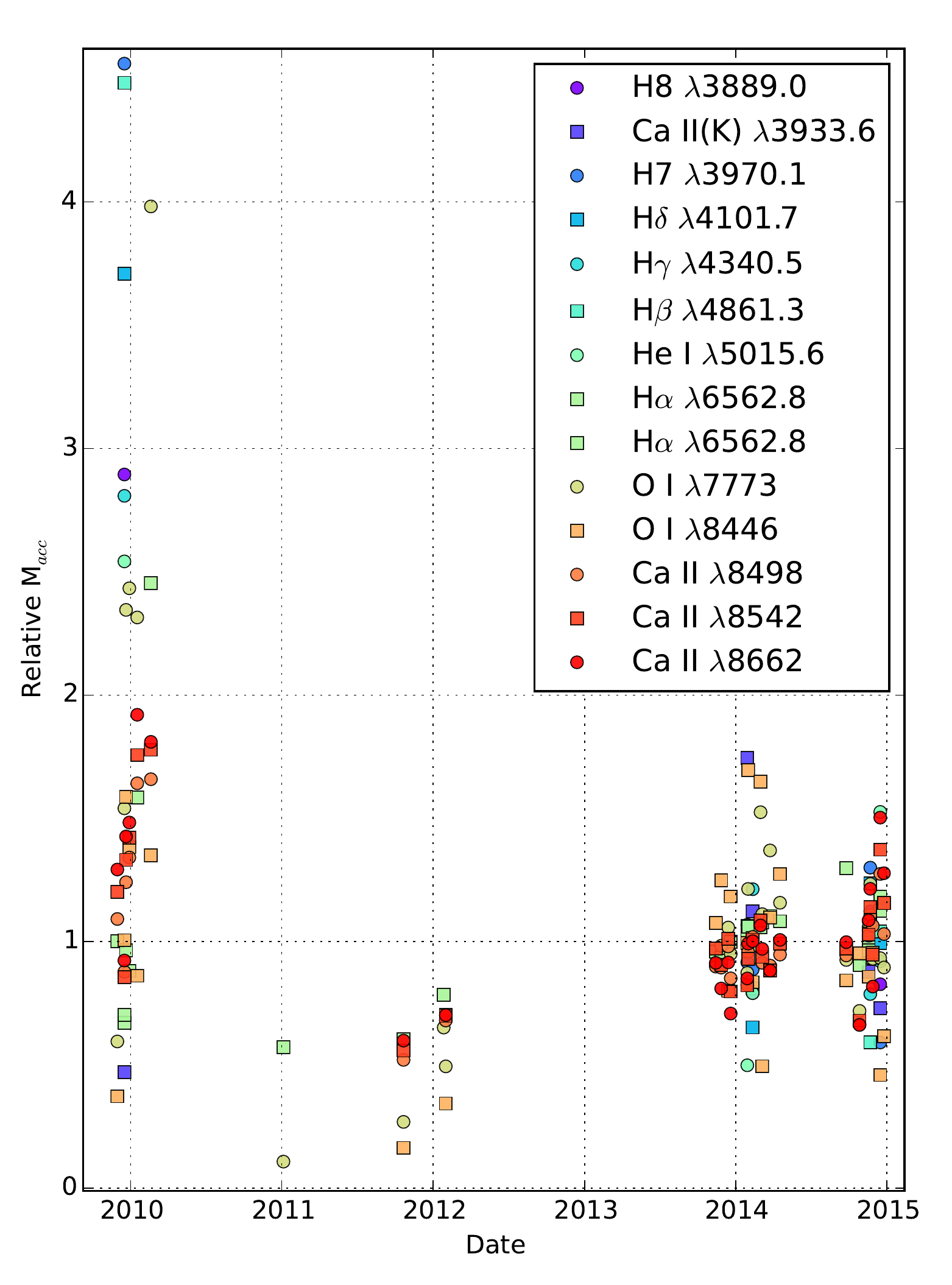}
 \caption{Relative change in accretion rates estimated from various lines during the transition from the first outburst phase to the quiescent phase and back to the second outburst.}
 \label{JDvsNormAccRatePlot}
 \end{figure}

Many of the spectral line fluxes originating near the magnetosphere have been found to correlate with the accretion luminosity. The correlation is empirically calibrated in the simple form : $log(L_{acc}/L_\odot) = a$ $log(L_{line}/L_\odot) + b$. For estimating accretion luminosity from Hydrogen Balmer series, He I, O I, Ca II IR triplet, Pa$\beta$ and Pa$\delta$ lines, we used the coefficients $a$ and $b$ from \citet{alcala14}. Distance to V899 Mon was taken to be 905 pc and the observed fluxes were corrected for an extinction of A$_V$ = 2.6 mag. To estimate the accretion rate from the accretion luminosity, we assumed the mass and age of V899 Mon to be 2.5 M$_\odot$ and 1 Myr, respectively (consistent with the CMD). Stellar radius was then estimated to be 4 R$_\odot$ based on the isochrone for these values of mass and age \citep{siess00}. Finally, using the formula $\dot{M}_{disk} = (L_{acc}R_*/GM) (1-R_*/R_i)^{-1}$ (where disk inner radius $R_i \sim 5 \times R_*$ \citep{gullbring98}), we obtained accretion rates from various lines along the spectrum. No veiling correction was done for lines which are in absorption, hence these accretion rate estimates can be slight underestimates. Figure \ref{AccRatePlot} shows all the estimates of accretion rates from various lines as a function of their wavelengths. Figure \ref{JDvsNormAccRatePlot} shows the normalized accretion rates obtained from different lines at various epochs. A clear reduction in relative accretion, by at least a factor of 2 during quiescent phase with respect to the second outburst phase, is seen in all the accretion rates obtained from different lines.

\subsubsection{Mass loss}
Ratio of the mass loss to mass accretion in YSOs is a key parameter predicted by various jet launch mechanism models. FUors/EXors family of objects give us a direct observational constrain on this ratio. 
Based on the SED fit in Section \ref{sed}, we obtained the L$_{bol}$ of V899 Mon as $\sim$ 150 $L_\odot$ during the outburst phase. Using the relation between mass loss and L$_{bol}$ in young HeBe and T-Tauri stars, viz. $log(\dot{M})= -8.6 + 0.7 \times log(L_{bol})$ by \citet{nisini95}, we get an estimate of the mass outflow in V899 Mon to be 1$\times$10$^{-7}$ M$_\odot$/yr. This rate is larger than typical T-Tauri stars (10$^{-8}$ M$_\odot$/yr) but less than classical FUors (10$^{-5}$ M$_\odot$/yr) \citep{hartmann96}.

We could also directly measure the mass loss rate from [OI] $\lambda$6300 flux using the relation \textit{A8} of \citet{hartigan95}. Mass outflow is directly proportional to the optically thin forbidden line emission fluxes originating in outflow. The mass loss estimate from [OI] $\lambda$6300 flux through 2\arcsec\, slit aperture, by assuming a distance of 905 pc and a typical sky plane component of the outflow velocity to be 150 km/s is 2.6$\times$10$^{-7}$ M$_\odot$/yr.

We could not detect any significant change in these forbidden line fluxes during quiescence. Our 2\arcsec\, slit aperture at 905 pc distance corresponds to 2.7$\times$10$^{11}$ km wide region. Even if we assume the sky plane velocity to be as large as 700 km/s (seen in P-Cygni outflow), the gap in outflow due to one year of quiescence will be only 2.2$\times$10$^{10}$ km, which is just one-tenth of the total aperture. Hence, 
our non-detection does not conclude whether the large scale low density outflow traced by forbidden lines was interrupted or remained uninterrupted during the quiescence phase.

\subsubsection{Outflow Temperature and Density}
Flux ratios of the optically thin forbidden emission lines provide a direct estimate of the density and temperature of the outflow region. For better signal to noise ratio, we estimated the average flux of the forbidden lines by combining all the second outburst phase spectra of V899 Mon. Flux ratio [S II] $\lambda$6716/$\lambda$6731 is a good tracer of density and is insensitive to temperature. Flux of [S II] $\lambda$6731 line is 2.4$\times$10$^{-14}$ $erg/cm^2/s$, whereas [S II] $\lambda$6716 line is not detected in our spectrum, so we have an estimate of upper limit to be $<$ 1$\times$10$^{-14}$ $erg/cm^2/s$. The ratio [S II] $\lambda$6716/$\lambda$6731 $<$ 0.45, implies the electron density n$_e$ to be $>$ 10$^4/cm^3$ (Figure 5.8 in \citet{osterbrock06}).  Ratio of [O I] $\lambda$5577/$\lambda$6300 in V899 Mon is $\sim$ 0.09. For a range of temperature from 9000 K to 20000 K, this ratio is consistent with n$_e$ ranging from 2$\times$10$^5 /cm^3$ to 4$\times$10$^6 /cm^3$ (Figure 6 in \citet{hamann94}). 

Once we have a density constraint, we can now use other line ratios to estimate the temperature. For instance, using the formula 5.5 in \citet{osterbrock06}, for a density of n$_e$ = 2$\times$ 10$^5 /cm^3$,  ratio [O I] ($\lambda$6300+$\lambda$6364)/$\lambda$5577 $\sim 15$ gives an estimate on the temperature to be $\sim$ 8500 K. 
Density insensitive line ratio [Ca II] $\lambda$7291 / [OI] $\lambda$6300 $\sim 0.1$ in V899 Mon, implies a temperature $<$ 9000 K (Figure 5 in \citet{hamann94}). But the line ratio [S II] $\lambda$6731 / [O I] $\lambda$6300 $\sim$ 0.08 implies a temperature $>$ 9000 K (Figure 5 in \citet{hamann94}). To be consistent with all these independent estimates, we shall take the temperature of the outflow to be $\sim$ 9000 K.

\subsubsection{Line profiles and variability}
\paragraph{H$\alpha$ $\lambda6563$}
Almost all prominent lines in the spectrum showed variability during our period of observations. The line which showed the most dramatic changes in line profile is H$\alpha$, which has a strong P-Cygni absorption component and is believed to be formed in magnetospheric accretion funnel. 

\begin{figure}
 \includegraphics[width=0.5\textwidth]{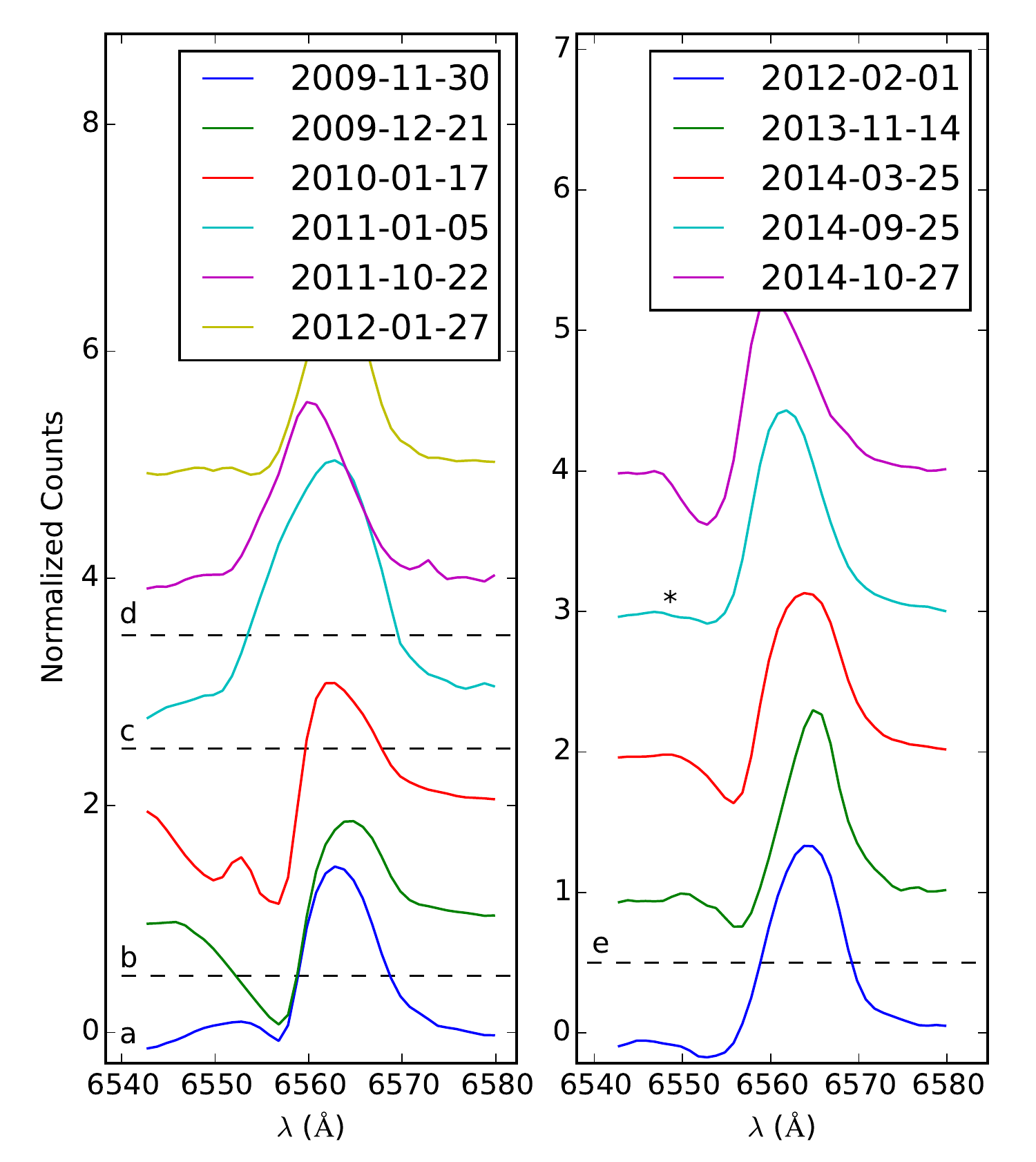}
 \caption{Selected sample of H$\alpha$ emission line profiles from different phases of V899 Mon. a) First outburst phase, b) Heavy outflow during the end of first outburst phase, c) Quiescent phase, d) Transition phase from quiescence phase to second outburst phase, e) Second outburst phase. *) shows a particular profile on 2014 September 25, when the P-Cygni became very weak for a very short duration. Spectrum taken immediately after one month shown on top right panel had strong P-Cygni profile.}
 \label{HalphaProfiles1D}
 \end{figure}

\begin{figure}
  \centering
  \begin{tabular}[b]{@{}p{0.45\textwidth}@{}}
    \centering\includegraphics[width=\linewidth]{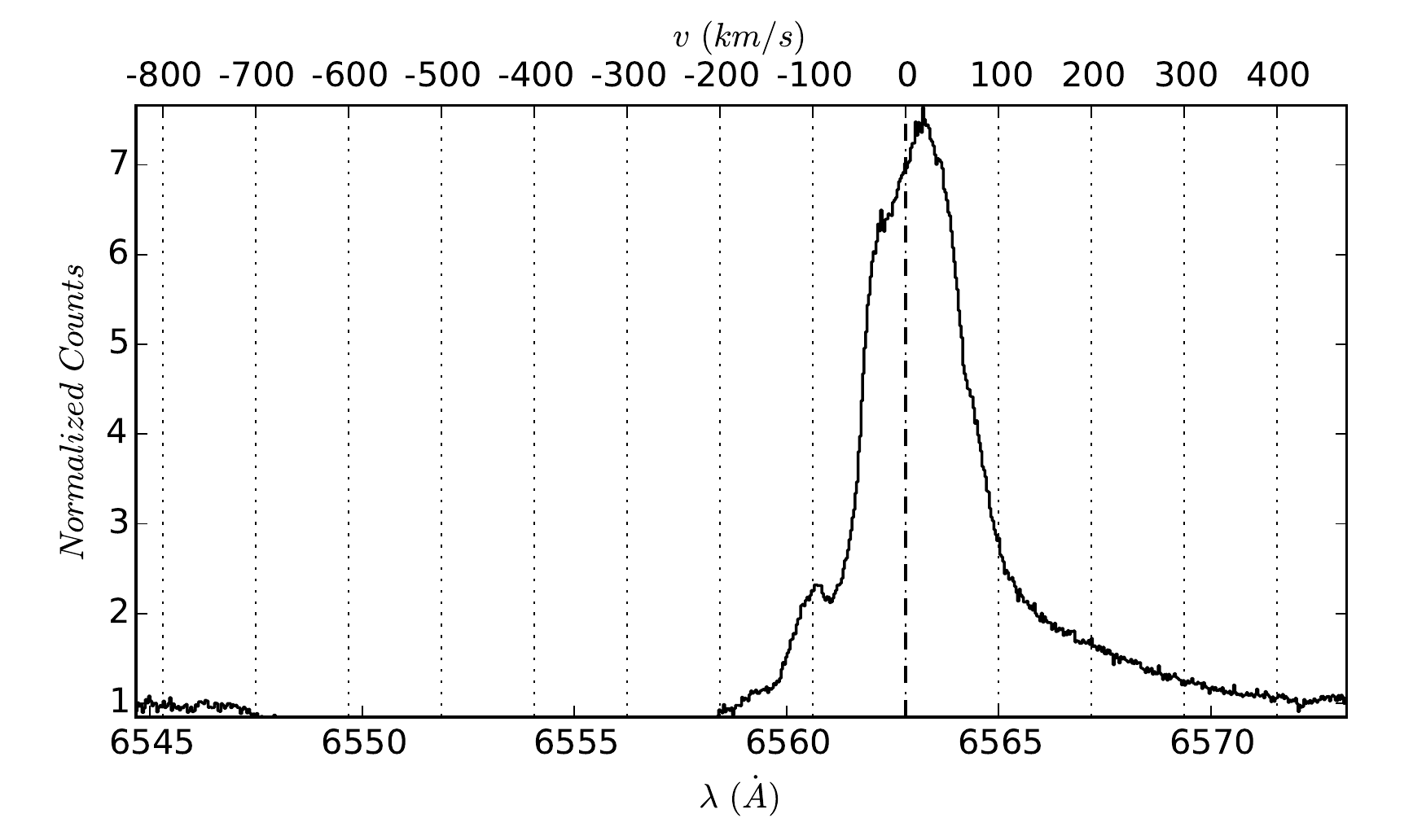} \\
    \centering\small (a) Emission component
  \end{tabular}%
  \quad
  \begin{tabular}[b]{@{}p{0.45\textwidth}@{}}
    \centering\includegraphics[width=\linewidth]{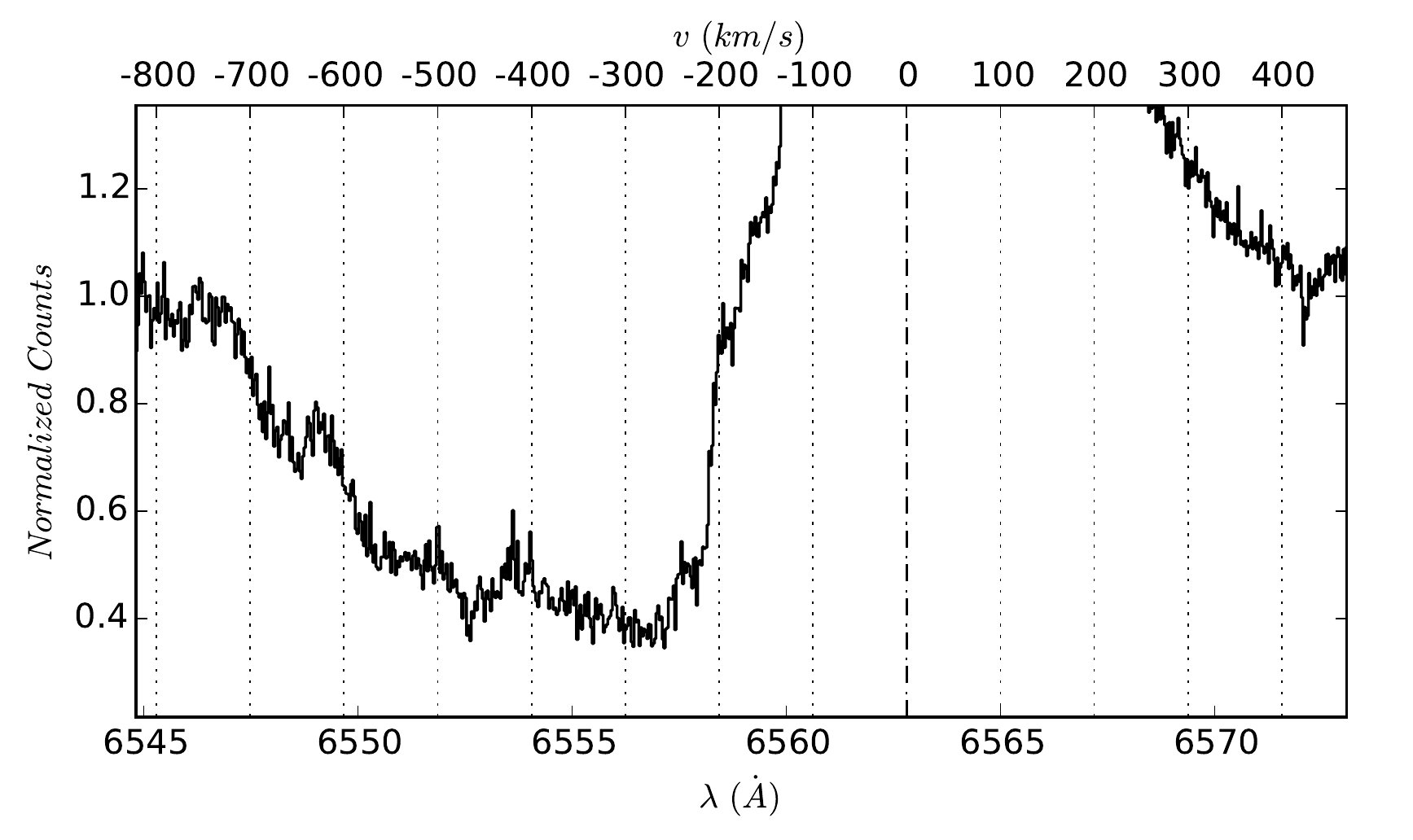} \\
    \centering\small (b) P-Cygni absorption component
  \end{tabular}
  \caption{H$\alpha$ (a) emission, (b) P-Cygni absorption line profiles of V899 Mon observed on 2014 December 22 using SALT-HRS.}
  \label{SALTHRSHalphaEmiAbs}
\end{figure}

 Figure \ref{HalphaProfiles1D} shows the evolution of H$\alpha$ P-Cygni profile during its first outburst, quiescent and the current ongoing second outburst phases. 
It is a selected representative sample of line profile plots from each epoch. The first spectrum published by \citet{wils09} shows no P-Cygni profile in H$\alpha$. Our initial spectrum observed 13 days later during the peak of first outburst shows a CI Tau profile typically seen in many T-Tauri stars \citep{stahler05}. In subsequent spectra, the absorption component grows considerably, transforming H$\alpha$ line's profile into a strong P-Cygni profile (Figure \ref{HalphaProfiles1D}). We also see complex structures in the absorption component of P-Cygni in the spectrum taken on 2010 January 17. All these evolution indicate the outflow wind increased dramatically towards the end of first outburst. By the onset of quiescent phase, the absorption component in P-Cygni completely disappears and the H$\alpha$ shows a near symmetric emission line. We also do not see any P-Cygni profile when the source was undergoing transition from the quiescent phase to the second outburst phase. P-Cygni profiles, much fainter in strength compared to the peak of the first outburst, only starts appearing after the full onset of the second outburst. Even though during the second outburst, outflow P-Cygni profile is more stable than during the last phase of the first outburst, spectrum taken on 2014 September 25 shows only a very weak P-Cygni in H$\alpha$. These delays and short pauses indicate a non-steady nature in the outflows from FUors/EXors. Table \ref{table:linefluxeqw} shows the fluxes and equivalent widths obtained by fitting a two-component Gaussian to the line profiles.
Our high resolution spectrum ($R \sim 37000$) taken using SALT-HRS on 2014 December 22 resolves the H$\alpha$ line and its multi-component profile. Figure \ref{SALTHRSHalphaEmiAbs} (a) shows the emission component of the H$\alpha$ line profile and Figure \ref{SALTHRSHalphaEmiAbs} (b) shows the absorption component in the spectrum taken using SALT-HRS. The red-shifted part of the profile peaks at +18 km/s and has smooth wings with an extra broad component reaching up to velocity of +420 km/s (Figure \ref{SALTHRSHalphaEmiAbs} (a)). On the other-hand blue part of the profile shows complicated multi-component velocity structures (Figure \ref{SALTHRSHalphaEmiAbs} (b)). The outflow absorption component extends up to -722 km/s. We could see structures in absorption at -648 km/s, -568 km/s, -460 km/s, -274 km/s, -153 km/s, -100 km/s and -26 km/s.

These are typically associated with bulk motion within the outflowing gas \citep{stahler05}. To understand how the outflow components evolve, it will be interesting to study the evolution of these components in velocity time space by carrying out multi-epoch high resolution spectroscopic observations.

\begin{figure*}
 \includegraphics[width=0.8\textwidth]{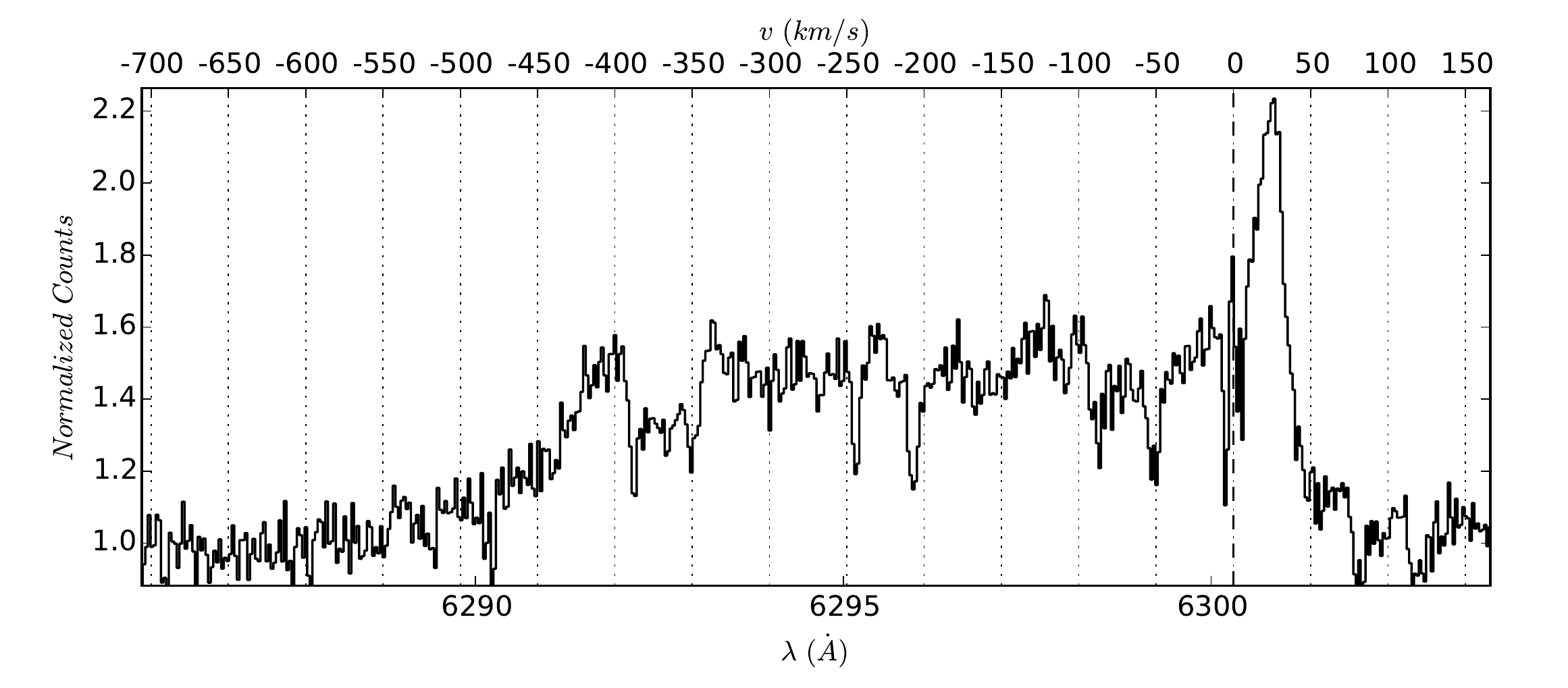}
 \caption{[OI] $\lambda$6300 emission line structure from V899 Mon observed on 2014 December 22 using SALT-HRS.}
 \label{SALTHRSOI6300}
 \end{figure*}

\begin{figure*}
 \includegraphics[width=0.8\textwidth]{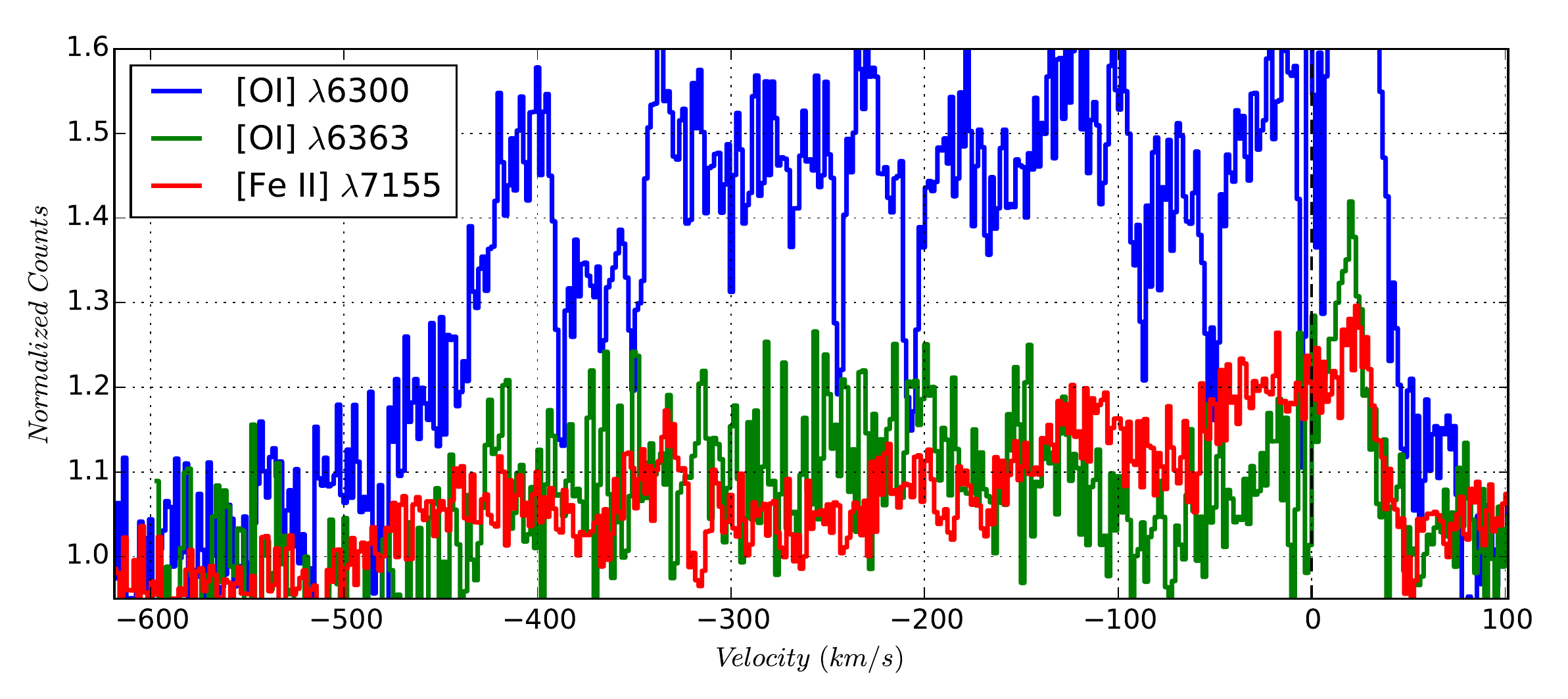}
 \caption{[OI] $\lambda$6363 and [FeII] $\lambda$7155 emission line profile structures plotted over [OI] $\lambda$6300 profile of V899 Mon observed on 2014 December 22 using SALT-HRS.}
 \label{SALTHRSAllFBL}
 \end{figure*}

\paragraph{Forbidden lines [OI] $\lambda$6300,  $\lambda$6363, [Fe II] $\lambda$7155}
Another important tracer of outflow is the forbidden line [OI] at 6300.304 $\mathring{A}$. Our medium resolution time evolution study does not show significant variation in this line, possibly due to the fact that they originate in low density jets/outflows. The SALT-HRS high resolution spectrum shows a very interesting plateau profile for [OI] $\lambda$6300.3 line (see Figure \ref{SALTHRSOI6300} for the resolved profile). Since we have not corrected our spectrum for telluric absorption, we shall not interpret the narrow absorption dips seen on the plateau structure. The profile has a strong red-shifted emission at +25 km/s with an FWHM of 21 km/s. The emission in blue-shifted part extends up to $\sim$ -450 km/s. Figure \ref{SALTHRSAllFBL} shows other two weaker forbidden lines [OI] $\lambda$6363 and [Fe II] $\lambda$7155 plotted over [OI] $\lambda$6300 profile. We see almost similar structure in [OI] $\lambda$6363.8, with a blue-shifted emission extending up to -450 km/s and a narrow peak emission at +20 km/s. 
[Fe II] $\lambda$7155 line also shows very similar structure with a blue part extending up to $\sim$ -500 km/s and a peak emission at $\sim$ +22 km/s. 
Since the forbidden line emissions are optically thin, their flux is directly proportional to the column density of emitting species. The plateau profile implies almost equal column density in the blue-shifted outflow with a velocity gradient in the jet. Since we are detecting only the velocity component of the jets along line of sight, the gradient can be either due to actual change in the outflow velocity or due to geometrical projection effect of the outflow. A cone shaped outflow in the direction of the observer can give rise to different projected velocities from different radial regions of the cone. Time evolution study of these profiles will give insight into the structure of the outflow. The red-shifted emission peak is more difficult to explain. This might be originating from the 
envelop to disk infall shock regions on the surface of the disk, or from the tail part of a bipolar outflow which is moving away from us and emerging out of the region occulted by the disk in our line of sight.

\paragraph{Ca II IR triplet $\lambda$8498, $\lambda$8542, $\lambda$8662}
Ca II IR triplet lines (8498.02 $\mathring{A}$, 8542.09 $\mathring{A}$, 8662.14 $\mathring{A}$) are believed to be originating inside and very near to the accretion column resulting in their tight correlation with accretion rate \citep{muzerolle98}. Our continuous spectral monitoring shows that the P-Cygni profile in Ca II IR triplet lines evolved similar to H$\alpha$. Outflow component got stronger towards the end of the first outburst and then it completely disappeared during the quiescent phase, and finally  reappeared after the full onset of the second outburst. As seen in typical T-Tauri stars and FUors/EXors, the line ratios of Ca II IR triplet lines in V899 Mon spectrum are found to be 1:1.01:0.77, representing an optically thick gas dominated by collisional decay. Unlike the emission components, the ratio of P-Cygni absorption components of  $\lambda$8542 and $\lambda$8662 measured from high resolution spectrum is 0.83:0.53 (= 1.57:1), which is more consistent with atomic transition strengths (1.8:1). This implies the absorption components in outflow are optically thin. These values are surprisingly similar to the P-Cygni profiles of Ca II IR triplet lines detected in the episodic winds of V1647 Ori \citep{ninan14}. Following the same arguments as in \citet{ninan14}, we could estimate the Ca II column density in the outflow to be 3.4 $\times$ 10$^{12}$ cm$^{-2}$ and the Hydrogen column density N$_H$ $\sim$ 3.8 $\times$ 10$^{20}$ cm$^{-2}$.

\begin{figure}
  \centering
  \begin{tabular}[b]{@{}p{0.45\textwidth}@{}}
    \centering\includegraphics[width=\linewidth]{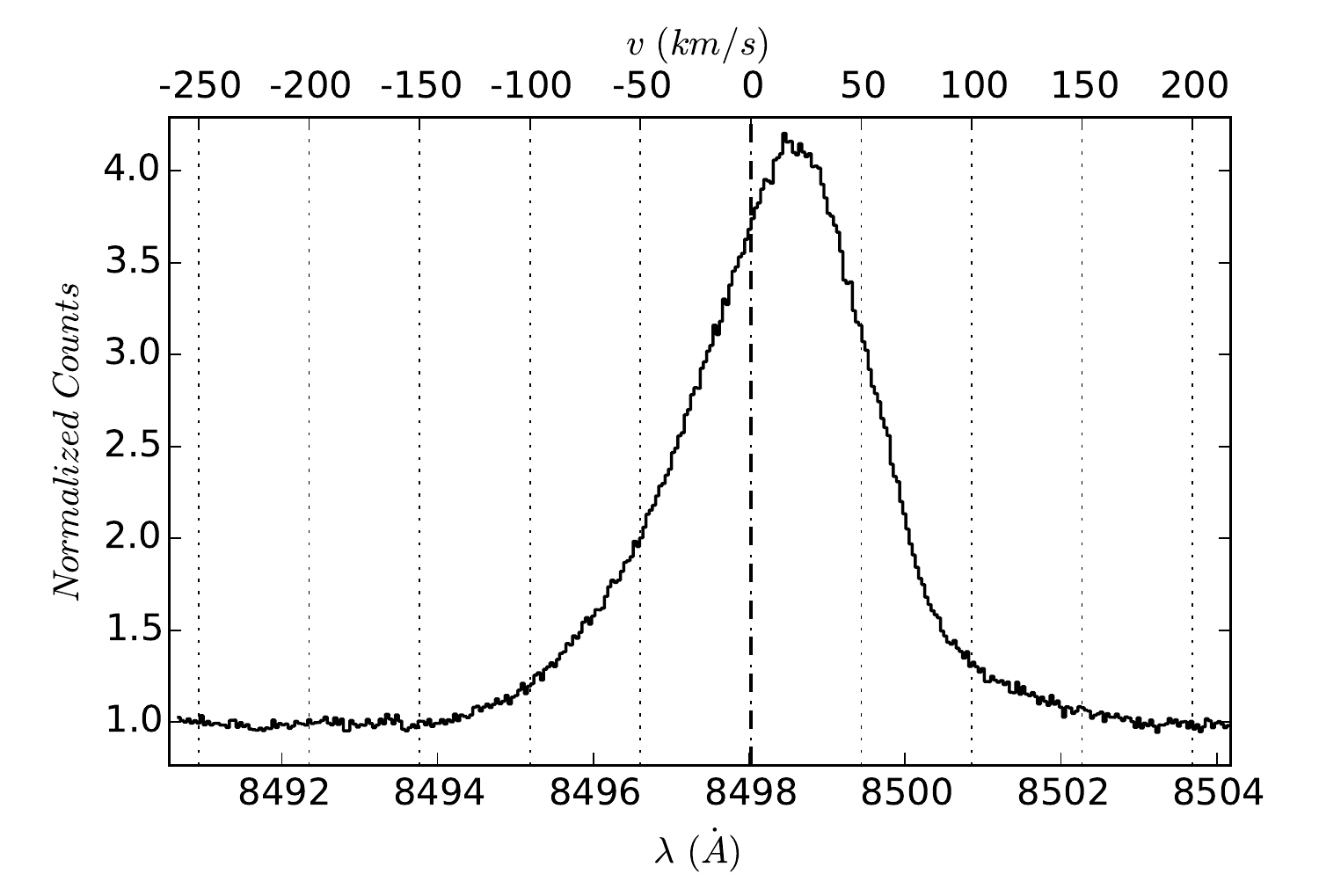} \\
    \centering\small (a) $\lambda$8498
  \end{tabular}%
  \quad
  \begin{tabular}[b]{@{}p{0.45\textwidth}@{}}
    \centering\includegraphics[width=\linewidth]{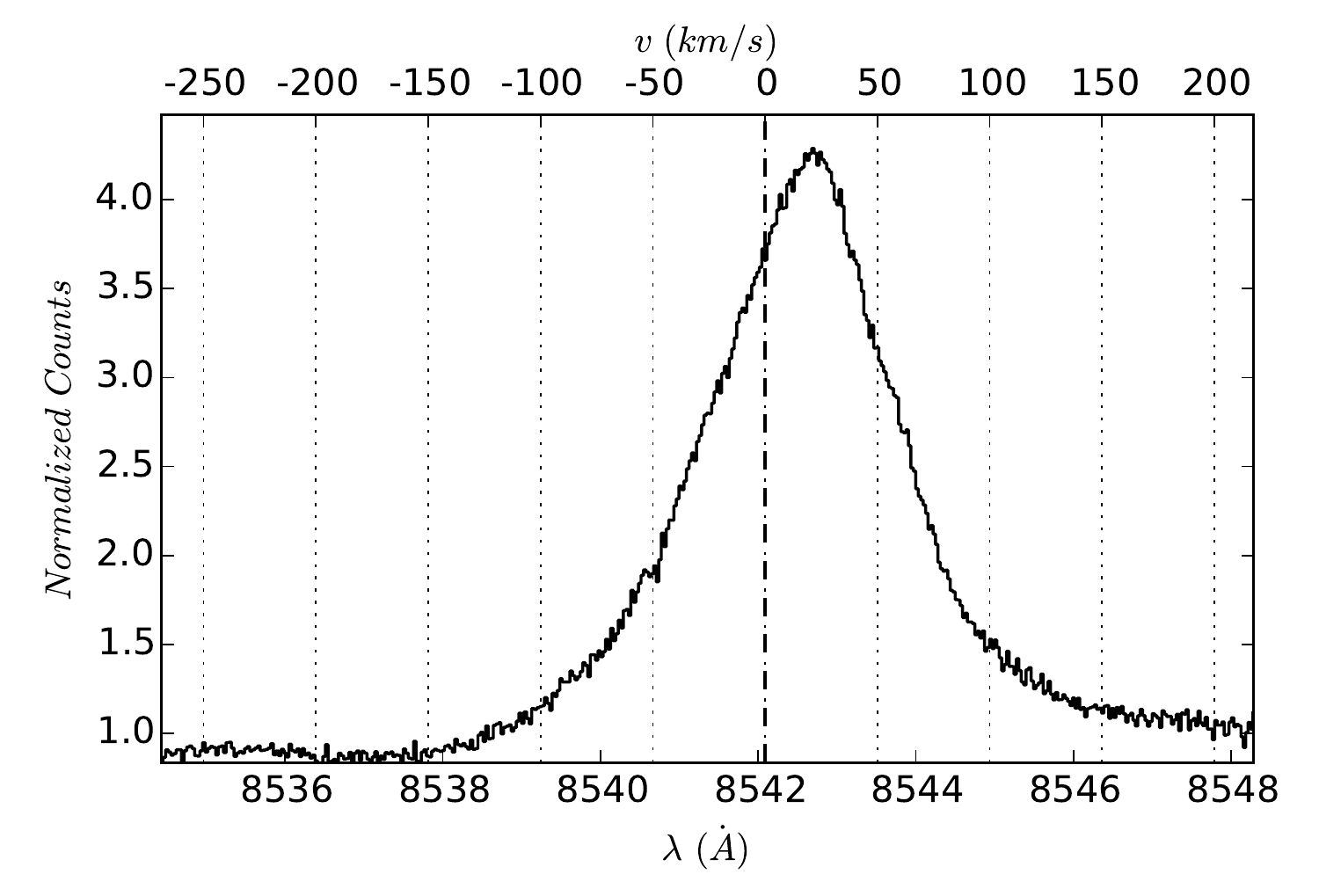} \\
    \centering\small (b) $\lambda$8542
  \end{tabular} \\
  
  \begin{tabular}[b]{@{}p{0.45\textwidth}@{}}
    \centering\includegraphics[width=\linewidth]{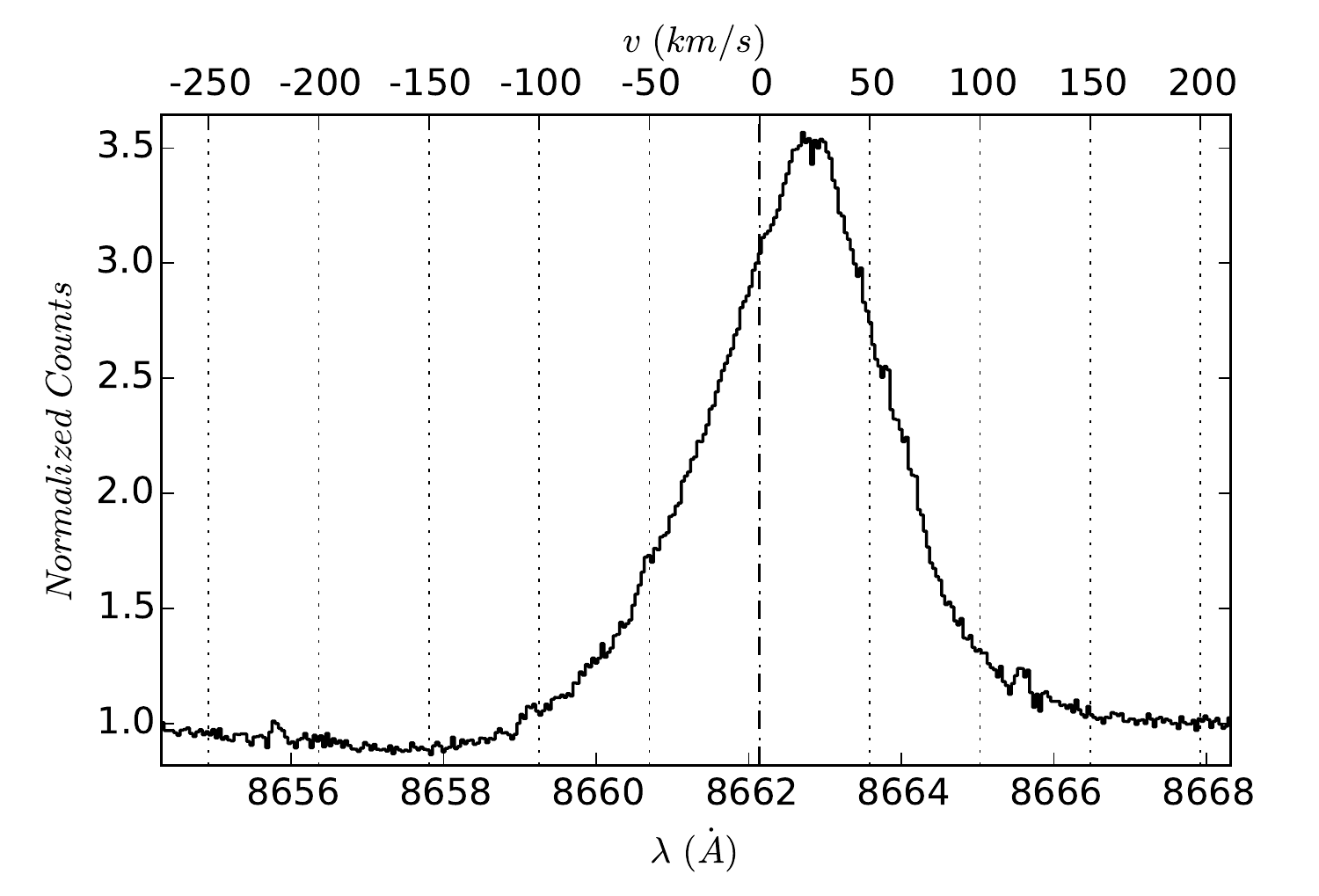} \\
    \centering\small (c) $\lambda$8662
  \end{tabular}
  
  \caption{Ca II IR triplet emission line profiles of V899 Mon observed on 2014 December 22 using SALT-HRS.}
  \label{SALTHRSCaII123Emi}
\end{figure}

\begin{figure}
  \centering
  \begin{tabular}[b]{@{}p{0.45\textwidth}@{}}
    \centering\includegraphics[width=\linewidth]{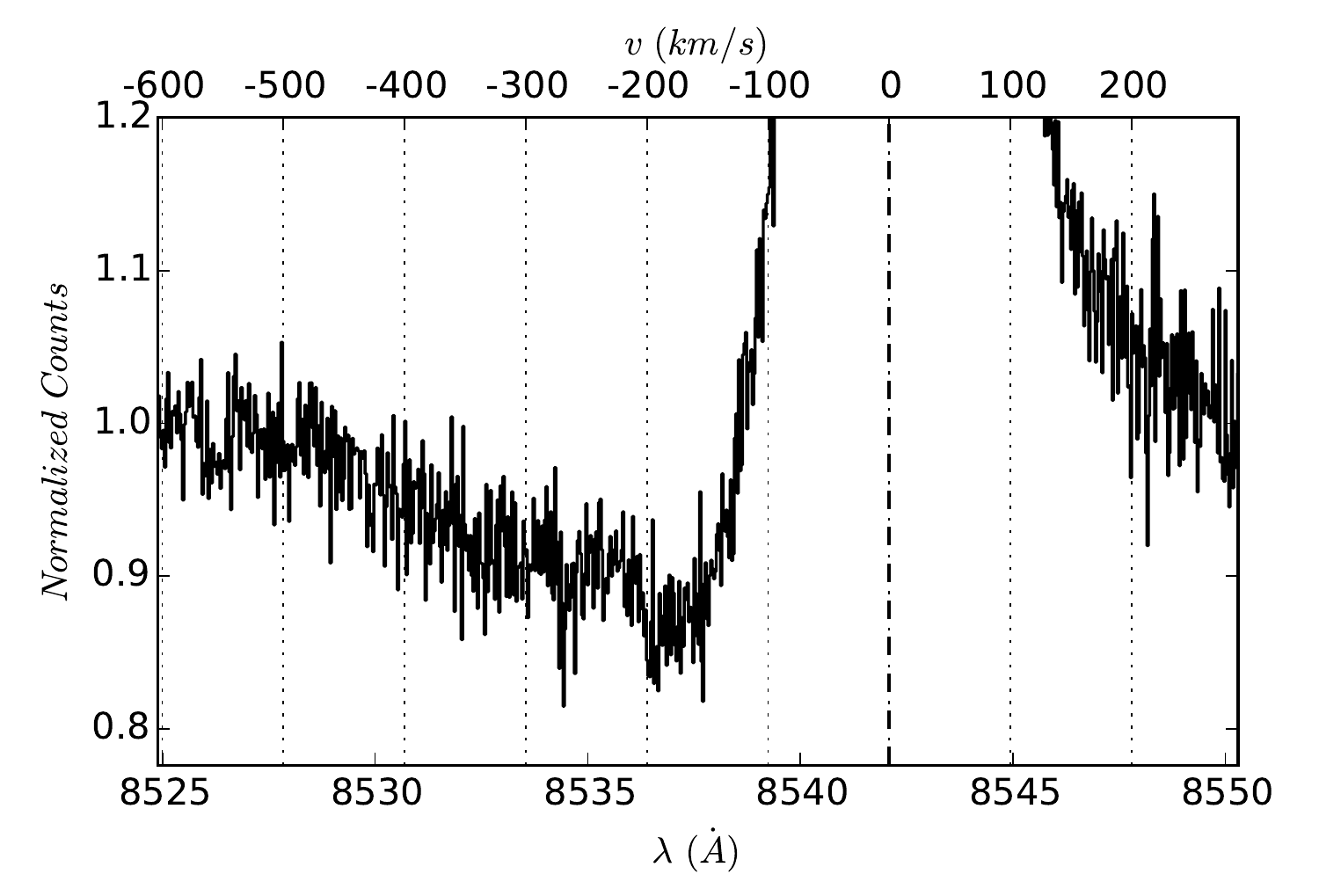} \\
    \centering\small (a) $\lambda$8542
  \end{tabular}%
  \quad
  \begin{tabular}[b]{@{}p{0.45\textwidth}@{}}
    \centering\includegraphics[width=\linewidth]{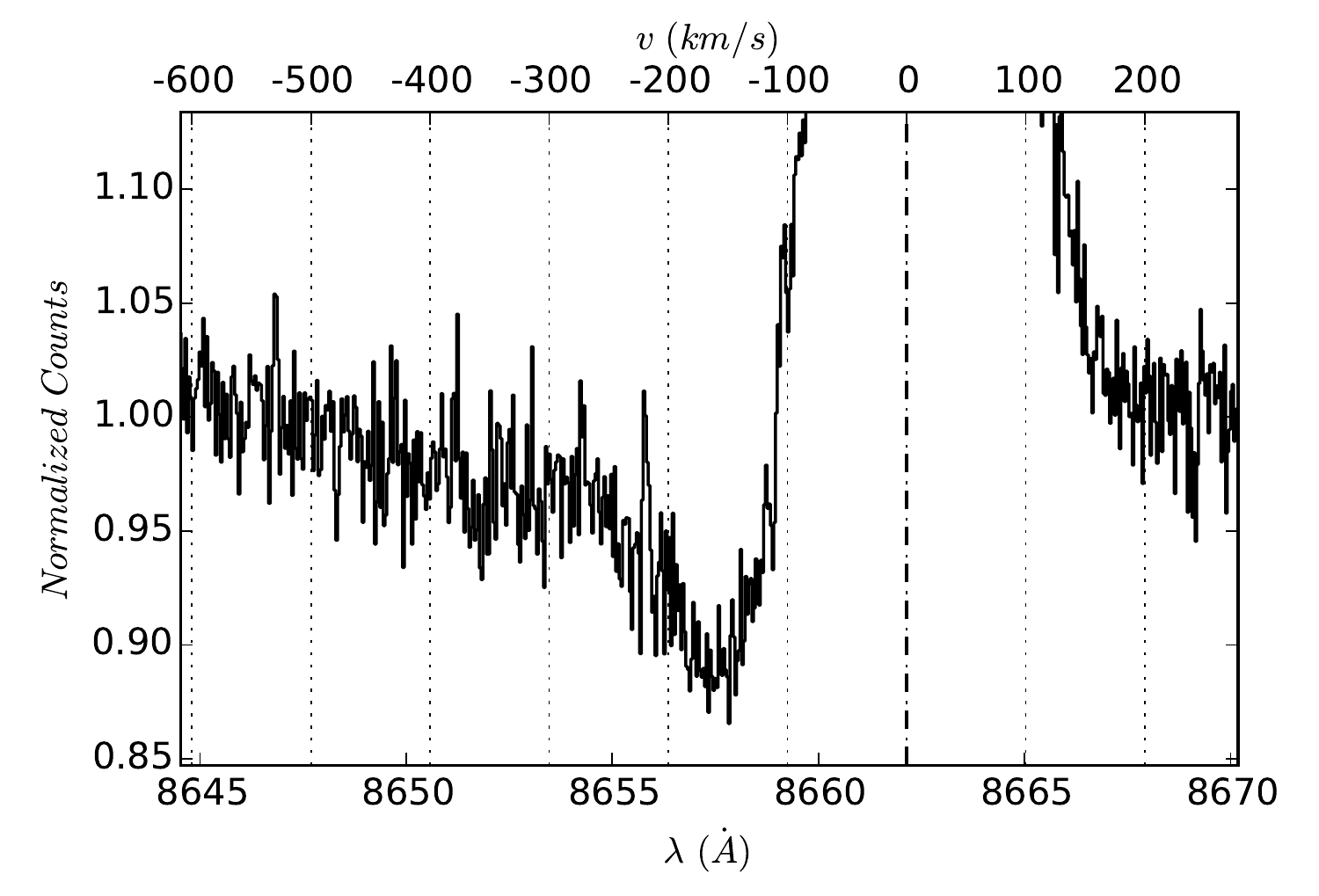} \\
    \centering\small (b) $\lambda$8662
  \end{tabular}
  \caption{Ca II IR triplet P-Cygni absorption line profile of V899 Mon observed on 2014 December 22 using SALT-HRS.}
  \label{SALTHRSCaII23Abs}
\end{figure}

Figure \ref{SALTHRSCaII123Emi} shows the resolved profiles of Ca II IR triple lines. All the three triplet lines ($\lambda$8498.02, $\lambda$8542.09, $\lambda$8662.14) show an asymmetric triangular profile with a steeper slope on the red side and a shallower slope on the blue side. The peaks of these lines are red shifted by +18.7 km/s, +21.4 km/s and +23.2 km/s, respectively. This increase in the red-shift is also seen in line center estimated by fitting a Gaussian profile to the lines. The $\sigma$ of the fitted Gaussian for each line is 1.177 $\mathring{A}$, 1.229 $\mathring{A}$ and 1.186 $\mathring{A}$, respectively. The ratio of line widths of $\lambda$8498 and $\lambda$8542 is 0.96, which implies $\lambda$8498 is $\sim$4\% thinner than $\lambda$8542, as seen in many other T-Tauri stars \citep{Hamann92}, though there is no statistically significant evidence for the peak of the former line to be larger than the later one. \citet{Hamann92} attributed these ratios to substantial opacity broadening of $\lambda$8542 or to lower dispersion velocity in the deeper part of the region from where Ca II IR triple lines originate. 
The velocities of the smooth broadened emission profiles extend upto $\pm$ 150 km/s.
Figure \ref{SALTHRSCaII23Abs} shows the P-Cygni absorption component in Ca II $\lambda$8542 and $\lambda$8662 lines. The absorption components are quite broad and extend up to $\sim$ -470 km/s in both the lines. The absorption component in $\lambda$8662 shows a prominent structure which extends only up to $\sim$ -250 km/s.

\begin{figure}
 \includegraphics[width=0.5\textwidth]{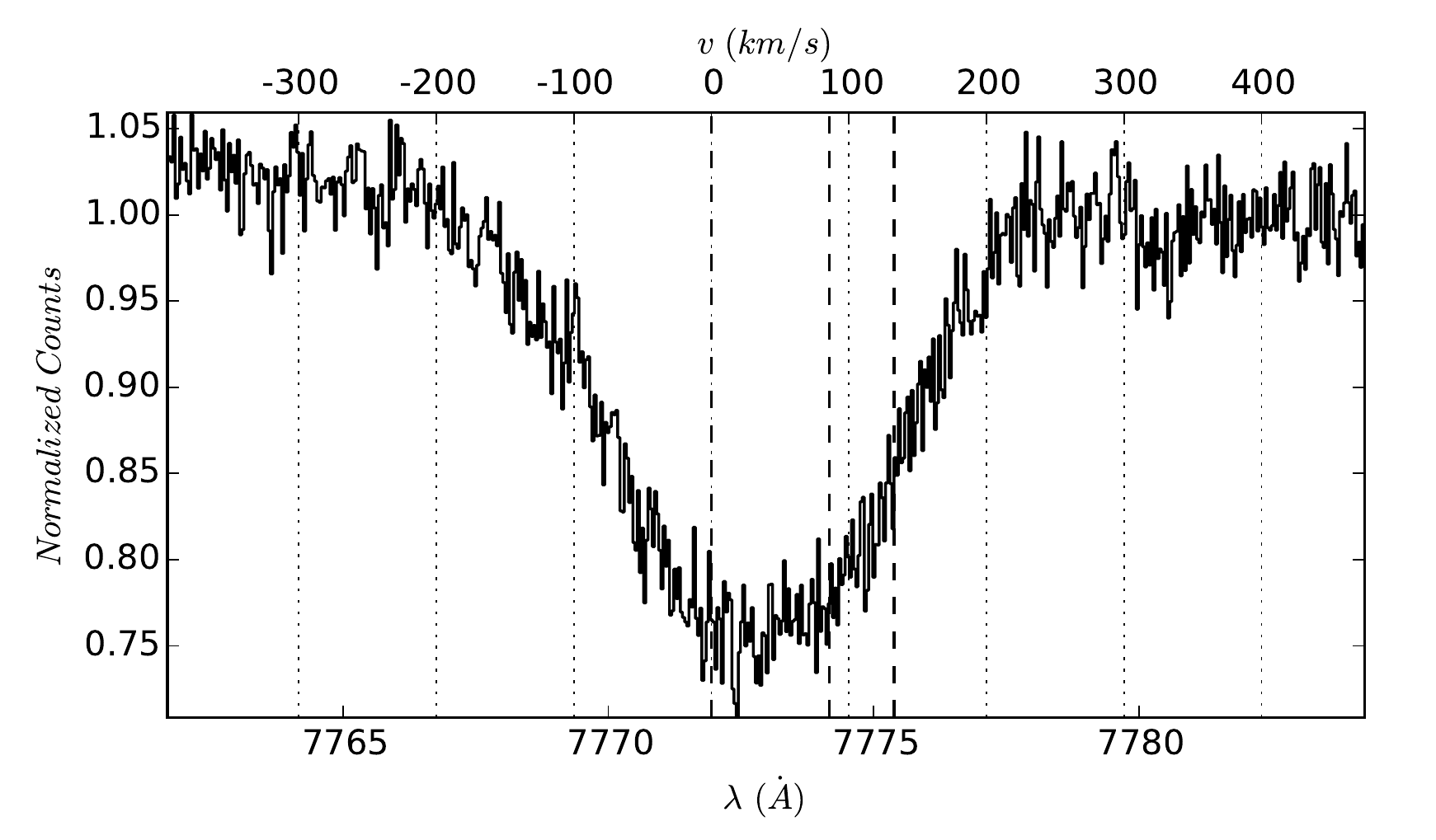}
 \caption{Blended OI 7773 triplet absorption line profile of V899 Mon observed on 2014 December 22 using SALT-HRS. The vertical dashed lines mark the central positions of its triplet components at 7771.94 $\mathring{A}$, 7774.17 $\mathring{A}$ and 7775.39 $\mathring{A}$ }
 \label{SALTHRSOI7773}
 \end{figure}

\begin{figure}
 \includegraphics[width=0.5\textwidth]{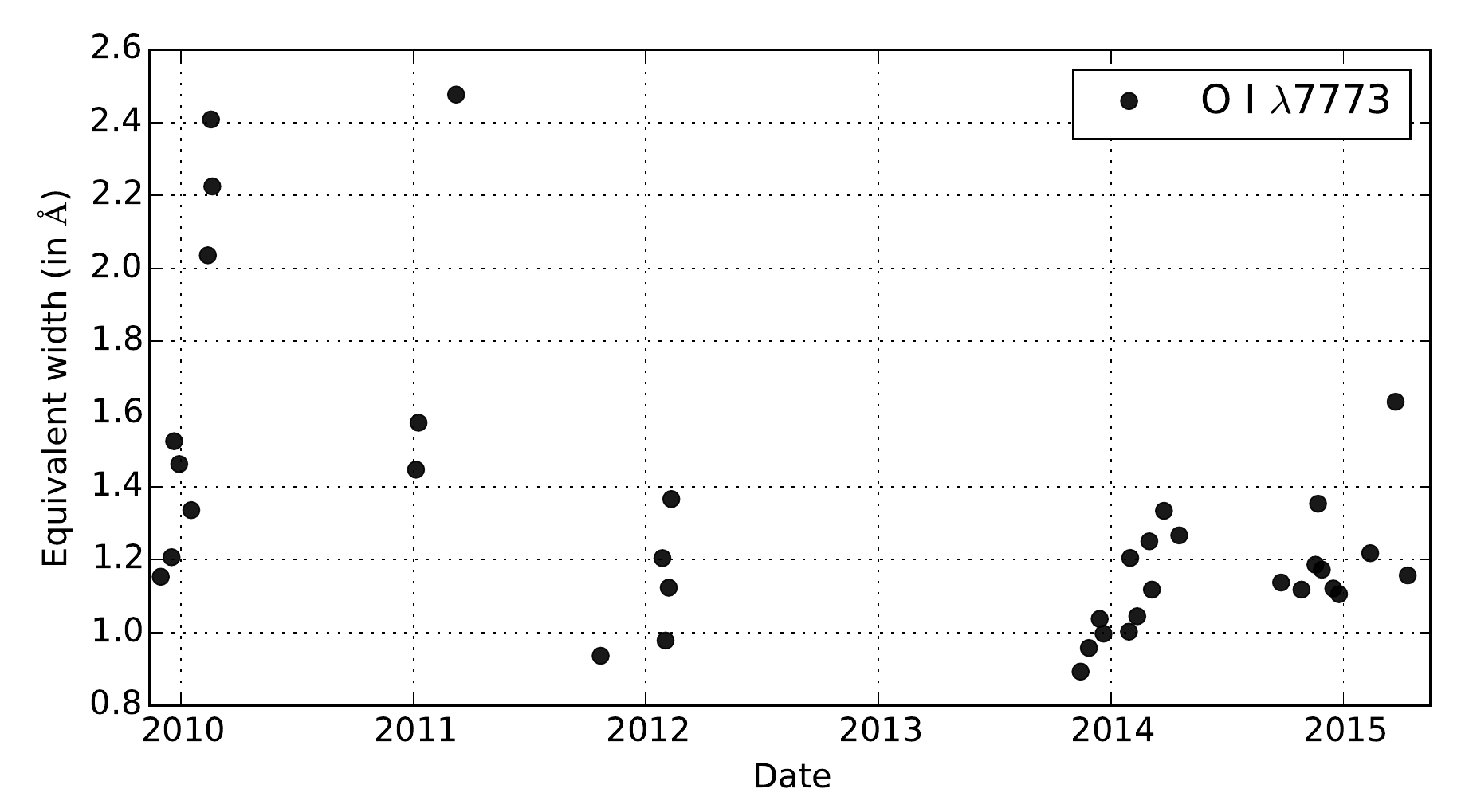}
 \caption{Evolution of equivalent width of O I $\lambda$7773 triplet, which traces the strength of turbulence in the disk of V899 Mon.}
 \label{OI7773eqw}
 \end{figure}

\paragraph{O I $\lambda$7773, $\lambda$8446}
The absorption component of O I triplet lines at 7773 $\mathring{A}$ (Figure \ref{SALTHRSOI7773}), which is not seen in the photosphere of cool stars, is believed to be formed in T-Tauri stars due to warm gas in the envelope or hot photosphere above the disk, and is an indicator of the turbulence \citep{Hamann92}. The equivalent widths of this line in V899 Mon spectra show a very strong increase and then sudden decrease in strength just before the source went into the quiescent phase (see Figure \ref{OI7773eqw}).  Due to turbulence or disk rotational broadening of the O I triplet lines at 7771.94 $\mathring{A}$, 7774.17 $\mathring{A}$ and 7775.39 $\mathring{A}$, our high resolution spectrum (see Figure \ref{SALTHRSOI7773}) shows a blended Gaussian absorption profile with $\sigma = 2.43$ $\mathring{A}$ (FWHM = 220 km/s). The velocities originating from turbulence or disk rotation extend up to $\sim$ $\pm$200 km/s.

Another O I triplet at $\lambda$8446 (containing 8446.25 $\mathring{A}$, 8446.36 $\mathring{A}$, 8446.76 $\mathring{A}$ triplet lines) is also seen in absorption in our SALT-HRS spectrum. These lines are also blended and have a combined FWHM = 174 km/s. The central position of the absorption line does not show any significant red-shift and if present can be constrained to be $<$ 5 km/s. This implies the $\sim$ +20 km/s (heliocentric velocity) red-shift component seen in emission line profiles like Ca II IR triplet and forbidden lines are not due to any uncorrected peculiar velocity of V899 Mon with respect to our sun.

\paragraph{Fe I $\lambda$8514, $\lambda$8387, $\lambda$8689, $\lambda$8675}
Many Fe I emission lines were detected in our medium resolution optical spectrum. All these lines were also resolved in SALT-HRS spectrum. Fe I $\lambda$8514 is a blend of two lines at 8514.07 $\mathring{A}$ and 8515.11 $\mathring{A}$; their blended profile has an FWHM of 76 km/s. Fe I 8387.777 $\mathring{A}$ has a +26.5 km/s red-shifted profile with an FWHM of 62.8 km/s, Fe I 8688.625 $\mathring{A}$ has a +26.74 km/s red-shifted profile with an FWHM of 63.3 km/s, whereas a weaker Fe I 8674.746 $\mathring{A}$ line has a +40 km/s red shifted profile with an FWHM of 59.2 km/s. The red-shifts in these lines are consistent with the red-shifts we have measured in other resolved lines (for example, in Ca II IR triplets lines and forbidden lines).

\subsection{Constrains from 1280 MHz Observation} \label{RadioResults}
V899 Mon was not detected in 1280 MHz GMRT radio continuum map observed during its second outburst. The local background noise $\sigma$ in the map after cleaning was $\sim$ 0.1 mJy. We could detect 8 other point sources in  the 28\arcmin\, $\times$ 28\arcmin\, FoV map centered at V899 Mon's position. 
For this study, we take a 5$\sigma$ $\sim$ 0.5 mJy to be a strict upper limit on V899 Mon's 1280 MHz flux.

The ionizing flux from the magnetospheric accretion can create an H II region around the central source. Even though Hydrogen Balmer lines detected in our optical spectra originate in the ionized region, since they are not optically thin, we cannot use them to constrain the extent of the H II region.

If we consider the H II region to be formed by a smooth, isothermal and isotropic stellar outflow, from our mass outflow rate and velocity estimates, we can obtain the expected flux using the following formula \citep{moran83}. \vbox{
\begin{eqnarray*}
S(\nu) &=& 1.3 \left(\frac{T_e}{10^4 K}\right)^{0.1} \left(\frac{\dot M}{10^{-6} M_\odot yr^{-1}}\right)^{4/3} \\
&&\left(\frac{v}{10^2 km \;s^{-1}}\right)^{-4/3} \left(\frac{D}{kpc}\right)^{-2} \left(\frac{\nu}{GHz}\right)^{0.6} \; mJy
\end{eqnarray*} }
where $S(\nu)$ is the expected flux density, $T_e$ is the electron temperature of the outflowing ionized wind, $\dot M$ is the mass outflow rate, $v$ is the terminal velocity of outflow, $D$ is distance to V899 Mon, and $\nu$ is the frequency of observation.
From our spectroscopic estimates of these quantities, we can substitute $\dot M$ = 0.26 $\times$ 10$^{-6}$ M$_\odot$/yr, outflow velocity $v$ = 1.5 $\times$ 10$^2$ km/s , T$_e$ = 0.9 $\times$ 10$^4$ K, D = 0.905 kpc, and $\nu$ = 1.28 GHz.
We obtain $S$(1.28 GHz) = 0.17 mJy, which is consistent with our observed upper limit of 0.5 mJy. Since the outflows we detected in optical spectrum are unlikely to be an isotropic outflow, our predicted estimate of 0.17 mJy is also an upper limit.

Instead of considering a radial density profile due to outflow, since we have an estimate of density from forbidden optical lines, we can calculate an upper limit on the maximum size of homogeneous, isotropic, spherical H II region around V899 Mon.
For an H II region of radius $R_s$, using a spherical volume emission measure, we have the expression \citep{moran83}.
\begin{eqnarray*}
S(\nu) &=& 3.444 \times 10^{-83} \left(\frac{10^4 K}{T_e}\right)^{0.35} \left(\frac{1 GHz}{\nu}\right)^{0.1} \\
&&\left(\frac{1 kpc}{D}\right)^2 \frac{4\pi R_s^3}{3} n_e^2 
\end{eqnarray*}
where, the radius of H II region $R_s$ is in cm, and electron density $n_e$ is per cm$^3$.
Taking 0.5 mJy as an upper limit on $S(\nu)$, and $n_e \sim 1 \times 10^6$/cm$^3$, we obtain $R_s < 20$ AU.
Since the flux of ionizing Lyman continuum inside an H II region is equal to the volume emission measure times the recombination rate, we have obtained the following relation for $N_{Lyc}$, 
\begin{eqnarray*} 
N_{Lyc} &=& 7.5487 \times 10^{43}  \times \left(\frac{S(\nu)}{mJy}\right) \left(\frac{T_e}{10^4 K}\right)^{-0.45} \\
&&\left(\frac{\nu}{1 GHz}\right)^{0.1} \left(\frac{D}{1 kpc}\right)^2 
\end{eqnarray*}
Using our upper limit of $S(\nu)$, we obtain an upper limit on the Lyman continuum flux from V899 Mon to be $N_{Lyc} < 5 \times 10^{43}$ photons/sec.

\section{Discussion} \label{discuss}

\subsection{Cause of break in the first outburst} \label{ReasonForBreak}
According to standard instability models of FUors/EXors, a critical disk surface density is required to sustain the instability. Disk transition occurs from outburst phase to quiescent phase when its surface density drops below that critical value \citep{bell94,zhu09}. Since the effective viscosity of disk is lower during the quiescent phase than the outburst phase, the surface density increases slowly during the quiescence than the rate at which it drained during the last outburst phase. Particularly in case of V899 Mon when the first outburst stopped, if the disk density had drained below the critical density, then it has to spend more time in the subsequent quiescent phase to replenish the disk before undergoing a second outburst. Such a scenario is also seen in the light curve of another famous and a very similar FUor candidate, V1647 Ori \citep{ninan13}. V582 Aur (FUor source) also showed multiple very brief quiescences during ongoing outburst \citep{semkov13}. In order to explain these short duration breaks in outbursts, we need to look for mechanisms during an outburst which can pause the accretion for a short duration before the disk gets critically drained. In such scenarios, the outburst can re-initiate as soon as the mechanism which was pausing the accretion disappears. 

The average rate of change in V899 Mon's magnitude during the onset of first outburst was 0.038 mag/month; on the other hand during the onset of second outburst it was larger than 0.15 mag/month. This difference implies the timescales of the mechanisms which triggered the first and the second outbursts are different.

Before V899 Mon transitioned to the quiescent phase, its spectroscopic observations showed an increase in the outflow activity (Section \ref{specresults}). Later the strong P-Cygni profiles of outflow suddenly vanished as the object entered the quiescent phase. Since outflows are believed to be driven by magnetic fields, it is possible that some magnetic field related mechanism is responsible for abruptly pausing the accretion during the active outburst phase. Spectral line fluxes which are proportional to accretion, as well as the continuum flux of V899 Mon increased suddenly during this short duration before the quiescence. In the final optical spectrum taken before the source transitioned from its first outburst phase to quiescent phase, we detect a sudden increase in equivalent width of OI $\lambda$7773 absorption line (an indicator of turbulence). They are indicative of a highly turbulent activity around the V899 Mon source just before it transitioned to quiescent phase.

Accretion in low-mass YSOs is generally accepted to be via magnetospheric accretion funnels from disk to star. Large rate of accretion can provide a negative feedback in many ways. There are various semi-analytic and numerical simulation studies in literature on instabilities which can break and restart magnetospheric accretion \citep{kulkarni08,orlando11,blinova15}. 
 For instance, the accretion can stop if the inner truncation radius of the disk moves outside the co-rotation radius \citep{dangelo10}. In another scenario, the differential rotation between the inner accretion disk and the star can lead to inflation of the funnel resulting in field lines opening and reconnecting, reducing accretion flow while enhancing outflow \citep{bouvier03}. These kinds of breaking mechanisms have the extra advantage that they can easily explain how a second outburst can re-start without having to replenish the depleted disk within the short quiescence phase.  Some other FUor sources like FU Ori, V1515 Cyg and V2493 Cyg also show a short duration dip in their light curves after they attained the initial peak of their outburst. All of them seem to indicate a complex negative feedback loop in magnetospheric accretion. 
 
It is interesting to note that, both V899 Mon and  V1647 Ori (discussed in \citet{ninan13}) spent shorter duration in the first outburst compared to their ongoing second outbursts. The reason for the second outburst being more stable than the first outburst could be because of the change in the inner disk's physical parameters by the heating or the draining of the disk during the first outburst (such as the extent of the interaction region between magnetosphere and inner disk, their coupling, inner radius of the disk, etc.). In case of V1647 Ori, from the stability in the phase of X-ray accretion spot, \citet{hamaguchi12} had shown that the location of the spot of accretion column shock on the star did not change significantly between the first and second outbursts. Hence, if the quiescent phase of V1647 Ori was due to disruption in magnetic funnel accretion, at least the base of the magnetic funnel on the star surface was not completely disrupted. 

To summarize, even though we do not have direct observational evidence to support increased magnetic activity, the outflow and turbulence signatures in the spectrum and continuum flux are consistent with instability in magnetic accretion funnel, and the timescales are strongly inconsistent with any other instability models which depend on critical disk surface densities to switch on and off the outburst.

\subsection{V899 Mon: FUor or EXor} \label{FUorEXorClassification}
The empirical classification between FUors and EXors is based on the similarity in observed properties with classical FUors (FU Ori, V1515 Cyg, etc.) and EXors (EX Lup, V1118 Ori, etc.). \citet{audard14} provide a detailed comparison of these objects from the literature. 
The $\sim$3 magnitude change in the brightness of V899 Mon during its outburst is typical of EXors family of outbursts. On the other hand, the duration of outbursts in V899 Mon, $\sim$4 years for first outburst and greater than 3.5 years for ongoing second outburst, is significantly more than the typical duration EXors spend in outburst (less than 1 - 3 years). Unlike V899 Mon, EXors also remain in quiescence for more duration than the time they spend in outburst phase. However, V899 Mon's outburst timescales are still much less than classical FUors.

Spectroscopically, classical FUors have all optical and NIR Hydrogen lines in absorption, while EXors have those lines in emission. In case of V899 Mon, H$\alpha$ and a small component of H$\beta$ are in emission, while all the other Hydrogen lines in optical and NIR are in absorption. CO band heads starting at 2.29 $\mu m$ are also in absorption. Hence, spectroscopically also V899 Mon lies between classical FUors and EXors. The L$_{bol}$ estimates of V899 Mon from SED are greater than that of typical EXors, and they are less than that of classical FUors. The mass accretion rate during outburst phase is typical of EXors, and is an order less than that of classical FUors.

Many of the recently discovered outbursts like V1647 Ori, OO Ser, etc. also show such intermediate properties between the classical bimodal classification of FUors and EXors. Discovery of more such intermediate type outburst sources like V899 Mon indicates that this 
family of episodic accretion outbursts probably have a \textquotedblleft continuum\textquotedblright\, distribution and not bimodal.

\section{Conclusions} \label{conclusion}

We have carried out a long-term monitoring of V899 Mon from September 2009 to April 2015. During this period, V899 Mon underwent transition from the first outburst phase to the quiescent phase and then back to the second outburst phase. Following are the main results from our study.
\begin{enumerate}
\item Optical and NIR spectroscopy of V899 Mon confirms it to be a member of the FUors/EXors family of outbursts. Photometrically and spectroscopically V899 Mon's properties lie between EXors and classical FUors. But it is probably more similar to EXors than classical FUors.
\item At the end of 2010, V899 Mon abruptly ceased its first outburst phase and transitioned to the quiescent phase for a duration of little less than a year, immediately after which it returned to the second outburst.
\item Just before the break in the first outburst phase of V899 Mon, its spectra showed heavy outflow activity (indicated by a strong P-Cygni) and increased turbulence.
\item The excess flux of outburst initially had a cooler temperature than V899 Mon's photosphere and slowly became hotter as V899 Mon transitioned from the quiescent phase to the second outburst phase. This was seen in the change of $V$-$R$ and $V$-$I$ color of V899 Mon, which had reddest values during the intermediate transition period.
\item The outflows indicated by P-Cygni profiles completely disappeared during the quiescent phase of V899 Mon, and reappeared gradually only after V899 Mon reached its peak of the second outburst.
\item High resolution spectrum of the forbidden emission line profiles originating in outflows/jets show blue-shifted velocity components extending up to -500 km/s, while the outflows traced by H$\alpha$ P-Cygni absorption profiles originating near the central star show complex multiple absorption components with velocities up to -722 km/s.
\item As expected from accretion outbursts, accretion rates estimated of V899 Mon from spectral lines show significant variation between first outburst, quiescent and second outburst phases. 
\item Since the A$_V$ estimate from far-infrared is inconsistent with optical observations, V899 Mon is not embedded inside the far-infrared clump detected at its location. It is possible that our line of sight is through a cavity cleared out in envelope. The clump is thermally influenced by the irradiation from V899 Mon since it showed variation in far-infrared flux between the first and the second outburst phases. 
\item Since the disk replenishes slowly in the quiescent phase, all instability models which depend on the disk surface density to cross a certain threshold to trigger/stop an outburst are incompatible with V899 Mon's light curve. The case is similar to V1647 Ori.
\item The heavy outflow activity we detected just before V899 Mon transitioned to the quiescent phase is consistent with various magnetic instabilities which can arise in magnetospheric accretion. Such instability driven breaks in accretion can restart outburst immediately once the magnetic accretion funnel stabilizes.
\end{enumerate}

\acknowledgments
We thank the anonymous referee for giving us invaluable comments and suggestions that improved the content and structure of the paper.
It is a pleasure to thank all the members of the Infrared Astronomy Group of TIFR for their support during the TIRCAM2 and TIRSPEC observations. The authors thank the staff of CREST at Bangalore and HCT at Hanle (Ladakh), operated by the Indian Institute of Astrophysics, Bangalore; IGO at Girawali, operated by Inter-University Centre for Astronomy and Astrophysics, Pune; and  GMRT operated by the National Center for Radio Astrophysics of the Tata Institute of Fundamental Research (TIFR) for their assistance and support during observations. High resolution spectrum reported in this paper was obtained with the Southern African Large Telescope (SALT), and we would like to thank Dr. Brent Miszalski, and the entire SALT team for conducting SALT observations. This publication makes use of data products from the Wide-field Infrared Survey Explorer (WISE), which is a joint project of the University of California, Los Angeles, and the Jet Propulsion Laboratory/California Institute of Technology, funded by the National Aeronautics and Space Administration (NASA). The CSS survey is funded by the NASA under Grant No. NNG05GF22G issued through the Science Mission Directorate Near-Earth Objects Observations Program.  The CRTS survey is supported by the U.S. National Science Foundation under grants AST-0909182.
All the plots are generated using the 2D graphics environment \textit{Matplotlib} \citep{hunter07}.



{\it Facilities:} \facility{HCT}, \facility{TIRSPEC (HCT)}, \facility{HFOSC (HCT)}, \facility{IGO}, \facility{HRS (SALT)}.

\bibliography{V899MonPaper2015}

\begin{thebibliography}{}
\expandafter\ifx\csname natexlab\endcsname\relax\def\natexlab#1{#1}\fi

\bibitem[{\'{A}brah\'{a}m {et~al.}(2009)\'{A}brah\'{a}m, Juh\'{a}sz, Dullemond,
  K\'{o}sp\'{a}l, van Boekel, Bouwman, Henning, Mo\'{o}r, Mosoni,
  Sicilia-Aguilar, \& Sipos}]{abraham09}
\'{A}brah\'{a}m, P., Juh\'{a}sz, A., Dullemond, C.~P., {et~al.} 2009, Nature,
  459, 224

\bibitem[{Alcal\'{a} {et~al.}(2014)Alcal\'{a}, Natta, Manara, Spezzi, Stelzer,
  Frasca, Biazzo, Covino, Randich, Rigliaco, Testi, Comer\'{o}n, Cupani, \&
  D'Elia}]{alcala14}
Alcal\'{a}, J.~M., Natta, A., Manara, C.~F., {et~al.} 2014, Astronomy \&
  Astrophysics, 561, A2

\bibitem[{Andr\'{e} {et~al.}(2010)Andr\'{e}, Men'shchikov, Bontemps, Könyves,
  Motte, Schneider, Didelon, Minier, Saraceno, Ward-Thompson, Di~Francesco,
  White, Molinari, Testi, Abergel, Griffin, Henning, Royer, Merín, Vavrek,
  Attard, Arzoumanian, Wilson, Ade, Aussel, Baluteau, Benedettini, Bernard,
  Blommaert, Cambrésy, Cox, Di~Giorgio, Hargrave, Hennemann, Huang, Kirk,
  Krause, Launhardt, Leeks, Le~Pennec, Li, Martin, Maury, Olofsson, Omont,
  Peretto, Pezzuto, Prusti, Roussel, Russeil, Sauvage, Sibthorpe,
  Sicilia-Aguilar, Spinoglio, Waelkens, Woodcraft, \& Zavagno}]{andre10}
Andr\'{e}, P., Men'shchikov, A., Bontemps, S., {et~al.} 2010, Astronomy and
  Astrophysics, 518, L102

\bibitem[{{Astropy Collaboration} {et~al.}(2013){Astropy Collaboration},
  Robitaille, Tollerud, Greenfield, Droettboom, Bray, Aldcroft, Davis,
  Ginsburg, Price-Whelan, Kerzendorf, Conley, Crighton, Barbary, Muna,
  Ferguson, Grollier, Parikh, Nair, Unther, Deil, Woillez, Conseil, Kramer,
  Turner, Singer, Fox, Weaver, Zabalza, Edwards, Azalee~Bostroem, Burke, Casey,
  Crawford, Dencheva, Ely, Jenness, Labrie, Lim, Pierfederici, Pontzen, Ptak,
  Refsdal, Servillat, \& Streicher}]{astropy13}
{Astropy Collaboration}, Robitaille, T.~P., Tollerud, E.~J., {et~al.} 2013,
  Astronomy and Astrophysics, 558, 33

\bibitem[{{Audard} {et~al.}(2014){Audard}, {{\'A}brah{\'a}m}, {Dunham},
  {Green}, {Grosso}, {Hamaguchi}, {Kastner}, {K{\'o}sp{\'a}l}, {Lodato},
  {Romanova}, {Skinner}, {Vorobyov}, \& {Zhu}}]{audard14}
{Audard}, M., {{\'A}brah{\'a}m}, P., {Dunham}, M.~M., {et~al.} 2014, Protostars
  and Planets VI, 387

\bibitem[{Baraffe {et~al.}(2012)Baraffe, Vorobyov, \& Chabrier}]{baraffe12}
Baraffe, I., Vorobyov, E., \& Chabrier, G. 2012, The Astrophysical Journal,
  756, 118

\bibitem[{Bell \& Lin(1994)}]{bell94}
Bell, K.~R., \& Lin, D. N.~C. 1994, The Astrophysical Journal, 427, 987

\bibitem[{Bessell \& Brett(1988)}]{bessell88}
Bessell, M.~S., \& Brett, J.~M. 1988, Publications of the Astronomical Society
  of the Pacific, 100, 1134

\bibitem[{Blinova {et~al.}(2015)Blinova, Romanova, \& Lovelace}]{blinova15}
Blinova, A.~A., Romanova, M.~M., \& Lovelace, R. V.~E. 2015, arXiv:1501.01948
  [astro-ph]

\bibitem[{Bouvier {et~al.}(2003)Bouvier, Grankin, Alencar, Dougados,
  Fernández, Basri, Batalha, Guenther, Ibrahimov, Magakian, Melnikov, Petrov,
  Rud, \& Osorio}]{bouvier03}
Bouvier, J., Grankin, K.~N., Alencar, S. H.~P., {et~al.} 2003, Astronomy \&
  Astrophysics, 409, 24

\bibitem[{{Bramall} {et~al.}(2010){Bramall}, {Sharples}, {Tyas}, {Schmoll},
  {Clark}, {Luke}, {Looker}, {Dipper}, {Ryan}, {Buckley}, {Brink}, \&
  {Barnes}}]{bramall10}
{Bramall}, D.~G., {Sharples}, R., {Tyas}, L., {et~al.} 2010, in Society of
  Photo-Optical Instrumentation Engineers (SPIE) Conference Series, Vol. 7735,
  Society of Photo-Optical Instrumentation Engineers (SPIE) Conference Series,
  4

\bibitem[{Cardelli {et~al.}(1989)Cardelli, Clayton, \& Mathis}]{cardelli89}
Cardelli, J.~A., Clayton, G.~C., \& Mathis, J.~S. 1989, The Astrophysical
  Journal, 345, 245

\bibitem[{Ciardi {et~al.}(1998)Ciardi, Woodward, Clemens, Harker, \&
  Rudy}]{ciardi98}
Ciardi, D.~R., Woodward, C.~E., Clemens, D.~P., Harker, D.~E., \& Rudy, R.~J.
  1998, The Astronomical Journal, 116, 349

\bibitem[{D'Angelo \& Spruit(2010)}]{dangelo10}
D'Angelo, C.~R., \& Spruit, H.~C. 2010, Monthly Notices of the Royal
  Astronomical Society, 406, 1208

\bibitem[{Evans~II {et~al.}(2009)Evans~II, Dunham, Jørgensen, Enoch, Merín,
  Dishoeck, Alcal\'{a}, Myers, Stapelfeldt, Huard, Allen, Harvey, Kempen,
  Blake, Koerner, Mundy, Padgett, \& Sargent}]{evans09}
Evans~II, N.~J., Dunham, M.~M., Jørgensen, J.~K., {et~al.} 2009, The
  Astrophysical Journal Supplement Series, 181, 321

\bibitem[{Greene {et~al.}(1994)Greene, Wilking, Andre, Young, \&
  Lada}]{greene94}
Greene, T.~P., Wilking, B.~A., Andre, P., Young, E.~T., \& Lada, C.~J. 1994,
  The Astrophysical Journal, 434, 614

\bibitem[{Gullbring {et~al.}(1998)Gullbring, Hartmann, Briceño, \&
  Calvet}]{gullbring98}
Gullbring, E., Hartmann, L., Briceño, C., \& Calvet, N. 1998, The
  Astrophysical Journal, 492, 323

\bibitem[{Hamaguchi {et~al.}(2012)Hamaguchi, Grosso, Kastner, Weintraub,
  Richmond, Petre, Teets, \& Principe}]{hamaguchi12}
Hamaguchi, K., Grosso, N., Kastner, J.~H., {et~al.} 2012, The Astrophysical
  Journal, 754, 32

\bibitem[{Hamann(1994)}]{hamann94}
Hamann, F. 1994, The Astrophysical Journal Supplement Series, 93, 485

\bibitem[{Hamann \& Persson(1992)}]{Hamann92}
Hamann, F., \& Persson, S.~E. 1992, The Astrophysical Journal Supplement
  Series, 82, 247

\bibitem[{Hartigan {et~al.}(1995)Hartigan, Edwards, \& Ghandour}]{hartigan95}
Hartigan, P., Edwards, S., \& Ghandour, L. 1995, The Astrophysical Journal,
  452, 736

\bibitem[{Hartmann(1998)}]{hartmann98}
Hartmann, L. 1998, Accretion {Processes} in {Star} {Formation} (Cambridge Univ.
  Press, Cambridge)

\bibitem[{Hartmann \& Kenyon(1996)}]{hartmann96}
Hartmann, L., \& Kenyon, S.~J. 1996, Annual Review of Astronomy and
  Astrophysics, 34, 207

\bibitem[{Herbig(1977)}]{herbig77}
Herbig, G.~H. 1977, Astrophysical Journal, Part 1, vol. 217, Nov. 1, 1977, p.
  693-715.

\bibitem[{Hunt {et~al.}(1998)Hunt, Mannucci, Testi, Migliorini, Stanga, Baffa,
  Lisi, \& Vanzi}]{hunt98}
Hunt, L.~K., Mannucci, F., Testi, L., {et~al.} 1998, The Astronomical Journal,
  115, 2594

\bibitem[{Hunter(2007)}]{hunter07}
Hunter, J.~D. 2007, Computing in Science \& Engineering, 9, 90

\bibitem[{Ioannidis \& Froebrich(2012)}]{ioannidis12}
Ioannidis, G., \& Froebrich, D. 2012, Monthly Notices of the Royal Astronomical
  Society, 425, 1380

\bibitem[{Jones {et~al.}(2001)Jones, Oliphant, Peterson, {et~al.}}]{jones01}
Jones, E., Oliphant, T., Peterson, P., {et~al.} 2001, {SciPy}: Open source
  scientific tools for {Python}, [Online; accessed 2015-03-03]

\bibitem[{Kenyon {et~al.}(1990)Kenyon, Hartmann, Strom, \& Strom}]{kenyon90}
Kenyon, S.~J., Hartmann, L.~W., Strom, K.~M., \& Strom, S.~E. 1990, The
  Astronomical Journal, 99, 869

\bibitem[{K\'{o}sp\'{a}l {et~al.}(2011)K\'{o}sp\'{a}l, \'{A}brah\'{a}m, Goto,
  Reg\'{a}ly, Dullemond, Henning, Juh\'{a}sz, Sicilia-Aguilar, \& van~den
  Ancker}]{kospal11}
K\'{o}sp\'{a}l, a., \'{A}brah\'{a}m, P., Goto, M., {et~al.} 2011, The
  Astrophysical Journal, 736, 72

\bibitem[{K\'{o}sp\'{a}l {et~al.}(2013)K\'{o}sp\'{a}l, \'{A}brah\'{a}m,
  Acosta-Pulido, Morales, Balog, Carnerero, Szegedi-Elek, Farkas, Henning,
  Kelemen, Kov\'{a}cs, Kun, Marton, M\'{e}sz\'{a}ros, Mo\'{o}r, P\'{a}l,
  S\'{a}rneczky, Szak\'{a}ts, Szalai, Szing, T\'{o}th, Turner, \&
  Vida}]{kospal13}
K\'{o}sp\'{a}l, A., \'{A}brah\'{a}m, P., Acosta-Pulido, J.~A., {et~al.} 2013,
  Astronomy \& Astrophysics, 551, A62

\bibitem[{Kulkarni \& Romanova(2008)}]{kulkarni08}
Kulkarni, A.~K., \& Romanova, M.~M. 2008, Monthly Notices of the Royal
  Astronomical Society, 386, 673

\bibitem[{Landolt(1992)}]{landolt92}
Landolt, A.~U. 1992, The Astronomical Journal, 104, 340

\bibitem[{Lombardi {et~al.}(2011)Lombardi, Alves, \& Lada}]{lombardi11}
Lombardi, M., Alves, J., \& Lada, C.~J. 2011, Astronomy \& Astrophysics, 535,
  A16

\bibitem[{McGehee {et~al.}(2004)McGehee, Smith, Henden, Richmond, Knapp,
  Finkbeiner, Ivezic, \& Brinkmann}]{mcgehee04}
McGehee, P., Smith, J., Henden, A., {et~al.} 2004, Arxiv preprint
  astro-ph/0408308

\bibitem[{Men'shchikov {et~al.}(2012)Men'shchikov, Andr\'{e}, Didelon, Motte,
  Hennemann, \& Schneider}]{menshchikov12}
Men'shchikov, A., Andr\'{e}, P., Didelon, P., {et~al.} 2012, Astronomy \&
  Astrophysics, 542, A81

\bibitem[{Meyer {et~al.}(1997)Meyer, Calvet, \& Hillenbrand}]{meyer97}
Meyer, M.~R., Calvet, N., \& Hillenbrand, L.~A. 1997, The Astronomical Journal,
  114, 288

\bibitem[{{Moran}(1983)}]{moran83}
{Moran}, J.~M. 1983, \rmxaa, 7, 95

\bibitem[{Muzerolle {et~al.}(1998)Muzerolle, Hartmann, \& Calvet}]{muzerolle98}
Muzerolle, J., Hartmann, L., \& Calvet, N. 1998, The Astronomical Journal, 116,
  455

\bibitem[{Naik {et~al.}(2012)Naik, Ojha, Ghosh, Poojary, Jadhav, Meshram,
  Sandimani, Bhagat, D'Costa, Gharat, Bakalkar, Ninan, \& Joshi}]{naik12}
Naik, M.~B., Ojha, D.~K., Ghosh, S.~K., {et~al.} 2012, Bull. Astr. Soc. India,
  40, 531

\bibitem[{Ninan {et~al.}(2013)Ninan, Ojha, Bhatt, Ghosh, Mohan, Mallick,
  Tamura, \& Henning}]{ninan13}
Ninan, J.~P., Ojha, D.~K., Bhatt, B.~C., {et~al.} 2013, The Astrophysical
  Journal, 778, 116

\bibitem[{Ninan {et~al.}(2014)Ninan, Ojha, Ghosh, D'Costa, Naik, Poojary,
  Sandimani, Meshram, Jadhav, Bhagat, Gharat, Bakalkar, Prabhu, Anupama, \&
  Toomey}]{ninan14}
Ninan, J.~P., Ojha, D.~K., Ghosh, S.~K., {et~al.} 2014, Journal of Astronomical
  Instrumentation, 1450006

\bibitem[{Nisini {et~al.}(1995)Nisini, Milillo, Saraceno, \& Vitali}]{nisini95}
Nisini, B., Milillo, A., Saraceno, P., \& Vitali, F. 1995, Astronomy and
  Astrophysics, 302, 169

\bibitem[{Ojha {et~al.}(2012)Ojha, Ghosh, D'Costa, Naik, Sandimani, Poojary,
  Bhagat, Jadhav, Meshram, \& Bakalkar}]{ojha12}
Ojha, D.~K., Ghosh, S.~K., D'Costa, S. L.~A., {et~al.} 2012, in Astronomical
  Society of India Conference Series, Vol.~4, 191

\bibitem[{Orlando {et~al.}(2011)Orlando, Reale, Peres, \& Mignone}]{orlando11}
Orlando, S., Reale, F., Peres, G., \& Mignone, A. 2011, Monthly Notices of the
  Royal Astronomical Society, 415, 3380

\bibitem[{{Osterbrock} \& {Ferland}(2006)}]{osterbrock06}
{Osterbrock}, D.~E., \& {Ferland}, G.~J. 2006, {Astrophysics of gaseous nebulae
  and active galactic nuclei} (CA: University Science Books)

\bibitem[{Rieke \& Lebofsky(1985)}]{rieke85}
Rieke, G.~H., \& Lebofsky, M.~J. 1985, The Astrophysical Journal, 288, 618

\bibitem[{Robitaille {et~al.}(2007)Robitaille, Whitney, Indebetouw, \&
  Wood}]{robitaille07}
Robitaille, T.~P., Whitney, B.~A., Indebetouw, R., \& Wood, K. 2007, The
  Astrophysical Journal Supplement Series, 169, 328

\bibitem[{Safron {et~al.}(2015)Safron, Fischer, Megeath, Furlan, Stutz, Stanke,
  Billot, Rebull, Tobin, Ali, Allen, Booker, Watson, \& Wilson}]{safron15}
Safron, E.~J., Fischer, W.~J., Megeath, S.~T., {et~al.} 2015, The Astrophysical
  Journal Letters, 800, L5

\bibitem[{Scholz {et~al.}(2013)Scholz, Froebrich, \& Wood}]{scholz13}
Scholz, A., Froebrich, D., \& Wood, K. 2013, Monthly Notices of the Royal
  Astronomical Society, 430, 2910

\bibitem[{Semkov {et~al.}(2013)Semkov, Peneva, Munari, Dennefeld, Mito,
  Dimitrov, Ibryamov, \& Stoyanov}]{semkov13}
Semkov, E.~H., Peneva, S.~P., Munari, U., {et~al.} 2013, Astronomy \&
  Astrophysics, 556, A60

\bibitem[{Siess {et~al.}(2000)Siess, Dufour, \& Forestini}]{siess00}
Siess, L., Dufour, E., \& Forestini, M. 2000, Astronomy and Astrophysics, 358,
  593

\bibitem[{{Stahler} \& {Palla}(2005)}]{stahler05}
{Stahler}, S.~W., \& {Palla}, F. 2005, {The Formation of Stars} (Wiley-VCH)

\bibitem[{Swarup {et~al.}(1991)Swarup, Ananthakrishnan, Kapahi, Rao,
  Subrahmanya, \& Kulkarni}]{swarup91}
Swarup, G., Ananthakrishnan, S., Kapahi, V.~K., {et~al.} 1991, Current Science,
  Vol. 60, {NO}.2/{JAN}25, P. 95, 1991, 60, 95

\bibitem[{van~der Walt {et~al.}(2011)van~der Walt, Colbert, \&
  Varoquaux}]{vanderWalt11}
van~der Walt, S., Colbert, S.~C., \& Varoquaux, G. 2011, Computing in Science
  \& Engineering, 13, 22

\bibitem[{van~der Walt {et~al.}(2014)van~der Walt, Schönberger,
  Nunez-Iglesias, Boulogne, Warner, Yager, Gouillart, \& Yu}]{vanderWalt14}
van~der Walt, S., Schönberger, J.~L., Nunez-Iglesias, J., {et~al.} 2014,
  PeerJ, 2, e453

\bibitem[{{Wils} {et~al.}(2009){Wils}, {Greaves}, {Drake}, \&
  {Catelan}}]{wils09}
{Wils}, P., {Greaves}, J., {Drake}, A.~J., \& {Catelan}, M. 2009, Central
  Bureau Electronic Telegrams, 2033, 1

\bibitem[{Wright {et~al.}(2010)Wright, Eisenhardt, Mainzer, Ressler, Cutri,
  Jarrett, Kirkpatrick, Padgett, McMillan, Skrutskie, Stanford, Cohen, Walker,
  Mather, Leisawitz, Iii, McLean, Benford, Lonsdale, Blain, Mendez, Irace,
  Duval, Liu, Royer, Heinrichsen, Howard, Shannon, Kendall, Walsh, Larsen,
  Cardon, Schick, Schwalm, Abid, Fabinsky, Naes, \& Tsai}]{wright10}
Wright, E.~L., Eisenhardt, P. R.~M., Mainzer, A.~K., {et~al.} 2010, The
  Astronomical Journal, 140, 1868

\bibitem[{Zhu {et~al.}(2009)Zhu, Hartmann, Gammie, \& McKinney}]{zhu09}
Zhu, Z., Hartmann, L., Gammie, C., \& McKinney, J.~C. 2009, The Astrophysical
  Journal, 701, 620

\end{thebibliography}

\clearpage
\begin{deluxetable}{lccll}
 
\tabletypesize{\footnotesize}
\tablecolumns{5} 
\tablewidth{0pt}
\tablecaption{  Observation log of V899 Mon \label{table:Obs_Log}}
\tablehead{ \colhead{Date} & \colhead{JD} & \colhead{FWHM\tablenotemark{a}} & \colhead{Filter(s)/Grism(s)}& \colhead{Instrument(s)}}
\startdata
2009 Nov 30 & 2455166 &  1\arcsec.9 & V,R,gr8            & HFOSC	\\
2009 Dec 04 & 2455170 &	 3\arcsec.1 & V,R,I				 & HFOSC	\\
2009 Dec 16 & 2455182 &	 1\arcsec.8 & V,R,I				 & HFOSC	\\
2009 Dec 17 & 2455183 &	    & gr7,gr8				 & HFOSC	\\
\enddata
\tablenotetext{a}{Measured average FWHM. This is a measure of the seeing.}
\tablecomments{(This table is available in its entirety in a machine-readable form in the online journal. A portion is shown here for guidance regarding its form and content.)}

\end{deluxetable}

\begin{deluxetable}{lcccccccc}
 
\tabletypesize{\footnotesize}
\tablecolumns{5} 
\tablewidth{0pt}
\tablecaption{  Magnitudes of V899 Mon \label{table:PhotMags}}
\tablehead{ \colhead{JD} & \colhead{$U$} & \colhead{$B$} & \colhead{$V$}& \colhead{$R$}& \colhead{$I$}& \colhead{$J$}& \colhead{$H$}& \colhead{$K_S$}}
\startdata
2455166 & \nodata & \nodata & 14.57 & 13.40 & \nodata & \nodata & \nodata & \nodata \\
2455170 & \nodata & \nodata & 14.22 & 13.10 & 12.08 & \nodata & \nodata &  \nodata \\
2455182 & \nodata & \nodata & 13.61 & 12.65 & 11.57 & \nodata & \nodata &  \nodata \\
2455186 & \nodata & \nodata & 13.51 & 12.37 & 11.37 & \nodata & \nodata &  \nodata \\

\enddata
\tablecomments{Errors on \textit{U, B, V, R} and \textit{I} magnitudes are $<$ 0.02 mag, TIRSPEC \textit{J, H and K$_S$} magnitudes have errors $<$ 0.02 mag, and TIRCAM2 and NIRCAM \textit{J, H, K} magnitudes have errors $<$ 0.06 mag.}
\tablecomments{(This table is available in its entirety in a machine-readable form in the online
journal. A portion is shown here for guidance regarding its form and content.)}
\end{deluxetable}

\begin{deluxetable}{lccc}
 
\tabletypesize{\footnotesize}
\tablecolumns{5} 
\tablewidth{0pt}
\tablecaption{  SED fit results of V899 Mon \label{table:SEDfitResults}}
\tablehead{ \colhead{Parameter} & \colhead{Quiescence } & \colhead{1$^{st}$ outburst} & \colhead{Far-IR alone}}
\startdata
A$_V$ (mag) & 4.5 $\pm$1.1 & 4.2 $\pm$0.34 & \nodata \\
distance (pc) & 870 $\pm$80 & 794 $\pm$30 & 891 $\pm$55 \\
Age$_*$ (Myr) & 4.84 $\pm$2.1 & 1.93 $\pm$1.8 & 0.0073 $\pm$.002 \\
Mass$_*$ (M$_\odot$) & 3.7 $\pm$0.3 & 5.1 $\pm$0.5 & 0.57 $\pm$0.04 \\
$\dot{M}_{envelope}$ (M$_\odot$) & 0 $\pm$5.7e-09 & 0 $\pm$7.9e-08 & 24.7e-05 $\pm$2.4e-05 \\
Mass$_{Disk}$ (M$_\odot$) & 1.2e-05 $\pm$0.0179 & 1.4e-03 $\pm$2.07e-03 & 0.048 $\pm$0.008 \\
$\dot{M}_{disk}$ (M$_\odot$/yr) & 1.54e-10 $\pm$4.9e-07 & 4.82e-09 $\pm$7.6e-08 & 5.48e-07 $\pm$2.7e-07 \\
L$_{total}$ (L$_\odot$) & 162 $\pm$79 & 419 $\pm$168 & 8.55 $\pm$1.54 \\
Mass$_{env}$ (M$_\odot$) & 8.13e-08 $\pm$0.027 & 1.52e-05 $\pm$0.07 & 23.43 $\pm$2.0 \\

\enddata
\tablecomments{Median and standard deviation of the best fitted 20 models in each case.}

\end{deluxetable}

\begin{deluxetable}{lccc}
 
\tabletypesize{\footnotesize}
\tablecolumns{5} 
\tablewidth{0pt}
\tablecaption{ Far-infrared fluxes of V899 Mon \label{table:herschelfluxs}}
\tablehead{ \colhead{Band} & \colhead{2010 Sep 14 } & \colhead{2013 Mar 06} & \colhead{$\Delta$ flux}}
\startdata
PACS 70 $\mu m$ & \nodata & 2.01 Jy & \nodata \\
PACS 160 $\mu m$  & \nodata & 7.02 Jy &  \nodata \\
SPIRE 250 $\mu m$ & 9.02 Jy & 6.86 Jy & 2.86 Jy \\
SPIRE 350 $\mu m$ & 6.82 Jy & 5.68 Jy & 1.48  Jy \\
SPIRE 500 $\mu m$ & 4.65 Jy & 4.06 Jy & 0.33  Jy \\

\enddata
\tablecomments{$\Delta$ flux between two epochs were estimated by aperture photometry on difference images using apertures 48\arcsec\,,60\arcsec\, and 70\arcsec\, for 250 $\mu m$, 350 $\mu m$ and 500 $\mu m$ respectively.}

\end{deluxetable}

\begin{deluxetable}{lccccccc}
 
\tabletypesize{\footnotesize}
\tablecolumns{8} 
\tablewidth{0pt}
\tablecaption{Spectral line measurements \label{table:linefluxeqw}}
\tablehead{ \colhead{Date} & \colhead{JD} & \multicolumn{4}{c}{H$\alpha$ $\lambda$6562.8}   &  \multicolumn{2}{c}{Ca II $\lambda$8498} \\
 \colhead{} & \colhead{} & \colhead{eqw   } & \colhead{ Flux } & \colhead{eqw$_{abs}$   } & \colhead{  Outflow Velocity } & \colhead{eqw   } & \colhead{ Flux } \\
 \colhead{} & \colhead{} &\colhead{ ($\mathring{A}$)}  & \colhead{ (erg/cm$^2$/s) }&\colhead{ ($\mathring{A}$)} &  \colhead\
{  (km/s) } &\colhead{ ($\mathring{A}$)}  & \colhead{ (erg/cm$^2$/s) } \\
}
\startdata
2009-11-30 & 2455166.33 & -17.5 & 1.9e-13 &  \nodata  &  \nodata  & -10.1 & 1.52e-13 \\
2009-12-17 & 2455183.21 & -6.88 & 1.38e-13 & 4.27 & -254 & -3.7 & 1.21e-13 \\
2009-12-21 & 2455187.27 & -7.77 & 1.83e-13 & 7.17 & -199 & -4.78 & 1.74e-13 \\
2009-12-29 & 2455195.22 & -6.82 & 1.69e-13 & 2.48 & -263 & -4.7 & 1.89e-13 \\

\enddata
 
\tablecomments{Table \ref{table:linefluxeqw} is published in its entirety in the electronic edition of the Astrophysical Journal. A portion is shown here for guidance regarding its form and content. Online version contains data of 28 lines.}

\end{deluxetable}


\end{document}